\documentclass[10pt]{iopart}

\usepackage{iopams}  

\newcommand{\bra}[1]{\left\langle#1\right|}
\newcommand{\ket}[1]{\left|#1\right\rangle}

\usepackage{epsfig,pstricks,graphicx}
\usepackage[mathscr]{eucal}
\usepackage{color}
\usepackage{comment}
\usepackage[normalem]{ulem}

\def\CC{{\rm\kern.24em \vrule width.04em height1.46ex depth-.07ex
\kern-.30em C}}
\def\id{{\rm 1\kern-.22em l}}

\begin{document}

\review[Quantifying entanglement resources
       ]
       {
        Quantifying entanglement resources
       }

\author{Christopher Eltschka}
\address{Institut f\"ur Theoretische Physik, 
         Universit\"at Regensburg, D-93040 Regensburg, Germany}
\ead{christopher.eltschka@physik.uni-regensburg.de}

\author{Jens Siewert}
\address{Departamento de Qu\'{\i}mica F\'{\i}sica, Universidad del 
         Pa\'{\i}s Vasco -- Euskal Herriko Unibertsitatea, 48080 Bilbao, Spain,
         and}
\address{IKERBASQUE, Basque Foundation for Science, 48011 Bilbao, Spain}
\ead{jens.siewert@ehu.es}

\begin{abstract}
We present an overview of the quantitative theory of 
single-copy entanglement
in finite-dimensional quantum systems. In particular we emphasize
the point of view that different entanglement measures
quantify different types of resources which leads to a natural
interdependence of entanglement classification and quantification.
Apart from the theoretical basis, we outline various methods for 
obtaining quantitative results on arbitrary mixed states.
\end{abstract}


\maketitle

\tableofcontents

\makeatletter
\@mkboth{Quantifying entanglement resources}{Quantifying entanglement resources}
\makeatother

\vspace*{1.5cm}

\section{Preamble}

The quantitative description of entanglement started with 
Bell's inequalites 50 years ago~\cite{Bell1964}. 
The next milestones from the theory point of view 
were Werner's seminal work on the 
precise mathematical characterization of entanglement for
mixed quantum states~\cite{Werner1989}, and the advent
of multipartite entanglement~\cite{Svetlichny1987,GHZ1989}. 
Since then, 
thousands of studies on quantum information have appeared 
for which entanglement is a central concept.
Therefore it may seem surprising that to date there is still no comprehensive 
quantitative theory of entanglement. Even worse, if a non-specialist
tries to get a quick overview how to quantitatively characterize
entanglement he will encounter a puzzling diversity of entanglement 
classifications and entanglement measures, based on different concepts
and methods without obvious connections between them.

However, the situation is much better than it might seem at first glance.
At least for the simplest subtopic of entanglement theory which comprises
the entanglement resources contained in a few systems with a finite number
of levels
there has been enormous progress during the past decade, so that
many features of the conceptual framework have already become apparent.
It is realistic to expect a reasonably complete theory of the basic
concepts within the next few years. 

In this review, we describe the state of affairs
in this subfield of quantitative entanglement theory.
Rather than mentioning  each and every detail,
our aim is to outline the well-established and commonly used concepts
for single-copy entanglement,
and to illuminate the logical connections between them. 
We put particular emphasis on the resource character of entanglement.
That is, entanglement is a necessary prerequisite to carry out a 
certain procedure or protocol. The more entanglement there is
(or the higher its quality is) the higher is the success probability 
of the protocol.
Different entanglement measures quantify distinct resources, and 
different resources require their specific entanglement measures.
The relation between measures, resources and the structure of the
quantum-mechanical state space is a central issue of this research.
It is needless to mention that at present we are able to mathematically 
distinguish many more types
of entanglement than we know protocols for which they might serve
as a resource.

More technically, we explain the basic concepts of the quantitative
theory for single-copy entanglement and then sketch the structure of 
entanglement
between two finite-dimensional systems as well as common entanglement
measures and methods how to evaluate (or estimate) them for arbitrary states.
Further, we discuss multipartite entanglement, in particular of three
and more qubits, and the polynomial measures that distinguish different
types of genuine multipartite entanglement. 

Due to this narrow focus, there are many topics that cannot be discussed
or even be mentioned. In particular, this article {\em does not} review
\begin{itemize}
\item
{\em asymptotic entanglement measures and protocols}, which we
consider very briefly, however, essentially only to mark down the
field of single-copy entanglement. Asymptotic measures and resources are
discussed in detail in the review by Plenio and Virmani~\cite{Plenio2007}, 
as well as in the review by the Horodecki family~\cite{Horodecki2009},

\item
{\em the relation between entanglement and nonlocality}. All relevant aspects
of this topic are extensively discussed by Brunner {\em et al.} in 
\cite{Brunner2013}. There is also a shorter overview of the main concepts 
                    by Werner and Wolf~\cite{Werner2001},

\item
{\em the entanglement of infinitely many degrees of freedom (continuous
     variables)}. There are up-to-date overviews of
     the relevant concepts by Adesso and Illuminati~\cite{Adesso2007},
     and by Braunstein and van Loock~\cite{Braunstein2005RMP},

\item
{\em entanglement for indistinguishible particles and in many-body systems.}
    There are two recent reviews of these subjects by 
    Tichy {\em et al.}~\cite{Tichy2011}
    and by
    Amico {\em et al.}~\cite{Amico2008},

\item
{\em entanglement for relativistic particles}. These concepts were reviewed, e.g.,
    by Alsing and Fuentes~\cite{Fuentes2012} as well as by Peres and 
    Terno~\cite{Peres2004}.
\end{itemize}

Apart from the sources mentioned so far there are excellent 
texts on quantum information and entanglement, e.g., 
A.\ Peres' textbook~\cite{Peres1993},
J.\ Preskill's lecture notes on quantum computation~\cite{Preskill1998}, 
the textbook by M.A.\ Nielsen and I.L.\ Chuang~\cite{Nielsen2000},
lecture notes by M.M.\ Wolf~\cite{MMW}, and the
textbook by I.\ Bengtsson and K.\ Zyczkowski~\cite{Bengtsson2006}.

\section{Entanglement as a resource}
\label{sec:resource}

\subsection{The LOCC paradigm}
\label{sec:defent}

At the very basis of entanglement theory lies the paradigm of Local
Operations and Classical Communication (LOCC) formulated by Bennett
\textit{et al.} \cite{Bennett1996}. Under this paradigm, a state is
distributed among different parties who can perform arbitrary local
operations (including measurements and operations involving additional
local systems, so-called ancillas), and in addition can communicate
with each other over a normal classical channel, however, they are not
able to exchange quantum systems. Under those restrictions, not all
transformations of states are possible. Especially it is not possible
to create arbitrary states from scratch this way. This gives rise to
the basic definition: A state is \textit{separable} if it can be
created using only local operations and classical communications, and
it is called \textit{entangled} otherwise \cite{Plenio1998,Bruss2002}.

The fact that entangled states cannot be generated locally makes them
a resource \cite{Lo1999}. This resource is used in different tasks in
quantum computation \cite{DiVincenzo1995}, quantum communication
\cite{Gisin2007} and quantum cryptography \cite{Gisin2002}. Those
tasks are usually written in the form of a protocol which uses the
entanglement as resource. While quantum computation protocols are
normally not written in this form, but instead in the language of
quantum gates, any quantum computation can also be done using
\textit{measurement-based quantum computation} \cite{Raussendorf2003},
also known as \textit{one-way quantum computation}. In this form, it
uses an LOCC protocol as well (where the local operations are
projective measurements) that consumes an entangled state, typically a
cluster state, as resource. However the question if entanglement is a
necessary resource for quantum computation is still open.

The resource character of entanglement immediately leads to two
questions. The first one is obvious: \emph{How much} of that resource
do we have? This is the question of \emph{entanglement
  quantification.} Answering this question leads to so-called
\textit{entanglement monotones} \cite{Vidal2000} or
\textit{entanglement measures}
\cite{Schlienz1995,Bennett1996,Vedral1997}, which are the topic of
this review. However, as we will see, entanglement monotones are not
only useful to answer this question, but also for answering the second
one.

The second question most naturally arises in situations where there
are more than two parties, so-called \textit{multipartite states}, but
turns out to be relevant even in \textit{bipartite states} (just two
parties): \emph{How many different types} of this resource do we
have? This is the question of \emph{entanglement classification.}

``The same resource'' in the previous paragraph means that we can use
those states as resource for the same tasks, although possibly with
lower efficiency. Obviously if we can convert the state
$\left|\psi\right>$ into the state $\left|\phi\right>$ using LOCC, then for
any task that works starting with the state $\left|\phi\right>$, also
works starting with the state $\left|\psi\right>$: Just convert
$\left|\psi\right>$ to $\left|\phi\right>$ and then run the corresponding
protocol on $\left|\phi\right>$. Also note that this conversion need not
be successful every time; as long as it is with non-vanishing
probability, we can still use it for the protocol, just with lower
efficiency: Only in those cases where the conversion succeeds, so does
the protocol.

Those operations which can be performed using LOCC but may fail are
known as \textit{Stochastic Local Operations and Classical
  Communication} (SLOCC)
\cite{VidalTarrach1999,Bennett2000,DVC2000,Vidal2000,Verstraete2001}.
Especially, states each of which can be transformed into one another,
are called \textit{SLOCC equivalent.} SLOCC
equivalent states therefore contain the same resource. The equivalence
classes under SLOCC equivalence are also called \textit{SLOCC classes}
or \textit{entanglement classes.} They are obviously invariant under
invertible SLOCC transformations.

Note that for a specific task, it is well possible that states from
different SLOCC classes can be used to perform it. The most obvious
example is when conversion by SLOCC is only possible one way, where
the given protocol works on the destination state. Therefore if one is
interested only in specific tasks, it makes sense to apply some
\emph{coarse graining} to the SLOCC classes which only distinguishes
the entanglement properties of interest. To distinguish the classes of
such a classification from the SLOCC classes, we speak of
\textit{families} in that case. Of course such a family is also
invariant under invertible SLOCC transformations, since any SLOCC
class is contained in it either completely or not at all.

\subsection{Single copy versus asymptotic entanglement properties}
\label{sec:scasym}

There is one point which we glossed over in the previous section:
Namely the question of what comprises a specific task. Here, two
different approaches exist. The first is to take a \emph{single} state
at a time, and operate on that. This approach is therefore called
\emph{single copy.} When looking at entanglement this way, the
relevant questions are whether one can convert a single copy of state
$\left|\psi\right>$ into a single copy of state $\left|\phi\right>$, and the
probability to do so. A typical protocol adhering to this restriction
is superdense coding~\cite{Bennett1992}: It uses a single shared Bell
state to allow the transmission of two classical qubits. To transfer
two additional qubits, an additional shared Bell pair can be used.
However, there is no operation involving both Bell pairs. Another such
protocol is quantum teleportation~\cite{Bennett1993} where each shared
Bell pair individually is used to teleport one qubit, and nothing
produced in that transmission (not even the classical information) is
used for transmission of further qubits.

The other approach, which is more in line with classical information
theory, is to take an unlimited number of copies of a state as
resource, and allow local transformations between those copies. In this
case, the relevant quantities are \emph{asymptotic} quantities (for
example, the number of generated systems in state $\left|\phi\right>$ per
initial system in state $\left|\psi\right>$, in the limit where the
number of initial systems goes to infinity). 
Also, it allows that the
final state is only approximated, as long as the error can be made
arbitrarily small. Typical protocols using this approach are quantum
compression \cite{Jozsa1994, Barnum1996} and entanglement distillation
protocols \cite{Bennett1996c}.

It is obvious that any single-copy protocol is also an asymptotic
protocol, therefore any transformation which is possible in the
single-copy is also possible asymptotically. However the reverse is
not true. For example, for three qubits, there are two inequivalent
types of three-qubit entanglement \cite{DVC2000}, called GHZ-type and
$W$-type entanglement, named after prominent states contained in those
classes, namely the Greenberger-Horne-Zeilinger (GHZ) state
$\left|\mathrm{GHZ}\right> = (\left|000\right> +
\left|111\right>)/\sqrt2$ resp. the $W$ state $\left|W\right> =
(\left|001\right> + \left|010\right> + \left|100\right>)/\sqrt3$. In
single-copy protocols, neither of those states can be transformed into
each other using SLOCC. However asymptotic protocols are able to
transform GHZ states into $W$ states and vice versa
\cite{Ji2004,Chitambar2008,Yu2014}.

This relation means that any entanglement classification based on
asymptotic protocols is also a coarse-grained single-copy
classification. Moreover it means that any entanglement measure
suitable for asymptotic protocols is also suitable for single-copy
protocols.

In this review, we adopt the single-copy viewpoint.

\subsection{Mathematical description of SLOCC operations and SLOCC equivalence}
\label{sec:SLOCC}

The basis of the mathematical description of SLOCC operations is the
concept of an instrument \cite{Davies1970,Ziman2008}. An instrument can be
thought of as a set of quantum channels $T_i$ with associated probability
$p_i$ that the corresponding channel is selected, and a
classical output giving the selected output. Since classical
communication is allowed, there is no need to explicitly distinguish
between the knowledge on different sites.

Every quantum channel has a Kraus decomposition
\cite{Hellwig1970,Kraus1983}, that is, any quantum channel $T$ can be
written as
\begin{equation}
  \label{eq:krausdecomposition}
  T(\rho) = \sum_k G_k \rho G_k^{\dagger} \mbox{\ with\ } 
            \sum_k G_k^\dagger G_k = \id
\end{equation}
where $\id$ is the identity operator. The operators $G_i$ are
called the Kraus operators of the channel. Now it is easy to see that
$T_k(\rho) = G_k \rho G_k^{\dagger} / 
              \tr(G_k \rho G_k^{\dagger})$ 
is also a (very special)
channel. Thus the channel can be modelled as applying the special
instrument giving the ``one-operator channel'' $T_k$ with probability
$p_k$ and subsequently discarding the result of the instrument. This
in turn means that any instrument can be considered such a
fine-grained instrument, followed by discarding part of the classical
information, which essentially causes a mixture of the corresponding
output states. Note that the one-operator channels always map pure
states onto pure states.

Since we are interested in local operations, we also need local
instruments. A local instrument only acts on a single subsystem,
therefore its Kraus decomposition consists of the tensor product of a
valid Kraus operator acting on that system and identities acting on
all the other systems.

An SLOCC protocol is built of a sequence of local instrument
applications, where the information gained from previous instruments
can be used to select further instruments. Unlike an LOCC protocol, an
SLOCC protocol may \emph{fail,} that is, it suffices if \emph{one} of
the possible outcomes is the desired one. The obtained classical
information therefore does not only tell which step to do next, but
also whether the protocol succeeded or failed. In the end, this means
that a state can be reached from another state by SLOCC iff there is a
local channel with postselection from one state to the other.

As described above, a single term of the Kraus decomposition defines,
after normalization, a channel by itself. However, for SLOCC
operations, those terms, while employing only local operators, are in
general not themselves describing local channels, that is, it is in
general not possible to implement that channel locally \emph{without
  postselection.} Only if the transformation can be done by LOCC
(that is, with certainty), the channel is local. Such a channel that
can be implemented with SLOCC through postselecion has been termed
local filtering
operation~\cite{Gisin1996,Verstraete2001,VerstraeteWolf2002}.

Note that local unitary transformations are also a specific type of
local channel.

As we have seen above, discarding information is equivalent to mixing
of the corresponding output states~\cite{Plenio2005}. However it has
to be stressed that for local operations, not \emph{all} mixtures can
be produced that way, since it is necessary to be able to produce each
of the states with a sequence of local channels from the original
state. However note that mixing with a separable state is always
possible because those can, by definition, be created from scratch
with SLOCC.

When looking at channels converting pure states to pure states, it is
easy to see that those have only one Kraus operator. Since 
concatenation of
local channels means to form products of the channels' Kraus
operators, this implies a simple rule to decide
whether one pure state can be
transformed into another~\cite{DVC2000}: The state $\left|\psi\right>$
can be transformed into the state $\left|\phi\right>$ if there is an
operator $G=G_1\otimes\ldots\otimes G_n$ (where the tensor product factors are for the
different subsystems) so that $G\left|\psi\right>=\left|\phi\right>$.

Obviously this leads also to a simple criterion of SLOCC equivalence
of pure states: The operator $G$ has to be invertible~\cite{DVC2000}.
In other words:

Two pure states $\left|\psi\right>$ and $\left|\phi\right>$ living in the
multipartite Hilbert space $\mathcal{H}_1\otimes\ldots\otimes\mathcal{H}_n$ are SLOCC
equivalent iff there exists an operator
$G\in \mathrm{GL}(\mathcal{H}_1)\otimes\ldots\otimes \mathrm{GL}(\mathcal{H}_n)$ so that
$G\left|\psi\right>=\left|\phi\right>$. We refer to such an operation as
local $\mathrm{GL}$ operation.

Obviously a local $\mathrm{GL}$ operation will map product states to
product states; therefore no non-product state can be produced from a
product state. On the other hand, obviously a pure product state can
be created locally from an arbitrary state by just doing a complete
projective measurement followed by an unitary operation depending on
the outcome.

The total set of separable states is therefore mathematically
characterized as the mixtures of product states \cite{Werner1989}.

\section{Properties of entanglement measures}
\label{sec:measures}

This section is about what conditions entanglement measures fulfil.
The first subsection (\ref{sec:mainprop}) gives the properties every
function needs to have in order to be called an entanglement measure
or an entanglement monotone, while the following sections describe
additional restrictions which one may require. Note that today, the
terms ``entanglement measure'' and ``entanglement monotone'' are
usually used synonymously for anything fulfilling the two properties
described in section \ref{sec:mainprop}.

\subsection{Main properties}
\label{sec:mainprop}

Given that the defining property of entanglement is that it cannot be
produced by SLOCC, at first it seems obvious that an entanglement
measure should be strictly non-increasing under SLOCC. Such measures
indeed do exist; one example is the Schmidt rank of bipartite states
\cite{Nielsen1999,Vidal1999} (see also section \ref{sec:schmidtrank}).
However while those measures can be quite useful for classification of
entanglement, they do not adequately quantify their resource
character. This is because, by definition, they are constant on every
SLOCC class. However not all states in a given SLOCC class are equally
good resources. For example, all states of the form $\alpha\left|00\right>
+ \beta\left|11\right>$ where $\alpha\neq0$ and $\beta\neq0$ are SLOCC equivalent, but
for example with superdense coding, only in the case $\alpha=\beta$ two
classical bits can be correctly transmitted with probability $1$ (up
to experimental errors).

For that reason, one only demands that the entanglement measure does
not increase \emph{on average} under SLOCC
\cite{Vedral1998,Vidal2000}. Not increasing  on 
average means that
if the protocol transforms the initial state $\rho$ into the final state
$\rho_n$ with probability $p_n$, then the measure $\mu$ must fulfil the
inequality
\begin{equation}
  \label{eq:monotonecondition}
  \sum_n p_n \mu(\rho_n) \leq \mu(\rho)
\end{equation}
Since such measures are necessarily constant on the
set of separable states, usually the trivial additional constraint
\begin{equation}
  \label{eq:separablemonotone}
  \rho \mbox{ separable } \Longrightarrow \mu(\rho) = 0
\end{equation}
is added.

Note that those properties imply that the set of states on which the
measure vanishes is invariant under SLOCC operations. That is, every
entanglement measure already implies a rough classification of states
into two classes: Those where the entanglement measure vanishes (which
includes all separable states, but also may include certain entangled
states), and those where it  does not vanish.

Sometimes authors only impose the even weaker condition that the
  measure has to decrease under (deterministic) LOCC.

\subsection{Convexity and the convex roof}
\label{sec:convexity}

Another common  requirement for 
entanglement measures is \textit{convexity}
\cite{Vidal2000}:
\begin{equation}
  \label{eq:convexity}
  p_1 + p_2 = 1 \Longrightarrow \mu(p_1\rho_1 + p_2\rho_2) \leq p_1\mu(\rho_1) + p_2\mu(\rho_2)
\end{equation}
Physically it means that mixing two states should never increase
entanglement. It seems intuitive because mixing certainly looks like a
local operation (and is indeed classified as local operation by
Vidal). Measures fulfilling conditions \eref{eq:monotonecondition},
\eref{eq:separablemonotone} and \eref{eq:convexity} were called
entanglement monotones by Vidal \cite{Vidal2000}.

Certainly whenever there is an SLOCC protocol to
generate the state $\rho_1$, and another SLOCC protocol to generate the
state $\rho_2$, there also exists an SLOCC protocol to generate any
mixture of both states, by randomly executing one of the two protocols
(with appropriately chosen probabilities), and then discarding the
information which of the protocols has been run. However, that is an
SLOCC operation from the original state to the mixture, not a mixing
operation by itself. Also, if \emph{both} states are available at the
same time, then of course it is easy to mix them by randomly selecting
one of them, discarding the other, and discarding the information
which one was chosen. However in that case, the mixing starts from the
product state $\rho_1\otimes\rho_2$, not from $\rho_1$ or $\rho_2$. Therefore convexity
is not a strictly necessary condition for entanglement measures.
Indeed, there are entanglement measures which are not convex, like the
logarithmic negativity \cite{Plenio2005}. Today, the term
``entanglement monotone'' is generally used for all measures
fulfilling conditions \eref{eq:monotonecondition} and
\eref{eq:separablemonotone}.

It is easy to see from \eref{eq:monotonecondition} and
\eref{eq:separablemonotone} that if $\mu_1(\rho)$ and $\mu_2(\rho)$ are
entanglement monotones, then the minimum of both, $\mu(\rho) =
\min\{\mu_1(\rho),\mu_2(\rho)\}$, is also an entanglement monotone. However, in
general the minimum of two convex functions is not itself a convex
function. For example, in the multipartite case, to be discussed
later, a bipartite entanglement measure may be applied to different
bipartitions. Then the minimum of that measure will also be an
entanglement measure, but in general it will not be convex, even if the
bipartite measure is.

Nonetheless, often convexity is taken as an additional 
requirement due to
the interpretation of mixing as loss of information, which of course
should not increase entanglement.
A general way to construct convex entanglement monotones is the convex
roof extension \cite{Uhlmann1998}. The convex roof extension takes a
measure $\mu$ that is defined only on the pure states and extends it to
the mixed states as
\begin{equation}
  \label{eq:convexroof}
  \mu(\rho) = \min_{\mathrm{decompositions}} \sum_i p_i \mu(\psi_i)
\end{equation}
where the minimum goes over all decompositions of $\rho$, that is, over
all sets $\{(p_i,\psi_i)\}$ so that $\sum_ip_i=1$ and
$\sum_ip_i\left|\psi_i\middle>\middle<\psi_i\right|=\rho$. Note that this
construction can be used to produce new measures $\tilde \mu$ for
measures $\mu$ which are also defined on mixed states, by defining for
pure states $\tilde \mu(\psi) = \mu(\psi)$ and convex roof-extending $\tilde \mu$.
For example, the negativity (see section \ref{sec:negativity}) is
defined for all bipartite mixed states, but its convex roof gives a
different measure known as the convex roof extended negativity (CREN)
(see section \ref{sec:cren}).

An important property of the convex roof is that it is the largest
convex function which agrees with the original function on the pure
states \cite{Uhlmann1998,Uhlmann2010}.

The convex roof extension
 has the advantage that it automatically
constructs a convex entanglement monotone on the mixed states from an
entanglement monotone defined only on the pure states. However it has
the disadvantage that it is in general hard to compute.

\subsection{Homogeneity and SL invariance}
\label{sec:hominv}

The close connection between SLOCC operations and local GL
transformations noted in section~\ref{sec:SLOCC} suggests to add
related conditions on the measures. Invariance under local GL
operations is obviously too strong, since it would only allow measures
which are invariant under SLOCC transformations. However by splitting
the GL transformations into SL transformations, that is invertible
transformations of determinant $1$, and a single prefactor, two very
useful conditions can be imposed.

The first condition is homogeneity. It means that for any state $\rho$
and any positive number $\lambda$, the measure $\mu$ has the property
\begin{equation}
  \label{eq:homogeneity}
  \mu(\lambda\rho) = \lambda^\alpha \mu(\rho)
\end{equation}
with some exponent $\alpha$, called the degree of the 
homogeneity. The same
condition can also be written down for pure states $\left|\psi\right>$,
but one has to be careful: The density matrix for the state
$\left|\psi\right>$ is the projector $\left|\psi\middle>\middle<\psi\right|$,
which is \emph{quadratic} in $\psi$. Therefore any measure which is
homogeneous of degree $n$ in the state vector is homogeneous of degree
$n/2$ in the density matrix. Therefore it is always important which
quantity the degree refers to.

Note that for convex roof extended measures, homogeneity on the pure
states automatically implies homogeneity on the mixed states as well.

The second condition is invariance under transformations $S=S_1\otimes\ldots\otimes S_n \in
\mathrm{SL}(d_1,\mathbb{C})\otimes\ldots\otimes\mathrm{SL}(d_n,\mathbb{C})$ 
where $S_j$ acts on the $j$-th 
subsystem and $\det S_j=1$. We refer to those
transformations as local SL transformations (or, for short,
LSL transformations). 
That is, if $S$ is a
local SL transformation, then
\begin{equation}
  \label{eq:SLinvariance}
  \mu(S\rho S^\dagger) = \mu(\rho).
\end{equation}

For convex roof extended measures, invariance under local SL
transformations of the measure on pure states generally does
\emph{not} imply invariance on the mixed states; however if the
measure is of homogeneous degree $1$ in the density matrix (degree $2$
in the state vector), the local SL invariance is carried over to the
mixed states as well \cite{Viehmann2012}.

Verstraete \etal~\cite{Verstraete2003} have shown that any local SL
invariant, convex roof extended measure of homogeneous degree $1$ is
automatically an entanglement monotone. Indeed, for systems of qubits,
any homogeneous measure in the state vector which is local SL
invariant on \emph{pure states} gives an entanglement monotone if and
only if the homogeneity degree in the state vector is nonnegative and
not larger than $4$ \cite{Eltschka2012PRA}.

Another reason why these two conditions are very useful is that there
exist methods to systematically build measures fulfilling 
them, based on local SL invariant polynomials in the state
coefficients~\cite{Verstraete2003,Osterloh2005,Gour2013}, see section
\ref{sec:inv2and4}.

\subsection{Dimension-independence and additivity}
\label{sec:additivity}

There are other common requirements for entanglement measures, which
are related to their use
for asymptotic protocols. The first one is that 
the measure must not depend on the
Hilbert space dimension: that is, the very same measure
can be applied to systems of arbitrary Hilbert space dimension, and
will give the same result for the same state embedded in a larger
Hilbert space. This is important for asymptotic protocols because they
do not work only on the state $\rho$, but on the state $\rho^{\otimes N}$ for the
limit of large $N$. Therefore to make any statements 
regarding such
measures, they have to be applicable for arbitrary $N$. Consequently,
all asymptotic measures are dimension-independent, but also the
$I$-concurrence (if defined with a dimension-independent prefactor), as
well as the negativity and related measures.

The second requirement, which rests on the first, is additivity. This
is the requirement that
\begin{equation}
  \label{eq:additivity}
  \mu(\rho\otimes\sigma) = \mu(\rho) + \mu(\sigma)
\end{equation}
where, in this case,
 the tensor product is not between different subsystems, but
between different states shared by the parties. However this equation
can be hard to prove. Generally, a less strict inequality can be
proved, the subadditivity \cite{Araki1970}
\begin{equation}
  \label{eq:subadditivity}
  \mu(\rho\otimes\sigma) \leq \mu(\rho) + \mu(\sigma)
\end{equation}
One quantity which is known to be subadditive
is the entanglement of formation.

\section{Connections between entanglement measures and other important concepts}
\label{sec:connections}

\subsection{Normal form}
\label{sec:normalform}

An important concept in entanglement is Verstraete's normal form
\cite{Verstraete2003}. The normal form of a state $\rho$ is a (generally
not normalized) state in the closure of the local SL orbit of $\rho$
whose reduced density matrices are all multiples of the unit matrix.
That is, there exists a set of local SL matrices $S(t)$ parameterized
by $t$ such that
\begin{equation}
  \label{eq:normalform}
  \rho_{\mathrm{NF}} = \lim_{t\to\infty} S(t)\rho S^\dagger(t),\ 
  \tr_{BC\ldots}\rho_{\mathrm{NF}}=\lambda\id_A,\ 
  \tr_{AC\ldots}\rho_{\mathrm{NF}}=\lambda\id_B,\ \ldots
\end{equation}
where $\id_X$ is the unit operator for system $X$.

In most cases, the normal form of a state is LSL-equivalent 
to the original
state, but in some cases the limit is explicitly needed. This is
especially the case for states where the normal form is zero.

There exists an explicit and efficient iterative algorithm to
calculate the normal form (or, in the case that it is only reached
asymptotically, a close approximation for it), which is also given in
\cite{Verstraete2003}.

All local SL invariant monotones reach their maximum value on a pure
state in normal form. This especially means that states whose normal
form is zero cannot  have their entanglement quantified
by any local SL invariant measure.

For pure states, also the reverse is true: If the normal form is
non-zero, the state is measured by at least one LSL-invariant
entanglement monotone. For mixed states this is not true, as can be
seen by the fact that the completely mixed state is already in normal
form, but as a separable state obviously cannot be measured by any
entanglement measure.

Since the normal form is obtained using local SL operations, 
it can be used
to calculate/estimate the value of homogeneous LSL-invariant measures,
as long as the value/an estimate is known on the normalized state for
the corresponding normal form: If the monotone $\mu$ is homogeneous of
degree $\alpha$ in the density matrix, then
\begin{equation}
  \label{eq:measurenormalform}
  \mu(\rho) = (\tr \rho_{\mathrm{NF}})^\alpha \mu\left(\frac{\rho_{\mathrm{NF}}}{\tr \rho_{\mathrm{NF}}}\right).
\end{equation}

\subsection{Entanglement witnesses}
\label{sec:witnesses}

An \textit{entanglement witness}
\cite{Horodecki1996,Terhal2000PLA,Lewenstein2000} is an observable
which has nonnegative expectation value on all separable states, but
a negative expectation value on at least some entangled state. A state
is said to be \textit{detected} by a witness if it has a negative
expectation value. While any detected state is, by construction,
entangled, the reverse is not true: For any entanglement witness,
there exist entangled states it does not
detect. However for every
entangled state, there exists an entanglement witness which detects it
\cite{Terhal2000PLA}. An extensive overview on entanglement detection
by witnesses was given by G{\"u}hne and
T{\'o}th~\cite{GuhneTothReview2009}.

An entanglement witness is called optimal, if there
is no entanglement witness which detects a proper superset of the
states detected by it.

It is also possible to define class-specific entanglement witnesses,
like Schmidt number witnesses \cite{Sanpera2001} or witnesses for
GHZ-type entanglement \cite{Acin2001}.

Another important concept is optimality relative to a subset of states
\cite{Lewenstein2000,Hulpke2004,Eltschka2013}, 
where only a subset of states is
considered for detecting optimality 
(of course 
the witness must not detect {\em any} 
unentangled states, including those outside
that subset -- otherwise it would not be an entanglement witness).

Entanglement witnesses are ultimately a geometric concept, since they
split the space of bounded operators (and especially the set of
positive operators of trace $1$, that is, the density matrices) into
two half-spaces, one positive and one negative. The hyperplane of zero
expectation value is a supporting plane of the set of unentangled
states iff the witness is optimal.

Entanglement detection through witnesses is particularly useful for
the assessment of experimental entanglement generation (cf., e.g.,
recent experiments such as
Refs.~\cite{JianWeiPan8bits,Blatt2011,Blatt2013}). In such experiments
the full density matrix (or parts of it) are determined. The question
whether experimental data of this kind are compatible with the
presence of entanglement was raised early on, e.g., in
Refs.~\cite{Horodecki-Jaynes1999,Audenaert2006}.

A general theory of quantifying entanglement by means of entanglement
witnesses was developed by Brandao and
coworkers~\cite{Brandao2005,Eisert2007} and by G\"uhne {\em et
  al.}~\cite{GuehneReimpellWerner2007,GuehneReimpellWerner2008}. For
example, given two arbitrary positive numbers $m$ and $n$, the
function
\begin{equation}
  \label{eq:brandaomeasure}
  E_{n:m} = \max \{0, -\min_{W \in \mathcal{M}_{n:m}} \tr{W\rho}\}
\end{equation}
is an entanglement monotone, where $\mathcal{M}_{n:m}$ is the set of
all entanglement witnesses $W$ fulfilling $-n\id \leq W \leq
m\id$~\cite{Brandao2005}.

As is discussed in section \ref{sec:threewit} also the converse is
true: entanglement measures (e.g., polynomial invariants) may be used
to derive well-known entanglement witnesses~\cite{Eltschka2012SciRep}.

\section{Bipartite entanglement}
\label{sec:bipartite}


\subsection{Schmidt decomposition and SLOCC classes}
We consider a quantum system consisting of
two subsystems $A$ and $B$ with dimensions
$\dim \mathcal{H}_A=d$ and $\dim \mathcal{H}_B=d'$, 
so that the Hilbert space of the composite system is 
$\mathcal{H}\equiv\mathcal{H}_A\otimes\mathcal{H}_B$. 
We will also call $\mathcal{H}$ a $d\times d'$-dimensional system.
Henceforth we assume $d\leqq d'$. With orthonormal bases
$\{\ket{a}\}$, $\{\ket{b}\}$ a  pure state 
$\psi\in \mathcal{H}_A\otimes\mathcal{H}_B$ can be written as
\begin{equation}
   \ket{\psi}\ =\ \sum_{a=1}^{d}\sum_{b=1}^{d'} \psi_{ab}\ket{a}\otimes\ket{b}\
               \equiv\ \sum_{a,b} \psi_{ab}\ket{ab}
\ \ .
\end{equation}
Pure bipartite states have the important property that there
are always orthonormal bases $\{\ket{j_A}\}$, $\{\ket{j_B}\}$
such that~\cite{Preskill1998,Nielsen2000}
\begin{equation}
   \ket{\psi}\ =\ \sum_{j =1}^{r(\psi)}\ 
              \sqrt{\lambda_j}\ket{jj}
\label{eq:Schmidt}
\end{equation}
with real positive numbers  $\lambda_{j}$, the so-called Schmidt
coefficients. The Schmidt rank $r(\psi)\leqq d$ corresponds
to the rank of the reduced density matrices of the subsystems
$r(\psi)=\mathrm{rank}(\tr_B\ket{\psi}\!\bra{\psi})
        =\mathrm{rank}(\tr_A\ket{\psi}\!\bra{\psi})$.
With this we note that a pure bipartite state is separable if its
Schmidt rank equals 1, otherwise it is entangled.
In particular, we may define the maximally entangled state
in $d$ dimensions
\begin{equation}
   \ket{\Psi_d}\ =\ \sum_{j =1}^d\ \frac{1}{\sqrt{d}}\ket{jj}
\ \ .
\label{eq:Bell-d}
\end{equation}

Evidently the rank of the reduced state
$\rho_A=\tr_B\ket{\psi}\!\bra{\psi}$
does not change under arbitrary invertible operations 
$A\in \mathrm{GL}(d,\CC)$ (and analogously for
subsystem $B$), so that $r(\psi)$ is an entanglement 
monotone~\cite{Nielsen1999,Vidal1999}. 
Hence, there are $d$ different SLOCC classes for
$\psi\in \mathcal{H}_A\otimes\mathcal{H}_B$, each of which is characterized 
by its Schmidt rank. 

A mixed state $\rho$ of the composite system is represented by a positive definite 
bounded Hermitian operator acting on the vectors $\psi\in\mathcal{H}$, i.e.,
$\rho\in\mathcal{B}(\mathcal{H})$. It has a decomposition 
into pure states $\{(p_j,\psi_j)\}$
\begin{equation}
           \rho\ =\ \sum_{j=1}^{\ell}\ p_j\ \pi_{\psi_j}\ \ \ , \ \
           \mathrm{with}\ \ \pi_{\psi_j}\ \equiv\ \ket{\psi_j}\!\bra{\psi_j}
\label{eq:decomp-rho}
\end{equation}
where $\ell \geqq \mathrm{rank}(\rho)$ is called the length of the decomposition.
The weights $p_j>0$ obey $\sum_j p_j=1$ and $\tr\pi_{\psi_j}=1$, 
if not stated otherwise.
That is, a mixed state can be regarded as a convex combination of
pure states.
Note that there are infinitely many ways to decompose a 
state~\cite{Schroedinger1936,HJW1993}: Given  $\{(p_j,\psi_j)\}$ 
and a unitary matrix $U$ with at least $\ell$ columns we find
another decomposition $\{(q_k, \varphi_k )\}$ where
\[
 \ket{\varphi_k}=\frac{1}{\sqrt{q_k}}\ket{\tilde{\varphi}_k}\ , \ \
  q_k=\langle\tilde{\varphi_k}|\tilde{\varphi}_k\rangle\ ,
\ \ \mathrm{and}\ \
 \ket{\tilde{\varphi}_k}  =  \sum_{j=1}^{\ell} U_{kj}\sqrt{p_j}\ket{\psi_j}
\ \ .
\]
This ambiguity  is at the origin of many difficulties in entanglement
theory.

As to the entanglement classes of bipartite states, we generalize 
the Schmidt rank to mixed states in a spirit similar to the convex roof:
The Schmidt number~\cite{Terhal2000PRA} is the smallest possible 
maximal Schmidt rank occurring in
any pure-state decomposition  of $\rho$
\begin{equation}
    r(\rho)\ =\ \min_{\{(p_j,\psi_j)\}} \max_j\ r(\psi_j)
\ \ .
\label{eq:Schmidt-number}
\end{equation}
As opposed to the convex roof, the maximum Schmidt rank gets
minimized, not the average.
Also $r(\rho)$ is an entanglement monotone~\cite{Terhal2000PRA},
see also~\cite{Sperling2011}. 
In particular,
we say $\rho$ is separable if $r(\rho)=1$, that is, if it can 
be decomposed into pure product states~\cite{Werner1989}
\begin{equation}
   \rho\ \ \mathrm{separable}\ \ \ \Longleftrightarrow\ \ \
\rho\ =\ \sum_{j=1}^{\ell} p_j\ \pi_{a_j}\otimes\pi_{b_j}
\end{equation}
with $a_j\in\mathcal{H}_A$ and $b_j\in\mathcal{H}_B$.

The states of a given Schmidt number $k$
form a compact convex set $\mathcal{S}_k$ which on their part build a hierarchy
$\mathcal{S}_1\subset\mathcal{S}_2\subset\ldots\subset\mathcal{S}_d$
\cite{Sanpera2001}. 
%
%
This hierarchy describes an SLOCC classification for the states 
of a bipartite system.
However, it is not the only one. Another example is the classification
of bipartite states with respect to the sign of the partial transpose 
(i.e., whether or not the partial transpose has negative eigenvalues). 
Note that this alternative classification
has little in common with the one based on Schmidt numbers: The 
class $\mathcal{S}_1$ belongs entirely to the PPT class and, moreover,
there is the conjecture that the class $\mathcal{S}_d$ does not contain
PPT-entangled states~\cite{Sanpera2001}.

These considerations provide a clear illustration of the fact 
that there is no such
thing like ``the'' SLOCC classification of a given system. 
Basically any SLOCC-invariant criterion (or a combination 
of several criteria)
induces an SLOCC classification.  The question is whether or not the 
criterion is appropriate to characterize a certain resource.

The concept of Schmidt decomposition can also be applied to mixed
states. In that case, except for $d=d'=2$, it is not possible to transform
the state into a state-independent basis. However with SL
transformations it is possible to transform it into the form
\begin{equation}
  \label{eq:leinaasnf}
  \rho = \frac{1}{dd'}\left(\id_d\otimes\id_{d'} + \sum_k\xi_kJ_k^A\otimes J_k^B\right)
\end{equation}
where the $J_k^X$ are traceless \cite{Leinaas2006}. Note that
\eref{eq:leinaasnf} is a special representation of the normal form (see
section \ref{sec:normalform}).

%
%


\subsection{The most important bipartite entanglement measures}
\label{sec:bipartitemeasures}

In this section, we look at the most important bipartite entanglement
measures for single copies.

\subsubsection{The Schmidt number}
\label{sec:schmidtrank}

The \textit{Schmidt number} \eref{eq:Schmidt-number} is an entanglement
monotone which is \emph{strictly} nonincreasing under SLOCC. As
explained in section \ref{sec:mainprop}, this implies that it does not
quantify a resource, but it allows classification of the entanglement.
Indeed, for pure states this classification is complete, that is, two
pure states are SLOCC equivalent iff they have the same Schmidt rank.
For mixed states, there exist entanglement criteria like bound
entanglement which are not covered by the Schmidt number.

Note that strictly speaking, one has to subtract $1$ from the
Schmidt number in order to get an entanglement monotone, since
otherwise it does not fulfil condition~\eref{eq:separablemonotone},
section~\ref{sec:mainprop}.

Sometimes, especially when studying asymptotic properties, the
logarithm of the Schmidt rank is used. This is because the Schmidt
rank of tensor products is the product of the Schmidt ranks, and
therefore the logarithm of the Schmidt rank is additive.

A generalization of the Schmidt number to multipartite states
(as the minimum number of product components) was studied in 
Ref.~\cite{Eisert2001}.

\subsubsection{The $k$-concurrences}
\label{sec:kconcurrence}

Gour \cite{Gour2005} introduced a hierarchy of entanglement monotones
for $d\times d$ systems, which measure Schmidt-rank specific entanglement,
called \textit{$k$-concurrence} (where $2\leq k\leq d$). The $k$-concurrence
is defined for pure states as
\begin{equation}
  \label{eq:kconcurrence}
  C_k(\psi) =
  N_k^{(d)}\left(\sum_{i_1<i_2<\ldots<i_k}\lambda_{i_1}\lambda_{i_2}\cdots\lambda_{i_k}\right)^{1/k},\quad 
  N_k^{(d)} = d{d \choose k}^{-1/k}
\end{equation}
where $N_k^{(d)}$ is a normalization factor chosen in a way that
$C_k(\psi)=1$ for the maximally entangled state, where $\lambda_1=\ldots=\lambda_d=1/d$.
For mixed states, it is defined by convex roof extension.

The $k$-concurrence is nonzero exactly if the Schmidt number of the
state is at least $k$. Unless $k=d$, it is not invariant under local
SL transformations, however it is homogeneous of degree $2$ in the
state coefficients resp. of degree $1$ in the density matrix.

The $k$-concurrences are ordered: If $k>k'$, then $C_k(\rho)<C_{k'}(\rho)$
for all $\ket{\rho}$. This was proven by Gour for pure states, but is
easily extended to mixed states using the properties of the convex
roof.

For pure states, the $k$-concurrences together completely determine
the Schmidt coefficients of the state.

Note that the choice of the normalization factors $N_k^{(d)}$ by Gour
means that the $k$-concurrence of a given state depends on the
dimension of the Hilbert space, even if the support of the state is a
true subspace. It is of course possible to choose normalization
factors $N'_k$ which are independent of $d$, which then makes the
$k$-concurrences independent of the Hilbert space dimension, and also
gives a natural way to apply them to bipartite systems with different
dimensions $d$ and $d'$ for both systems.

The next two measures are special cases of the $k$-concurrence.

\subsubsection{$I$-concurrence}
\label{sec:iconcurrence}

The \textit{$I$-concurrence} was defined by Rungta
\etal~\cite{Rungta2001,Rungta2003} using the concept of an ``universal
inverter'' defined by $S_d(\rho) = (\tr(\rho)\id_d-\rho)$ (they allowed an
arbitrary normalization factor, which they later set to $1$). They
then defined the pure state $I$-concurrence as
\begin{equation}
  \label{eq:iconcurrenceinverter}
  C_I(\psi) = \sqrt{\bra{\psi}S_d\otimes S_{d'}(\ket{\psi}\bra{\psi})\ket{\psi}}
\end{equation}
resulting in the square root of the linear entropy of the reduced
density matrix
\begin{equation}
  \label{eq:iconcurrence}
  C_I(\psi) = \sqrt{2((\tr \rho_A)^2-\tr(\rho_A^2))}.
\end{equation}
For mixed states the $I$-concurrence is defined by convex roof extension.

This is actually the $2$-concurrence of section
\ref{sec:kconcurrence}, except for the different normalization factor:
Eq. \eref{eq:kconcurrence} leads to a prefactor $d/(d-1)$ instead if
$2$ under the square root for $k=2$.

Usually the $I$-concurrence is referred to just as the concurrence,
without further qualification. The $I$-concurrence is nonzero iff the
state is entangled.

\subsubsection{$G$-concurrence}
\label{sec:gconcurrence}

The \textit{$G$-concurrence} is the $k$-concurrence for $k=d$. It is the
geometric mean of the Schmidt coefficients times a constant factor
($d$ in Gour's normalization) \cite{Gour2005},
\begin{equation}
  \label{eq:gconcurrence}
  C_G(\psi) = d(\lambda_1\cdots\lambda_d)^{1/d}
\end{equation}
Alternatively it can be defined by the determinant of the reduced
density matrix:
\begin{equation}
  \label{eq:gconcurrencedensity}
  C_G(\psi) = d(\det \rho_A)^{1/d}
\end{equation}

It is nonzero exactly for states of maximal Schmidt rank. Unlike all
other $k$-concurrences it is invariant under local SL operations.
Indeed, it is the only SL-invariant convex-roof extended bipartite
monotone.

Although for each dimension, the $G$-concurrence is a $k$-concurrence,
it is not dimension-independent even with a dimension-independent
normalization factor, because it is a different $k$-concurrence for
different dimensions. Note that for $d\neq d'$, there exists no
SL-invariant measure at all.

\subsubsection{Negativity}
\label{sec:negativity}

The \textit{negativity} is defined both for pure and mixed states as
\cite{Zyczkowski1998,Lee2000,Eisert2001PhD,Vidal2002}
\begin{equation}
  \label{eq:negativity}
  \mathcal{N}=\case12\left(\|\rho^{T_A}\|_1-1\right)
\end{equation}
where $\|\textperiodcentered\|_1$ denotes the trace norm, $\|A\|_1=\tr\sqrt{A^\dagger A}$, and
$\rho^{T_A}$ the partial transpose, $(A\otimes B)^{T_A} = A^T\otimes B$. Especially the
negativity is \emph{not} a convex-roof extended measure.

Another way to describe the negativity is that it is the absolute
value of the sum of the negative eigenvalues of $\rho^{T_A}$. Note that
this means that the negativity is zero iff the partial transpose
$\rho^{T_A}$ is positive, that is, has no negative eigenvalues. In that
case, the state $\rho$ is also called a PPT state (PPT = Positive Partial
Transpose). The set of PPT states is invariant under SLOCC. The fact
that a non-positive partial transpose implies entanglement was
actually discovered earlier by Peres \cite{Peres1996} and is also
known as Peres condition.

The negativity can be zero even for entangled states.
\cite{Horodecki1997}. Such states are called PPT-entangled states. All
PPT-entangled states are bound entangled, that is, their entanglement
cannot be distilled \cite{Horodecki1998b}. Whether the reverse is also
true is still an open question.

\subsubsection{Convex roof extended negativity (CREN)}
\label{sec:cren}

The \textit{convex-roof extended negativity} is defined as the convex roof
extension \eref{eq:convexroof} of the negativity \eref{eq:negativity}
on pure states \cite{Lee2003}:
\begin{equation}
  \label{eq:cren}
  \mathcal{N}^{\mathrm{CREN}}(\rho) = \min_{\mathrm{decompositions}}\sum_ip_i\mathcal{N}(\psi_i)
\end{equation}
Given that the negativity is convex, it is always a lower
bound to the CREN:
\begin{equation}
  \label{eq:crenbound}
  \mathcal{N}(\rho) \leq \mathcal{N}^{\mathrm{CREN}}(\rho).
\end{equation}
The CREN is nonzero iff the state is entangled.

\subsubsection{Logarithmic negativity}
\label{sec:logneg}

The \textit{logarithmic negativity} is defined as \cite{Plenio2005}
\begin{equation}
  \label{eq:logneg}
  LN(\rho) = \log_2\left\|\rho^{T_A}\right\|_1.
\end{equation}
with the same definitions as in section \ref{sec:negativity}.

Like the negativity, the logarithmic negativity vanishes exactly on
the PPT states. It has been linked to the cost of entanglement under
PPT-preserving operations \cite{Audenaert2003}, an asymptotic measure
based on a slightly different set of operations than SLOCC.

The logarithmic negativity is not convex.

\subsubsection{The geometric measure of entanglement and other
  distance-based measures}
\label{sec:geoment}

The geometric measure of entanglement for pure states is
defined as \cite{Shimony1995,Barnum2001}
\begin{equation}
  \label{eq:geomentpure}
  E_G(\psi) = 1 -
  \max_{\ket{\phi_1}\otimes\ket{\phi_2}}\left|(\bra{\phi_1}\otimes\bra{\phi_2})\ket{\psi}\right|^2.
\end{equation}
For mixed states, it is usually defined by convex roof extension. 
Note that it can be made homogeneous if we replace the number $1$ on the
right-hand side of Eq.~\eref{eq:geomentpure} by
$\langle\psi\ket{\psi}$.

The geometric measure is non-zero iff the state is entangled.

Another distance-based measure is the \textit{robustness}
\cite{VidalTarrach1999}. The robustness is
\begin{equation}
  \label{eq:robustness}
  R(\rho) = \min_{\sigma\mathrm{\ separable}} R(\rho\|\sigma)
\end{equation}
where the quantity
\begin{equation}
  \label{eq:relativerobustness}
  R(\rho\|\sigma) = \min \left\{s\geq0 : (\rho+s\sigma)/(1+s) \mbox{ is separable} \right\}
\end{equation}
is the robustness relative to the separable state $\sigma$. The geometrical
meaning of $R(\rho\|\sigma)$ is that of the ``mixing line'' from $\rho$ to $\sigma$,
the fraction $1/(1+R(\rho\|\sigma))$ consists of separable states.

\subsubsection{Entanglement of formation and other entropy-based measures}
\label{sec:eof}

In this section, we shortly review some measures based on the von
Neumann entropy. Those measures are generally connected with the
asymptotic viewpoint, therefore we will not consider them in detail,
but only list them for completeness.

Historically, the first well-established entanglement measure is the
\textit{entanglement of formation} (EoF) \cite{Bennett1996}. Its
definition is motivated from the asymptotic protocol viewpoint,
however it can be calculated directly on the single state.

The entanglement of formation is defined for pure states as
the von-Neumann entropy of the reduced density matrix
\begin{equation}
  \label{eq:eofpure}
  E_F(\psi) = S(\rho_A)
\end{equation}
where $S(\rho) = \tr(-\rho\log \rho)$ is the von Neumann entropy and $\rho_A =
\tr_B(\ket{\psi}\bra{\psi})$ is the reduced density matrix of system $A$.
Note that $S(\rho_A)=S(\rho_B)$, therefore the choice of subsystem does not
matter. For pure states, the entanglement of formation is also called
entanglement entropy.

For mixed states, the entanglement of formation is defined by convex
roof extension. It is nonzero iff the state is entangled.

Further, the entanglement of formation is subadditive
\cite{Araki1970,Audenaert2004} but not additive \cite{Hastings2009}.

It is conjectured to equal the \textit{entanglement cost} which is
defined as the number of Bell states per copy needed to create
asymptotically many copies of the state and is given by the explicitly
asymptotic expression \cite{Horodecki1998,Hayden2001}
\begin{equation}
  \label{eq:entanglementcost}
  E_C(\rho)=\lim_{n\to\infty}\frac{E_F(\rho^{\otimes n})}{n}.
\end{equation}

Another measure that is explicitly defined asymptotically is the
\textit{distillable entanglement} \cite{Bennett1996}. It is defined as
the asymptotic number of Bell states which can be extracted per copy
of the given state. For pure states, entanglement of formation and
distillable entanglement agree \cite{Bennett1996b}.

The \textit{relative entropy of entanglement,} introduced by Vedral
\etal \cite{Vedral1997,Vedral1998} is defined as
\begin{equation}
  \label{eq:relentropy}
  E_R(\rho) = \min_{\sigma\mathrm{\ separable}}
  \tr\left(\rho\ln\frac{\rho}{\sigma}\right)
\end{equation}

Another entropic measure is the \textit{squashed entanglement}
\cite{Christandl2004}. It is defined as
\begin{equation}
  \label{eq:squashed}
  E_{sq}(\rho) = \inf \{\case12 I(A;B|E): \rho_{AB} = \tr_E\rho_{ABE}\}
\end{equation}
where the infimum goes over \emph{all} such extensions of $\rho_{AB}$
with unbounded dimension of $E$, and
\begin{equation}
  \label{eq:condmutinfo}
  I(A;B|E) = S(AE) + S(BE) - S(ABE) - S(E)
\end{equation}
is the quantum conditional mutual information of $\rho_{ABE}$.

The squashed entanglement is convex, additive on tensor products,
upper bounded by the entanglement of formation, and lower bounded by
distillable entanglement.

\subsection{Two-qubit entanglement}
\label{sec:twoqubit}

The simplest system that can be entangled consists of two qubits. It
is up to now the only system whose entanglement properties have been
completely characterised both for pure and mixed states.

Two-qubit systems generally have quite unique properties
\cite{Vollbrecht2000}. For pure states, there are just two SLOCC
classes, that is, there exists only one type of entanglement (the
other SLOCC class contains the separable states). Consequently, for
pure states there is effectively only one entanglement monotone, that
is, different entanglement monotones, when restricted to pure states,
can be written as function of each other. Indeed, as it turns out,
even for mixed states this generally holds, even in cases where it is
not obvious.

Every two-qubit state is SLOCC-equivalent to a Bell-diagonal state
\cite{Verstraete2001},
that is, a mixture of the four Bell states
$\ket{\phi^\pm} = (\ket{00}\pm\ket{11})/\sqrt{2}$, $\ket{\psi^\pm} =
(\ket{01}\pm\ket{10})/\sqrt{2}$ which also is its normal form.

\subsubsection{Wootters' analytical solution for the concurrence}
\label{sec:woottersconcurrence}

For two qubits we have $d=2$, therefore the whole family of
$k$-concurrence consists of only one concurrence, the
$2$-concurrence, which in this case is both the $I$-concurrence (and
thus is non-zero exactly if the state is entangled) and the
$G$-concurrence (and thus is invariant under local SL operations).

Indeed, the concurrence was originally defined for two-qubit systems
\cite{Bennett1996,Hill1997,Wootters1998}, and the concurrences of sections
\ref{sec:kconcurrence} to \ref{sec:gconcurrence} are generalizations
of that.

For pure states $\ket{\psi} = \psi_{00}\ket{00} + \psi_{01}\ket{01} +
\psi_{10}\ket{10} + \psi_{11}\ket{11}$ the concurrence is
\cite{Wootters1998,Albeverio2001}
\begin{equation}
  \label{eq:concurrence}
  C(\psi) = 2\left|\psi_{00}\psi_{11}-\psi_{01}\psi_{10}\right|.
\end{equation}
Alternatively, it can be written as expectation value of an antilinear
operator \cite{Wootters1998}:
\begin{equation}
  \label{eq:concurrencesigma22}
  C(\psi) = \left|\bra{\psi}\sigma_y\otimes\sigma_y\ket{\psi^*}\right|
\end{equation}
with the Pauli matrix $\sigma_y =
\left(\begin{array}{cc}0&-\rmi\\\rmi&0\end{array}\right)$.

For mixed states, the concurrence is defined by convex roof extension.
It is one of the rare cases where a method to calculate the convex
roof without optimization is known \cite{Wootters1998}. To calculate
it, one needs the eigenvalues $r_1 \geq r_2 \geq r_3 \geq r_4$ of the matrix
\begin{equation}
  \label{eq:Rmatrix}
  R = \rho(\sigma_y\otimes\sigma_y)\rho^*(\sigma_y\otimes\sigma_y)
\end{equation}
where $\rho^*$ denotes the complex-conjugated density matrix, that is,
the matrix where \emph{in the computational basis} all matrix elements
are the complex conjugate of the corresponding matrix element in $\rho$.
Then the mixed-state concurrence is
\begin{equation}
  \label{eq:mixedstateconcurrence}
  C(\rho) = \max \{0,\sqrt{r_1} - \sqrt{r_2} - \sqrt{r_3} - \sqrt{r_4}\}.
\end{equation}

A special case is the mixture of a Bell state with an orthogonal
separable state, like $\sqrt{p}\ket{\phi^+} + \sqrt{1-p}\ket{01}$. For
such states, Abouraddy \etal~\cite{Abouraddy2001} found that the
concurrence simply equals the weight $p$ of the Bell state.

\subsubsection{Other measures related to the concurrence}
\label{sec:concrel}

Since for two qubits, the pure state negativity equals the concurrence
\cite{Verstraete2001JPA}, the CREN also agrees with the concurrence.

For two qubits, the geometric measure of entanglement can be
calculated from the concurrence by \cite{Wei2003} as
\begin{equation}
  \label{eq:geomconc}
  E_G(\rho) = \case12(1-\sqrt{1-C(\rho)^2})
\end{equation}
The EoF can be calculated from the concurrence as
\cite{Hill1997,Wootters1998}
\begin{equation}
  \label{eq:eofconc}
  E_F(\rho) = H\left(\case12(1+\sqrt{1+C(\rho)^2})\right)
\end{equation}
with $H(p)=-p\log_2p-(1-p)\log_2(1-p)$. Remarkably, these relations
hold not only for pure, but also for mixed states. This is because for
qubits, concurrence, geometric entanglement and entanglement of
formation have optimal decompositions in which all states have equal
measure.

Another relation that has been shown in~\cite{Eltschka2012SciRep}
relates the concurrence to the fully entangled
fraction introduced in~\cite{Bennett1996} and the standard projection
witness
\begin{equation}
  \label{eq:twobitprojectionwitness}
  \mathcal{W}_{\mathrm{2\ qubits}}=\case12-\ket{\phi^+}\bra{\phi^+}.
\end{equation}
Applying the witness $\mathcal{W}_{\mathrm{2\ qubits}}$ and optimizing
the state over local SL operations gives, up to a sign, the
concurrence for entangled states:
\begin{equation}
  \label{eq:optconcurrence}
  C(\rho^{\mathrm{2qb}}) = \max{\left(0,\max_{S=S_1\otimes S_2}{\left[
          2\bra{\Phi^+} S \rho^{\mathrm{2qb}}
          S^{\dagger}\ket{\Phi^+}
          - \tr \left(S \rho^{\mathrm{2qb}}
            S^{\dagger}\right)
        \right]}
    \right)}
\end{equation}
where $S_{1/2}\in SL(2,\mathbb{C})$. But the fully entangled fraction
is~\cite{Bennett1996}
\begin{equation}
  \label{eq:fef}
  f(\rho^{\mathrm{2qb}}) = \max_{\ket{e}}\bra{e}\rho^{\mathrm{2qb}}\ket{e}
  = \max_{U=U_1\otimes U_2}\bra{\phi^+}U\rho^{\mathrm{2qb}}U^\dagger\ket{\phi^+}
\end{equation}
where $\ket e = U\ket{\phi^+}$ is optimized over the maximally entangled
states, and $U_{1/2}\in SU(2)$. This differs from the first term in
\eref{eq:optconcurrence} only in the more restricted group. Therefore
the concurrence is basically the local-SL optimized fully entangled
fraction.

\subsubsection{Two-qubit monotones that do not depend on the concurrence}
\label{sec:qubitmeasures}

The above discussion seems to imply that the concurrence is the only
entanglement monotone for two-qubit mixed states. However, this is not
the case.

The mixed state negativity, while strictly positive for entangled
states, does not agree with the concurrence \cite{Verstraete2001JPA}.

Verstraete et al \cite{Verstraete2002} identified another monotone
based on the Lorentz singular values of the density matrix, namely
\begin{equation}
  \label{eq:verstraetemonotone}
  M(\rho) = \max\{0,-s_0+s_1+s_2\}
\end{equation}
where $s_0 \geq s_1 \geq s_2 \geq \left|s_3\right|$ are the (local SL
invariant) Lorentz singular values, that is, the state has the normal
form $s_0\id\otimes\id + s_1\sigma_x\otimes\sigma_x + s_2\sigma_y\otimes\sigma_y + s_3\sigma_z\otimes\sigma_z$, where $\sigma_x$,
$\sigma_y$ and $\sigma_z$ are the Pauli matrices.

Liang \etal \cite{Liang2008} derived a complete set of monotones for
Bell-diagonal entangled states. If $p_1>p_2>p_3>p_4$ are the mixing
coefficients of the Bell states, the entanglement monotones are given
by them as
\begin{eqnarray}
  \label{eq:liangmonotones}
  E_1(\rho_{\rm Bell}) = p_1,\\
  E_2(\rho_{\rm Bell}) = \frac{1-2p_2}{p_3+p_4},\\
  E_3(\rho_{\rm Bell}) = \frac{1-2p_2-2p_3}{p_4}.
\end{eqnarray}
Obviously $E_2$ is only defined if $p_3>0$ and $E_3$ only if $p_4>0$.
An entangled Bell diagonal state $\rho$ is convertible to another Bell
entangled state $\rho'$ if and only if all $E_i(\rho)\geq E_i(\rho')$.

As such, the functions do not fulfil the conditions on monotones written
above. The first obvious point is that they are only valid on
entangled states and do not vanish on separable states. However due to
the simple structure of Bell-diagonal states this is easy to fix: The
border of separable states is at $p_1=\frac12$, at which all three
$E_i$ are constant: $E_1(p_1=\case12)=\frac12$,
$E_2(p_1=\case12)=E_3(p_1=\case12)=2$. Therefore, $\tilde E_i(\rho_{\rm
  Bell}) := \max \{0, E_i(\rho_{\rm Bell})-E_i(p_1=\case12)\}$ satisfies that condition.
Note that $2\tilde E_1$ turns out to be the concurrence.

The second point is that they are only defined on
Bell-diagonal states. Given that each SL orbit corresponds to a specific Bell
state, one obvious way to do it is to replace $1$ by $p_1+p_2+p_3+p_4$
and then extend them on the SL orbit according to their degree of
homogeneity. If this is combined with the previous change, $E_1$,
being homogeneous of degree $1$ in the density matrix, gives exactly
the concurrence, and $E_2$ and $E_3$, being of degree $0$, give
entanglement measures that are constant on the complete orbit and
strictly non-increasing under SLOCC. The latter are therefore no
resource measures.

A final problem is that $E_2$ and $E_3$ diverge when $p_3$ resp. $p_4$
go to zero (that is, if the rank of the density matrix goes below $2$
resp. $3$). But given that they are not resource measures anyway, this
seems a minor point (and can easily be fixed by applying an
appropriate monotonic function).


\subsection{Partial transpose, concurrence, and negativity}
\label{sec:PTconcneg}


Historically, one of the first criteria to check whether or not a
generic mixed state is separable was the partial-transpose 
criterion~\cite{Peres1996,Horodecki1996}, that is, whether or not
the partial transpose (cf.~section \ref{sec:negativity})  
of a given state is positive. The
definition of the negativity measures is built
on the partial transpose. On the other hand,
there are the concurrence-based entanglement quantifiers which seem
to bear little relation with the partial transpose. In the following
we sketch the connection between them, as well as with other interesting
concepts. 

Consider  a pure product state 
$\psi\in \mathcal{H}_{A}\otimes\mathcal{H}_{B}$ (with $\dim\mathcal{H}_A=d$,
$\dim\mathcal{H}_B=d'$)
%
\begin{equation}
      \ket{\psi}\ =\ \ket{a}\otimes\ket{b}\ =\ \sum_{jk}\ \psi_{jk}\ \ket{j}\ket{k} 
                \ =\ \sum_{jk}\ a_j b_k\ \ket{j}\ket{k}
\end{equation}
where $jk$ can be read as a joint (two-digit) index on the wavefunction $\psi$.
The elements of the density matrix are
\begin{equation}
      \rho\equiv
      \ket{\psi}\!\bra{\psi} =  \sum_{jklm} a_j b_k a^*_l b^*_m \ket{jk}\!\bra{lm}
\  \longrightarrow\  \rho_{jk,lm} = a_j b_k a^*_l b^*_m
\ \ .
\end{equation}
In matrix representation, the off-diagonal element $\rho_{jk,lm}$
is located in the same row (or column) as the diagonal element
$\rho_{jk,jk}$ (or $\rho_{lm,lm}$, respectively).
At that position, the partial transpose $\rho^{T_B}$ has the element
\begin{equation}
(\rho^{T_B})_{jk,lm}\ 
                      =\ a_j b_{m} a^*_l b^*_{k}\
                      =\ \psi_{jm}\psi_{lk}^*\ 
\ \ .
\end{equation}
Consequently, a product state $\psi$ obeys the condition
\begin{eqnarray}
          \left|(\rho^{T_B})_{jk,lm}\right|^2-
          (\rho^{T_B})_{jk,jk}(\rho^{T_B})_{lm,lm}
        & = & |\psi_{jm}\psi_{lk}|^2-|\psi_{jk}\psi_{km}|^2
\nonumber\\
        & = & |\psi_{jm}\psi_{lk}-\psi_{jk}\psi_{lm}|^2 = 0\ \ .
\label{eq:condPPT}
\end{eqnarray}
Violation of condition (\ref{eq:condPPT}) for any pair of levels
$\{j,l\}$ of party $A$ and $\{k,m\}$ in $B$, respectively,
 means that $\psi$ cannot be a product state. 
Correspondingly, we may define 
\begin{equation}
         C(\psi)^2\ =\ \sum_{jklm} |\psi_{jm}\psi_{lk}-\psi_{jk}\psi_{lm}|^2 
\label{eq:concveclength}
\end{equation}
as a measure of the total violation of the product-state condition for $\psi$
on the bipartition $\mathcal{H}_A\otimes\mathcal{H}_B$.
It turns out~\cite{Albeverio2001} that $C(\psi)$ 
is invariant under local unitaries in $\mathcal{H}_A$, $\mathcal{H}_B$
and that it is
an alternative definition 
for the $I$-concurrence (\ref{eq:iconcurrence})
\begin{equation}
    C(\psi)\ =\ \sqrt{\sum_{jklm} |\psi_{jm}\psi_{lk}-\psi_{jk}\psi_{lm}|^2}
           \ =\ \sqrt{2\left[(\tr\rho_A)^2
                             -  \tr\rho_A^2\right]}
\label{eq:concvecconc}
\end{equation}
where $\rho_A=\tr_B\ket{\psi}\!\bra{\psi}$.

One may also note that $C(\psi)$ corresponds to the Euclidean 
length of a vector with components $C_{jk,lm}$, the so-called
concurrence vector~\cite{Audenaert2001,Badziag2002,Akhtarshenas2005,Li2008}.
On the other hand, (\ref{eq:concveclength}) may be viewed as
a 2-norm on the concurrence vector. Accordingly one might
expect that the corresponding 1-norm
\[
         \tilde{\mathcal{N}}(\psi)\ =\ 
           \sum_{jklm} |\psi_{jm}\psi_{lk}-\psi_{jk}\psi_{lm}| 
\]
is an entanglement measure as well~\cite{Ma2012-QIC}. However,
this expression is not invariant under local unitaries. Nonetheless,
it is interesting that the minimum of $\tilde{\mathcal{N}}(\psi)$ 
is obtained 
for the Schmidt decomposition of $\psi$ and that, with the restriction
$j<l, k<m$, it is equal to the negativity $\mathcal{N}(\psi)$. 
 This implies that for pure states
\begin{eqnarray}
2\mathcal{N}(\psi)\ \geqq\  C(\psi) &\ =\ & \sqrt{ 4 \sum_{j<l,k<m} 
               \left| \psi_{jk}\psi_{lm}-\psi_{jm}\psi_{lk}\right|^2
                                             }
\label{eq:neg_vs_conc1}
\\
    &\ \geqq\ & 2\sqrt{\frac{2}{d(d-1)}}\ \mathcal{N}(\psi)
\ \ .
\label{eq:neg_vs_conc2}
\end{eqnarray}
%
For the second inequality we have used that the quadratic 
always exceeds the arithmetic mean.

\subsection{Bounds on the Schmidt number and the concurrence
            of mixed states}
\label{sec:boundsconcmix}
It is a common feature of most entanglement measures that they
are easy to compute on pure states, but---due to their definitions
via optimization procedures---it is hard to evaluate or even just to estimate them 
on mixed states. The negativity measures are a notable exception to this
rule. Therefore it is an important task to find reliable lower bounds for 
the entanglement measures and the Schmidt number
in order to characterize the entanglement resources of a given
mixed state. A special case of the Schmidt-number estimation is the
separability problem, that is, the question whether or not the Schmidt 
number of a state is at least two~\cite{Terhal2000PRA}.

The negativity of the maximally
entangled state of Schmidt rank $k$ is 
$\mathcal{N}(\Psi_k)=k(k-1)/(2k)=\case12(k-1)$. Therefore,
for a state $\rho=\sum_j p_j\pi_{\psi_j}$ (with defintion~(\ref{eq:decomp-rho})) 
of Schmidt number $k$ (i.e., $r(\psi_j)\leqq k$)
we have~\cite{Eltschka2013-PRL} 
\[
        \mathcal{N}(\rho)\ \leqq\ \mathcal{N}^{\mathrm{CREN}}(\rho)\ 
                           \leqq\ \sum_j p_j \mathcal{N}(\psi_j)\ \leqq\ 
        \frac{k-1}{2}
\]
so that 
\begin{equation}
        r(\rho)\ \geqq\ 2\mathcal{N}(\rho)+1
\ \ ,
\end{equation}
i.e., the negativity provides a lower bound for the Schmidt number
(for alternative bounds, see~\cite{Vogel2011}).
Analogously we obtain $C(\Psi_k)=\sqrt{2(k-1)/k}$ and hence
\begin{equation}
        r(\rho)\ \geqq\ \frac{2}{2-C(\rho)^2}
\end{equation}
(with the normalization in (\ref{eq:concvecconc})), so non-zero concurrence
implies entanglement. 
Note that for mixed states there is no relation analagous to
(\ref{eq:neg_vs_conc1}), since the negativity vanishes for PPT-entangled states
while the concurrence does not. 
However, (\ref{eq:neg_vs_conc2}) leads to
\begin{equation}
        C(\rho)\ \geqq\ 2\sqrt{\frac{2}{d(d-1)}}
                        \mathcal{N}^{\mathrm{CREN}}(\rho)
               \ \geqq\ 2\sqrt{\frac{2}{d(d-1)}}
                        \mathcal{N}(\rho)
\ \ .
\label{eq:concestimate_mixed}
\end{equation}
Note that for two-qubit states~\cite{Audenaert2001-JMA} two times the CREN
equals the concurrence, so the resulting inequality for mixed states
seems to invert the pure-state inequality~(\ref{eq:neg_vs_conc1}).
In the general case it is not easy to estimate $C(\rho)$ and there is a vast
literature on lower bounds of the concurrence (e.g.,
\cite{Rungta2003,Grudka2004,Albeverio2005,Breuer2006,BuchMintert2007,Zhang2007,Liu2009,Augusiak2009,Gittsovich2010,ZhaoFei2011,MaGuhne2012}).
Here we focus on a powerful analytical method that is both simple
and easily generalizable to the multipartite case. 
It is based on ideas by G\"uhne and Seevinck~\cite{GuhneSeevinck2010} and
Huber {\it et al.}~\cite{Huber2010,MaHuber2011,WuHuber2012,Huber2013-PRA}.

Consider a subset $\mathcal{M}$ of $\eta$ level pairs $\{jk,lm\}$ 
$(j<l,k<m)$
of the sum for the concurrence in (\ref{eq:concvecconc}).
By using the inequality $\sqrt{a_1^2+\ldots +a_n^2}\geqq (a_1+\ldots +a_n)/\sqrt{n}$ for real
nonnegative numbers $a_1,\ldots , a_n$ as well as the triangle inequalities one finds 
\begin{eqnarray} 
           C(\psi) & \geqq & \frac{2}{\sqrt{\eta}}\sum_{jklm\in\mathcal{M}} 
               \left| \psi_{jk}\psi_{lm}-\psi_{jm}\psi_{lk}\right|
\nonumber\\
                   & \geqq & \frac{2}{\sqrt{\eta}}\sum_{jklm\in\mathcal{M}} \left(
               \left| \psi_{jk}\psi_{lm}\right| -\sqrt{\left|\psi_{jm}\right|^2
                                                       \left|\psi_{lk}\right|^2}
                                                                  \right)
\nonumber\\
                   & \geqq & \frac{2}{\sqrt{\eta}}\sum_{jklm\in\mathcal{M}} \left(
               \left| \rho_{jk,lm}\right| -\sqrt{\rho_{jm,jm}\rho_{lk,lk}}
                                                                  \right)
\ \ .
\label{eq:huber1}
\end{eqnarray}
In the last line of (\ref{eq:huber1}) we have used $\rho=\pi_{\psi}$.
Due to the
convexity of the concurrence and of the 
functions on the right-hand side of (\ref{eq:huber1}) this relation
can directly be extended to mixed states $\rho=\sum_j p_j\pi_{\psi_j}$
\begin{equation}
           C(\rho)\ \geqq\  
                   \frac{2}{\sqrt{\eta}}\sum_{jklm\in\mathcal{M}} \left(
               \left| \rho_{jk,lm}\right| -\sqrt{\rho_{jm,jm}\rho_{lk,lk}}
                                                                  \right)
\ \ .
\label{eq:huber2}
\end{equation}
This relation may be regarded as a set (for different level subsets 
$\mathcal{M}$) of witness inequalities for bipartite entanglement
in terms of density matrix elements, as for separable states 
the right-hand side of 
(\ref{eq:huber2}) cannot be positive. At the same time, 
the matrix element difference 
provides a lower bound for the concurrence of the mixed state $\rho$. 
Note that there is
also a simple method to optimize this bound: Local unitary operations 
on the parties $A$ and
$B$ do not change the concurrence, however, they may change 
the value of the right-hand side in (\ref{eq:huber2}). 
That is, the estimate can be improved by maximizing the
right-hand side over local unitaries.
%

\subsection{Axisymmetric states}
\label{sec:axisym}

\subsubsection{Definition}
\label{sec:defofaxisym}
As is the case in many other areas of physics, 
exact solutions of nontrivial problems
provide a testbed for the concepts of the theory, the models for the observed
phenomena, and for approximate methods.
In entanglement theory, this role is played by exact solutions which sometimes
are possible for special states of high symmetry and/or problems of reduced 
rank~\cite{Horodecki1999,Vollbrecht2001,Osborne2005}.
An early example are the Werner states for two qudits that are invariant
under $U\otimes U$ operations where $U$ denotes an arbitrary one-qudit unitary
transformation. From symmetry one concludes that $d\times d$ Werner states 
can be reprensented by a linear combination of the
projectors onto the symmetric and antisymmetric
subspaces. For two qubits (where the antisymmetric subspace has only one dimension),
the Werner states are locally equivalent to 
\begin{equation}
            \rho^{\mathrm{W}}(p)\ =\   p \ket{\Psi_2}\!\bra{\Psi_2}\ +\ \frac{1-p}{4}\id_4
\ \ .
\label{eq:Werner2x2}
\end{equation}
The states $\rho^{\mathrm{W}}(p)$ are particularly interesting because they
are mixtures of a maximally entangled state with the completely mixed 
state.  The latter is often used as a model of white noise and 
serves to characterize the robustness of the entanglement.
In fact, in Ref.~\cite{VidalTarrach1999} the minimal noise admixture $\tilde{p}$ at
which the mixed state $\frac{\tilde{p}}{d^2}\id_{d^2}+(1-\tilde{p})\pi_{\psi}$ 
becomes separable 
was termed random robustness of $\psi$ (note, however, that this quantity is
not an entanglement monotone~\cite{HarrowNielsen2003}).

If one attempts to obtain a state analogous to (\ref{eq:Werner2x2}) in
higher dimensions from symmetry considerations, the appropriate requirement
is invariance under $U\otimes U^{\ast}$ and the resulting one-parameter family
of states
\begin{equation}
            \rho^{\mathrm{iso}}(p)\ =\   p \ket{\Psi_d}\!\bra{\Psi_d}\ 
                                      +\ \frac{1-p}{d^2}\id_{d^2}
\ \ .
\label{eq:isotropic}
\end{equation}
is called isotropic states~\cite{Horodecki1999}. Various interesting exact results
were obtained for isotropic states, such as the separability 
criterion~\cite{VidalTarrach1999,Rungta2001Book}, the entanglement of formation~\cite{VollbrechtTerhal2000}
and the convex roofs of both $I$-concurrence and the square of the 
$I$-concurrence~\cite{Rungta2003}. An arbitrary $d\times d$ state $\rho$
can be projected onto the isotropic states by a so-called twirling operation
\begin{equation}
         \mathbb{P}_{\mathrm{iso}}(\rho)\ =
         \ \int \mathrm{d}U (U\otimes U^*)\rho(U\otimes U^*)^{\dagger}
\label{eq:twirling-iso}
\end{equation}
where $\mathrm{d}U$ is the Haar measure. Note that $\mathbb{P}_{\mathrm{iso}}$
combines local operations and mixing, that is, it is an LOCC operation and therefore
the entanglement cannot increase in the mapping $\mathbb{P}_{\mathrm{iso}}:\rho \rightarrow
\rho^{\mathrm{iso}}$. Although (\ref{eq:twirling-iso}) involves
a continuous average the twirling operation can be represented by a finite sum
(this is a consequence of the Krein-Milman and Caratheodory theorems~\cite{Lang1987}
for the finite-dimensional compact group of $(U\otimes U^*)$ transformations).

\begin{figure}
\centering
\includegraphics[width=.77\linewidth]{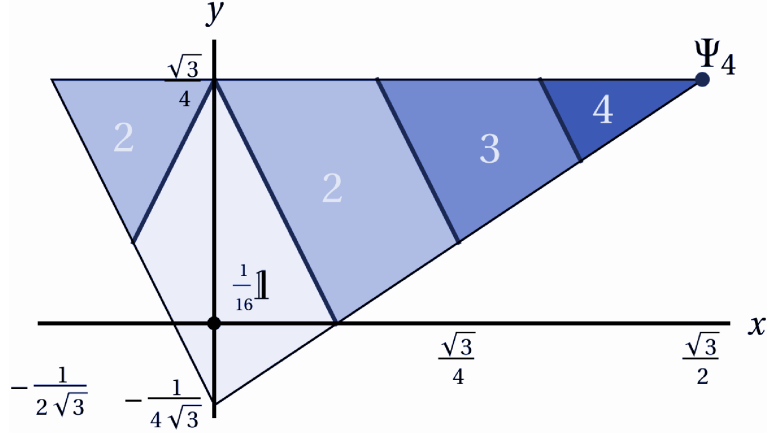}
    \caption{The family of $d\times d$ 
             axisymmetric states $\rho^{\mathrm{axi}}$ (here for $d=4$).
             It is characterized by two real parameters (for all $d$).
             The only pure state in the family is $\Psi_d$ in the
             right upper corner. The completely mixed state
             $\case{1}{d^2}\id_{d\times d}$ is located at the origin.
             One can clearly identify the hierarchy of convex sets
             $\mathcal{S}_k$ of increasing Schmidt number $k$
             $(1\leqq k\leqq d)$. 
            }
\label{fig:axi}
\end{figure}

By modifying the symmetry requirements it is possible to generate and study
other families of symmetric states. Another  interesting example 
are the rotationally
invariant states, i.e.,
they  do not change under rotations $R\in\ $SO(3)~\cite{ManneCaves2008}.
Here we consider in some detail a different option. 
Recently it was noticed that the symmetry requirement for isotropic states
can be relaxed so that one obtains 
a two-parameter family for all finite-dimensional $d\times d$ systems, the
axisymmetric states~\cite{Eltschka2013-PRL}. We expect that these
states will be instrumental in the further development of methods for bipartite
mixed states. Axisymmetric states have the same symmetries as the
maximally entangled state $\Psi_d$, that is,
\\ (i) exchange of two qudits,
\\ (ii) simultaneous permutations of the basis states for both parties
 {\em e.g.}, $\ket{1}_A\leftrightarrow\ket{2}_A$ and $\ket{1}_B\leftrightarrow\ket{2}_B$,
\\ (iii) simultaneous phase rotations
  \begin{equation}
    \label{eq:zrotdxd}
    V(\varphi_1,\ldots,\varphi_{d-1})=
   \rme^{\rmi\sum_j\varphi_j\mathfrak{g}_j}
   \otimes\rme^{-\rmi\sum_k\varphi_k\mathfrak{g}_k}
   \nonumber
  \end{equation}
where $\mathfrak{g}_j$ ($j=1,\ldots,d-1$) are the diagonal 
generators of the group SU($d$).
The resulting states $\rho^{\mathrm{axi}}$ for a $d\times d$-dimensional system
have the diagonal matrix elements
\[
          \rho^{\mathrm{axi}}_{jj,jj} \ =\ \frac{1}{d^2} + a \  ,\ \ \ 
          \rho^{\mathrm{axi}}_{jk,jk} \ =\ \frac{1}{d^2}-\frac{a}{d-1}\ \ (j\neq k)
\]
($j,k=1,\ldots,d$) and off-diagonal entries
\[
          \rho^{\mathrm{axi}}_{jl,km} \ =\ \left\{ \begin{array}{ll}
                                                b \ \ \ & \mathrm{for}\ l=j\ ,\ m=k
                                                \\
                                                0 \ \ \ & \mathrm{otherwise}\ \ .
                                                \end{array}
                                         \right.
\]
Each $\rho^{\mathrm{axi}}$ is a mixture of three states, therefore the family
can be represented by a triangle in a plane, see figure~\ref{fig:axi}. 
The only pure state
of this family is $\ket{\Psi_d}\!\bra{\Psi_d}$ in the
right upper corner of the triangle. 
We choose the 
coordinates $x$ and $y$ such that the Euclidean metric 
coincides with the Hilbert-Schmidt
metric (where we define $D_{\mathrm{HS}}^2(A,B)\equiv \tr[A-B][A-B]^{\dagger}$).
Then, the coordinate are in the ranges
\begin{eqnarray}
\label{eq:range-y}
   -\frac{1}{d\sqrt{d-1}}\ & \leqq & \ y\ \leqq\ \frac{\sqrt{d-1}}{d}\\
\label{eq:range-x}
   -\frac{1}{\sqrt{d(d-1)}} \ & \leqq & \ x\ \leqq\ 
    \sqrt{\frac{d-1}{d}}    \ \ ,
\end{eqnarray}
hence $a=y\sqrt{d-1}/d$ and $b=x/\sqrt{d(d-1)}$.
The fully mixed state $\frac{1}{d^2}\id_{d^2}$ lies at the origin.
Thus the isotropic states form a subset of the
axisymmetric family: They are located on the straight line connecting
the right upper triangle corner with the origin.

Clearly, there is a twirling operation 
$\mathbb{P}_{\mathrm{axi}}:\rho\rightarrow \rho^{\mathrm{axi}}$
analogous to (\ref{eq:twirling-iso}) also for axisymmetric states
\begin{equation}
         \mathbb{P}_{\mathrm{axi}}(\rho)\ =
         \ \int \mathrm{d}V V\rho V^{\dagger}
\label{eq:twirling-axi}
\end{equation}
where now the integral denotes an average over both
discrete and continuous symmetry operations (i)--(iii).

\subsubsection{Entanglement properties of axisymmetric states}
\label{sec:entofaxisym}
The SLOCC classes of axisymmetric states in the
Schmidt-number classification can be determined by using optimal
Schmidt-number witnesses~\cite{Sanpera2001}. The result is illustrated
in figure~\ref{fig:axi}. 
The borders of the classes $\mathcal{S}_k$ are straight lines
parallel to the left lower border of the triangle and divide the right lower
border into $d$ equal parts. Moreover, there is always a part of Schmidt
number 2 close to the left upper corner. 

Now we study the entanglement resources by calculating both the
negativity and the concurrence for $\rho^{\mathrm{axi}}(x,y)$ 
for $x\geqq 0$.
The negativity  is obtained straightforwardly as
\begin{equation}
    \mathcal{N}(\rho^{\mathrm{axi}}(x,y)) = \max\left\{ 0 , \frac{1}{2}\left[ \sqrt{d(d-1)}x+\sqrt{d-1}y-\frac{d-1}{d}
                                             \right]
                       \right\}\ .
\label{eq:axiN}
\end{equation}
For the concurrence we use the lower bound (\ref{eq:concestimate_mixed})
including all off-diagonal matrix elements of $\rho^{\mathrm{axi}}(x,y)$.
The result is
\begin{eqnarray}
    C(\rho^{\mathrm{axi}}(x,y)) & \geqq & \max\left\{ 0 ,  \right.
\nonumber\\
           && \left.  \!\! \sqrt{\frac{2}{d(d-1)}}\left[ \sqrt{d(d-1)}x+\sqrt{d-1}y-\frac{d-1}{d}
                                             \right]
                       \right\} .
\label{eq:axiC}
\end{eqnarray}
We note that both measures depend linearly on $x$ and $y$, so 
their graphs are planes. They both
vanish on the line $y=\sqrt{d-1}/d-x\sqrt{d}$, that is, the
border of $\mathcal{S}_1$. Moreover, they both assume their 
exact maximum for $\Psi_d$. Since, on the other hand, the measures
are convex, they both represent the largest possible
convex functions with the corresponding exact behavior for $\mathcal{S}_1$
and for $\Psi_d$. Hence, the right-hand sides of (\ref{eq:axiN})
and (\ref{eq:axiC}) give the exact convex-roof extended
negativity and the exact concurrence for the axisymmetric states, and
they coincide (up to a normalization factor). We mention that the
axisymmetric states share this property
with the rotationally invariant states discussed by Manne
and Caves~\cite{ManneCaves2008}.

It is remarkable that both $\mathcal{N}(\rho^{\mathrm{axi}})$
and $C(\rho^{\mathrm{axi}})$ are constant along the borders between the
SLOCC classes. This means that for the axisymmetric states 
$\mathcal{N}$ and $C$ are class-specific entanglement measures,
i.e., they measure the Schmidt number $k\leqq d$
\begin{equation}
   r(\rho^{\mathrm{axi}}) = k\ \
   \Longleftrightarrow\ \
   \left\{ \begin{array}{l}
           k-2\leqq 2\mathcal{N}(\rho^{\mathrm{axi}})\leqq 
           k-1
\nonumber\\[2mm]
           k-2\leqq \sqrt{\frac{d(d-1)}{2}}C(\rho^{\mathrm{axi}})
                          \leqq k-1\ \ .
\nonumber
           \end{array}
   \right.
\label{eq:SchmidtNC}
\end{equation}
Note that axisymmetric states have positive partial 
transpose iff they are separable.

We conclude this section with two remarks. First, the fact that the exact
concurrence (or more precisely, the 2-concurrence, 
see section~\ref{sec:kconcurrence})
has the shape of a plane for the axisymmetric states, gives rise to the
conjecture that each $k$-concurrence can be determined exactly in this
case and is represented by a plane, with its zero line coinciding with
the border between the classes $\mathcal{S}_{k-1}$ and $\mathcal{S}_k$.
While for $d=3$ this has indeed been proven~\cite{ES-unpublished}, 
for $d>3$ it is an open question.

The second remark regards the observation that the fidelity $F$ of an arbitrary
$d\times d$ state with the maximally entangled state $\Psi_d$ remains unchanged
when $\rho$ is projected onto the axisymmetric states
\begin{equation}
               F\ =\ \bra{\Psi_d} \rho \ket{\Psi_d}\ =\ 
                     \bra{\Psi_d} \mathbb{P}_{\mathrm{axi}}(\rho) 
                     \ket{\Psi_d}
\ \ .
\label{eq:proj_to_axi}
\end{equation}
On the other hand, for axisymmetric states 
the entanglement measures $C$ and $\mathcal{N}^{\mathrm{CREN}}$
depend only on the fidelity with $\Psi_d$. Consequently, we obtain an alternative
method to estimate these measures for arbitrary states: Project $\rho$ to the
axisymmetric states (which does not increase the amount of entanglement) and
use the entanglement measure of the image as a lower bound. Clearly, this 
bound can be improved by unitary optimization before the projection:
\begin{equation}
   C(\rho) \geqq C(\rho^{\mathrm{axi}}(\rho^{\mathrm{opt}})) = 
                   \max_{U_A\otimes U_B} 
              C\left(
              \mathbb{P}_{\mathrm{axi}}\left( [U_A\otimes U_B]\rho[U_A\otimes U_B]^{\dagger}
                                       \right)
               \right) .
\label{eq:axi-proj-conc}
\end{equation}
As long as the optimization is only over local unitaries, this method is
equivalent to the estimate (\ref{eq:huber2}) taking into account the
off-diagonal matrix elements $\rho_{jj,kk}$ ($j< k$). In fact, 
(\ref{eq:huber2}) can be rewritten for this choice of $\mathcal{M}$
as
\begin{equation}
         C(\rho)\ \geqq\ \sqrt{\frac{2d}{d-1}}
                     \tr\left(\rho\left[\ket{\Psi_d}\!\bra{\Psi_d}-\frac{1}{d}
                                              \id_{d^2}
                                  \right]
                        \right)
\end{equation}
which just links the optimal witness for Schmidt number 2 by Sanpera
{\em et al.}~\cite{Sanpera2001} with a lower bound for the concurrence.

\section{Quantification of multipartite entanglement}
\label{sec:quantificME}

\subsection{$k$-separability and genuine entanglement of multipartite states}
\label{sec:ksepandgenuine}

Multipartite quantum systems contain more than two individual
subsystems. Their pure states are vectors in the Hilbert space
$\mathcal{H}=\mathcal{H}_{A_1}\otimes\mathcal{H}_{A_2}\otimes
    \cdots\otimes\mathcal{H}_{A_N}$ with $\dim \mathcal{H}_{A_j}=d_j$. 
In analogy with bipartite states 
the pure state $\ket{\psi}$ of an $N$-partite system is called
{\em fully separable} (often also just {\em separable})
if it is a product of states of the individual
systems
\begin{equation}
     \ket{\psi}\ =\ \ket{\phi_1}\otimes\ket{\phi_2}\otimes\cdots\otimes
                    \ket{\phi_N}
\ \ .
\end{equation}
If this is not the case then $\psi$ contains entanglement. 
Then, if it can be written as a product of $k$ factors ($1<k<N$)
\begin{equation}
     \ket{\psi}\ =\ \bigotimes_{j=1}^k
                   \ket{\phi_j}
\label{eq:ksep}
\end{equation}
it is called {\em k-separable}. If a state
is not a product (\ref{eq:ksep}) for any $k>1$, it is truly
$N$-partite entangled. It is now common to say it is {\em genuinely 
entangled}.

Alternatively, a state is said
to be {\em k-party entangled} if none of the factors in
\begin{equation}
     \ket{\psi}\ =\ \bigotimes_{j=1}^m
                   \ket{\phi_j}
\label{eq:kent}
\end{equation}
contains genuine entanglement of more than $k$ 
parties~\cite{GuhneTothReview2009}. For mixed states we have analogous
definitions for the corresponding convex combinations.
In particular,
a mixed $N$-party state is fully separable if it can be written as
\begin{equation}
     \rho\ =\ \sum_{j=1}^N p_k \rho_k^{(1)}\otimes\rho_k^{(2)}\otimes
                               \cdots\otimes
                    \rho_k^{(N)}
\end{equation}
where $\rho_k^{(j)}$ is a state of the $j$-th system only, and it is
genuinely entangled if it cannot be written as a convex combination
of biseparable (2-separable) states. 

The subdivision into $k$-separable states does not change
under SLOCC, so it induces an entanglement classification.
While this classification is somewhat obvious, states of more than two
parties are far more subtle: As D{\"u}r {\em et al.} noticed~\cite{DVC2000},
for three qubits there a two inequivalent genuinely entangled types of
states, the Greenberger-Horne-Zeilinger (GHZ) state and the $W$ state. 
For four and more qubits there are infinitely
many inequivalent classes of genuinely entangled states. It is a far more
difficult question how these states and their entanglement can be classified.
Whenever we refer to the classification of multipartite entanglement,
we have in mind the classification of genuinely entangled states, rather
than the one according to separability classes.

As to the quantitative characterziation 
of multipartite entanglement, to date the most relevant questions 
are quantification of genuine entanglement, and of entanglement in
bipartitions of multipartite states. For the important three-qubit case,
this covers all separability classes. Therefore, we restrict our discussion
to measures of bipartite and genuine entanglement. Further, as
the vast majority of studies has been carried out for qubit systems,
we focus on measures for multi-qubit entanglement and mention higher-dimensional
systems where it is appropriate.

\subsection{Measures for bipartite entanglement in multipartite states}
\label{sec:measuresbipartitinmulti}

\subsubsection{Concurrence, 1-tangle, and negativity}
\label{sec:conc1tangleneg}

Consider a pure $N$-party state 
$\psi\in \mathcal{H}=\mathcal{H}_{A_1}\otimes\cdots\otimes\mathcal{H}_{A_N}$
and a single-party bipartition $A_j|A_1,\ldots,A_{j-1},A_{j+1},\ldots A_N$.
The local reduced state of party $A_j$ is
\begin{equation}
   \rho_{A_j}\ =\ \tr_{A_1\ldots A_{j-1},A_{j+1}\ldots A_N}\ket{\psi}\!\bra{\psi} 
\end{equation}
that is, we take the trace over the degrees of freedom of all the other parties
(analogously for mixed multipartite states $\rho=\sum_k p_k\ket{\psi_k}\!\bra{\psi_k}$).
In principle, any bipartite entanglement measure from 
section~\ref{sec:bipartitemeasures} that is
defined on the local state can be used
to describe the entanglement in such bipartitions of multipartite states.
The appropriate choice depends on the resource that is to be described.

Often, the concurrence (section~\ref{sec:iconcurrence}) is considered
\begin{equation}
        C_{A_j|A_1\ldots A_{j-1}A_{j+1}\ldots A_N}(\psi)\ \equiv
        \ C_{A_j}(\psi)\ =
        \ \sqrt{2\left(1-\tr\rho^2_{A_j}\right)}
\ \ .
\label{eq:1partyconc}
\end{equation}
If the $j$-th party is a qubit, one often defines the 
{\em 1-tangle}~\cite{Coffman2000}
\begin{equation}
        \tau_{A_j|A_1\ldots A_{j-1}A_{j+1}\ldots A_N}(\psi)
        \ \equiv\ \tau_{A_j}(\psi)
        \ =\ 
        \left(C_{A_j}(\psi)\right)^2
\ \ .
\label{eq:tau1}
\end{equation}
We mention that if all parties are qubits,
the average of this quantity over all parties was termed
{\em global entanglement} and considered
in Refs.~\cite{MeyerWallach2002,Brennen2003}
\begin{equation}
      Q(\psi)\ =\ \frac{1}{N}
                      \sum_{j=1}^N \tau_{A_j}
\ \ .
\label{eq:MeyerWallach}
\end{equation}
Note that, essentially, 
there is no difference between the definitions (\ref{eq:1partyconc})
and (\ref{eq:tau1}) as long as the multipartite state is pure.
However, it is relevant for mixed states since, in general,
$\left(\mathrm{convex\ roof}\left[\sqrt{\tau_{A_j}}\right]\right)^2
\neq \mathrm{convex\ roof}\left[\tau_{A_j}\right]$.

Alternatively, one may
use, e.g., the negativity to quantify the entanglement in $\rho_{A_j}$
\begin{equation}
        \mathcal{N}_{A_j}(\psi)\ =
        \ \case12\left(\|\rho^{T_{A_j}}\|_1-1\right)
\label{eq:1partyneg}
\end{equation}
which is particularly convenient for mixed multipartite
states just because this quantity is easily calculated.

We mention that the measures discussed by Emary~\cite{Emary2004}
amount to an application of the $k$-concurrence
to multipartite states.

\subsubsection{Concurrence of two qubits in a multipartite state}
\label{sec:concof2bitsinmulti}

Another method of analyzing bipartite entanglement in a 
multipartite state $\psi$ is to consider the 
bipartition $A_jA_k|A_1\ldots A_{j-1}A_{j+1}\ldots A_{k-1}A_{k+1}\ldots A_N$
and the corresponding two-party reduced
density matrix 
\begin{equation}
        \rho_{A_jA_k}\ =\ \tr_{A_{l\neq j,k}}\ket{\psi}\!\bra{\psi}
\ \ .
\end{equation}
This is particularly interesting for multi-qubit states, because
then the Wootters concurrence  
$C(\rho_{A_jA_k})\equiv C_{A_jA_k}(\psi)\equiv C_{A_jA_k}$ 
can be evaluated
exactly using the method of section~\ref{sec:woottersconcurrence}
(for an important application to condensed matter theory see, e.g.,
Refs.~\cite{Osterloh2002,Osborne2002}).

With the quantities defined so far a profound law of quantum correlations
can be stated: It is the  monogamy of bipartite qubit entanglement
which was conjectured by Coffman {\em et al.}~\cite{Coffman2000} and proved by
Osborne and Verstraete~\cite{Osborne2006}
\begin{equation}
        \tau_{A_1}\ \geqq\ 
                   C_{A_1A_2}^2+
                   C_{A_1A_3}^2+\ldots +
                   C_{A_1A_N}^2
\ \ ,
\label{eq:OsborneVerstraete}
\end{equation}
where $\rho_{A_1}$ and $\rho_{A_1 A_j}$ are the reduced one-qubit and two-qubit
states, respectively, of a pure $N$-qubit state $\psi$. Note that on the right-hand side
in (\ref{eq:OsborneVerstraete}) it does not matter whether the convex roof is
taken for $C$ or $C^2$. This is because for two qubits it is always possible
to find an optimal pure-state decomposition with the same concurrence 
for all its elements.

We mention also a variation of (\ref{eq:OsborneVerstraete})
for the negativities of the bipartitions~\cite{Ou2007}
\begin{equation}
        \mathcal{N}_{A_1}^2\ \geqq\ 
                   \mathcal{N}_{A_1A_2}^2+
                   \mathcal{N}_{A_1A_3}^2+\ldots +
                   \mathcal{N}_{A_1A_N}^2
\label{eq:OuFan}
\end{equation}
which follows immediately taking into account (\ref{eq:neg_vs_conc2}),
(\ref{eq:concestimate_mixed}) in Section~\ref{sec:PTconcneg}.
Other interesting monogamy inequalities related to the Osborne-Verstraete
relation~(\ref{eq:OsborneVerstraete}) were found for the concurrences
of states with higher-dimensional
local systems~\cite{Fei-Monog2008} and for the
entanglement of formation~\cite{Oliveira2014,Bai2014}.

\subsubsection{Concurrence of assistance}
\label{sec:concofassist}

Instead of asking, as in the previous section, about the minimum
entanglement contained in a two-qubit reduced state   
$\rho_{A_j A_k}$, one might wonder what the possible {\em maximum}
concurrence compatible with this state is. 
Thus one can define {\em concurrence of assistance}~\cite{Laustsen2003}
\begin{equation}
     C^{\sharp}_{A_jA_k}(\psi)\ =\ \max_{\mathrm{decompositions}}
                                   \sum_j p_j C(\psi_j)
\label{eq:CoA}
\end{equation}
where the optimization is over the decompositions of 
$\rho_{A_jA_k}=\sum_jp_j\ket{\psi_j}\!\bra{\psi_j}$.
In the multipartite case, this concept is also termed
{\em localizable entanglement}~\cite{VerstraetePopp2004,Popp2005}.
Operationally,
$C^{\sharp}_{A_jA_k}$ 
corresponds to the maximum average single-copy entanglement the two parties
$A_j$ and $A_k$ can achieve through local operations and
classical communication by the other
parties in the multipartite state. According to the
Hughston-Jozsa-Wootters theorem~\cite{HJW1993},
$C^{\sharp}_{A_jA_k}$ depends only on the reduced two-qubit state.
The calculation of 
$C^{\sharp}_{A_jA_k}$ is simple: one follows the procedure
of calculating the eigenvalues $r_j$ of
the $R$ matrix in (\ref{eq:Rmatrix}) for $\rho_{A_jA_k}$
and obtains
\begin{equation}
     C^{\sharp}_{A_jA_k}(\psi)\ =\ \sum_{j=1}^4 \sqrt{r_j} \ \ .
\label{eq:CoAr}
\end{equation}
The concept of concurrence of assistance can also be generalized
to higher-dimensional systems by replacing Wootters' concurrence with
the $G$ concurrence~\cite{Gour2006}.

\subsection{Measures for genuine multipartite entanglement}
\label{sec:measforGME}

If one simply wants to make sure that a 
multipartite state is genuinely entangled
it suffices to check that it is not separable on any
bipartition (or, conversely, that it has a minimum amount of entanglement
in each bipartition).
Consequently any bipartite measure can be used for this purpose,
however, clearly one would go back to those measures which can
easily be calculated or estimated. The obvious choices---the concurrence
and the negativity---indeed lead to the methods that have been
most successful in recent years. 

\subsubsection{Concurrence of genuine multipartite entanglement}
\label{sec:GMEconc}
The concept of minimum entanglement on all bipartitions in
a multipartite state was first applied by Pope and 
Milburn~\cite{PopeMilburn2003} invoking the von Neumann entropy.
Subsequently, Scott~\cite{Scott2004} and Love {\em et al.}~\cite{Love2007}
studied a generalization to the global entanglement~\eref{eq:MeyerWallach}
by considering the set of averaged linear entropies for general bipartite
splits. We mention that, for very large numbers of parties (and therefore
bipartitions), one can also study the statistics of purities
over bipartitions~\cite{Facchi2006,Parisi2010}. However, since purities
are not entanglement monotones, the relation of the results to entanglement
properties of the states is not clear.

A technically simpler quantity is the concurrence of genuine multipartite
entanglement, for short GME concurrence~\cite{MaHuber2011}. Consider
all possible bipartitions $\gamma_j= \{P_j|Q_j\}$ of a pure multipartite
state $\psi$. Then
\begin{equation}
    C_{\mathrm{GME}}(\psi)\ =\ \min_{\gamma_j} \sqrt{2\left(1-\tr \rho_{P_j}^2
                                                \right)}
\ \ .
\label{eq:GMEconc}
\end{equation}
The extension of GME concurrence to mixed states is via the
convex-roof extension (\ref{eq:convexroof}), Section~\ref{sec:convexity}.
Note the difference of this definition with Akhtarshenas' 
work~\cite{Akhtarshenas2005} where all linear entropies are added,
so that the corresponding concurrence is nonzero as soon as a single
bipartition has entanglement. 
   By considering such sums for all possibilities of
   $k$ partitions, this idea can be used to quantify the
   $k$-nonseparability of a pure state~\cite{Gao2012}.
Definition~(\ref{eq:GMEconc}) has the
advantage that lower bounds can readily be found by straightforward
extension of the lower bound of Sec.~\ref{sec:boundsconcmix} 
to the multipartite setting~\cite{MaHuber2011,WuHuber2012,Huber2012-PRA,Huber2013-PRA}.
To this end, one selects a subset $\mathcal{M}$ of $\eta$ 
multilevel index pairs $\{M,M'\}$
as in Sec.~\ref{sec:boundsconcmix}, only that now these index pairs
may originate from any bipartition $\gamma_j$
of $\mathcal{H}$, not just a single one.
That is, if we consider a particular bipartition $\gamma_j=\{A_j|B_j\}$,
the indices $M$, $M'$ may be decomposed in parts $M\rightarrow K_jL_j$,
$M'\rightarrow K_j'L_j'$ where $K_j,K_j'\in A_j$ and $L_j,L_j'\in B_j$.
Then the bound reads
\begin{eqnarray}
    C_{\mathrm{GME}}(\rho)\ \geqq\ & \frac{2}{\sqrt{\eta}} &
                  \sum_{M=K_jL_j,M'=K_j'L_j'\in\mathcal{M}} 
    \Bigl(|\rho_{K_jL_j,K_j'L_j'}|\ -  \Bigr.
\nonumber\\
     &&    -\ {\sum_{\gamma_j}}^{\prime} 
                        \sqrt{\rho_{K_jL_j',K_jL_j'}
                              \rho_{K_j'L_j,K_j'L_j}}
               \Bigl.                   \Bigr)
 \ .
\label{eq:GMEconcbound}
\end{eqnarray}
The prime in the last sum means that not necessarily all terms
have to be summed up. This is because by summing over all choices
$\{M,M'\}$ and corresponding bipartitions $\gamma_j$ repetitions of diagonal
elements may occur. It suffices to take into account the
maximum number of repetitions that occur for a single bipartition.
In practice, it may
be somewhat subtle how to select the subset $\mathcal{M}$ in order
to produce a good bound,
so it is difficult to establish a general rule.

The simplest example for this method is a biseparability criterion
and GME concurrence bound based on the three-qubit GHZ state.
For the off-diagonal index pair we choose the $\{000,111\}$,
so $\eta=1$. For three qubits ($A$, $B$, $C$) there are three bipartitions
$\gamma_1=\{A|BC\}$, $\gamma_2=\{B|AC\}$ and $\gamma_3=\{C|AB\}$. 
The corresponding permuted index combinations $\{K_jL_j',K_j'L_j\}$ are
$\{100,011\}$, $\{010,101\}$, $\{001,110\}$ which all occur once.
Hence we find
\begin{eqnarray}
    C_{\mathrm{GME}}(\rho)\ \geqq\  &  2\left( \right. &  
    |\rho_{000,111}|  -  \sqrt{\rho_{100,100}\rho_{011,011}}
\nonumber\\
                  &&  - \sqrt{\rho_{010,010}\rho_{101,101}}
                    - \sqrt{\rho_{001,001}\rho_{110,110}}
       \left.                               \right)
\ \ .
\label{eq:GMEconcGHZ3}
\end{eqnarray}
This entanglement detection criterion
was first derived by G{\"u}hne and Seevinck~\cite{GuhneSeevinck2010}
and generalized by Huber {\em et al.}~\cite{Huber2010}. The method
is rather powerful and enables entanglement detection and quantification
in numerous situations that were inaccessible before. An immediate
conclusion in Ref.~\cite{GuhneSeevinck2010} was, for example,
that GHZ-diagonal states contain entanglement as soon as the fidelity
of one GHZ state exceeds $\case12$. The exact GME concurrence of GHZ-diagonal
states was found in Ref.~\cite{HuberEberly2012}. It is worth noting that this
approach can be extended to $k$-separability, as well as combined with permutation
invariance to produce lower bounds on entanglement for the permutation-invariant
part of a state~\cite{vEnk2014} that apply to arbitrary multipartite states,
very much in the spirit of section~\ref{sec:threearbitrary}.

It is interesting to reflect upon the structure of \eref{eq:GMEconcGHZ3}
which lower bounds the concurrence by a difference between off-diagonal
and diagonal elements of the density matrix. In fact, this is the essential
idea  behind various detection and classification schemes of entanglement:
Given a particular entanglement class, the modulus of an 
off-diagonal element of a density matrix~\footnote{%
        An approach for entanglement detection based on
             the maximization of  (generalized) off-diagonal matrix elements
             was considered in~\cite{Zyczkowski2011}.
                              }
in that class cannot exceed a certain value that depends
on related diagonal elements (here: determined via the partial transpose).

\subsubsection{Genuine multipartite negativity}
\label{sec:GMEneg}

As an alternative to GME concurrence one may consider
the genuine multipartite negativity (GMN) which can 
be defined directly on mixed states~\cite{Jungnitsch2011,Hofmann2014}
\begin{equation}
    \mathcal{N}_{\mathrm{GME}}(\rho)\ =\ \min_{\rho=\sum p_k\rho_k} 
                                               \sum_k p_k \mu(\rho_k)\ ,
\ \ \
           \mu(\rho) = \min_{\gamma_j} 
                         \case12 \left(\|\rho^{T_{A_j}}\|_1 -1 \right)
\label{eq:GMEneg}
\end{equation}
where $\rho^{T_{A_j}}$ is the partial transpose with respect to party
$A_j$ in the bipartition $\gamma_j$ of the multipartite state $\rho$,
so that $\mu(\rho)$ is the bipartite negativity, 
minimized over all bipartitions of $\rho$.
The first equation defines $\mathcal{N}_{\mathrm{GME}}$ as the
convex hull of $\mu(\rho)$ and guarantees its convexity.
Rather than the border of the biseparable states this quantity detects
whether or not a state is a mixture of PPT states. The advantage, however,
is that it can be implemented favorably in a semidefinite 
program~\cite{Boyd1996}. Via this method, various interesting
bounds for {\em white-noise tolerance} could be found, i.e., 
the maximum weight $1-p^*$ 
for which an entangled $N$-qubit state $\psi_{\mathrm{ent}}$ remains
genuinely entangled in a mixture
\begin{equation}
             p\ \ket{\psi_{\mathrm{ent}}}\!\bra{\psi_{\mathrm{ent}}}\ +\
             \frac{1-p}{2^N}\id_{2^N}
\ \ .
\end{equation}
For example, it could be established that for $N$-qubit linear cluster
states~\cite{Briegel2001} the white-noise tolerance tends to 1
for large $N$~\cite{Jungnitsch2011}.

%

\subsection{Measures based on polynomial invariants}
\label{sec:polyinvmeas}

In research on entanglement classification and quantification, 
invariant polynomial functions were investigated early 
on~\cite{Schlienz1995,Schlienz1996,Grassl1998,Linden1998,Linden1999,Nielsen1999,Carteret1999,Carteret2000,Rains2000,Sudbery2001,Kus2001}.
The initial focus was on invariants under local unitaries. 
This view changed with the investigation of SLOCC~\cite{VidalTarrach1999,Vidal2000,Bennett2000,DVC2000,Verstraete2001} and
in particular at the point when Coffman {\em et al.} discovered
the residual tangle~\cite{Coffman2000}, which redirected the attention
towards the special linear group~\cite{Brylinski2002a,Brylinski2002b,Verstraete2001,Verstraete2002,Klyachko2002}. 
Local unitary invariance is characterized
by many more parameters than SL invariance, therefore it appears to
describe more subtle resources. In particular, it is linked to
deterministic interconvertibility of states, see, e.g., 
 \cite{Nielsen1999,Kraus2010,deVicente2013}, but also, for example,
to topological properties of multipartite 
states~\cite{Johansson2012,Johansson2013,Johansson2014}.
Here we exclusively consider invariance properties with
respect to the special linear group, and their relation with entanglement.

\subsubsection{Three-tangle}
\label{sec:3tangle}
It is a salient feature of multipartite entanglement that there
are locally inequivalent classes of genuinely entangled states~\cite{DVC2000}.
For example, for any qubit number $N$ the corresponding 
GHZ state
\begin{equation}
    \ket{\mathrm{GHZ}_N}\ =\ \frac{1}{\sqrt{2}}\left( \ket{00\ldots 0}
                                                    + \ket{11\ldots 1}
                                               \right)
\label{eq:GHZN}
\end{equation}
cannot be transformed by local invertible operations into
the $W$ state
\begin{equation}
    \ket{W_N} = \frac{1}{\sqrt{N}}\left( \ket{0\ldots 001}
                                                    + \ket{0\ldots 010}
+\ldots
                                                    + \ket{1\ldots 000}
                                               \right)
\label{eq:WN}
\end{equation}
where all basis states are understood to contain $N$ entries. 
Similarly, the $N$-qubit cluster state is locally inequivalent
to any $N$-qubit GHZ state (or $W$ state)~\cite{Wu2001,Hein2006}.
By applying the entanglement measures in the preceding sections  it would
be difficult, if not impossible, to distinguish such inequivalent classes.
However, there is an elegant way out of this problem. It was noticed
by Coffman {\em et al.} that a polynomial function of the coefficients
in a quantum state may help to distinguish  the GHZ from the $W$ state
for three qubits~\cite{Coffman2000}. They termed it the
{\em residual tangle} of the three-qubit state $\psi\in
\mathbb{C}^2\otimes\mathbb{C}^2\otimes\mathbb{C}^2$
\begin{eqnarray}
  \tau_{\mathrm{res}}(\psi) &=& 4\left|d_1 - 2 d_2 + 4 d_3\right|,\nonumber\\
  d_1 &=& \psi_{000}^2\psi_{111}^2 + \psi_{001}^2\psi_{110}^2 + \psi_{010}^2\psi_{101}^2
  + \psi_{011}^2\psi_{100}^2\nonumber\\
  d_2 &=& \psi_{000}\psi_{001}\psi_{110}\psi_{111} + \psi_{000}\psi_{010}\psi_{101}\psi_{111} +
 \nonumber\\ &&
  + \psi_{000}\psi_{011}\psi_{100}\psi_{111}
  + \psi_{001}\psi_{010}\psi_{101}\psi_{110} + 
 \nonumber\\ &&
  + \psi_{001}\psi_{011}\psi_{100}\psi_{110} +
  \psi_{010}\psi_{011}\psi_{100}\psi_{101}\nonumber\\
  d_3 &=& \psi_{000}\psi_{110}\psi_{101}\psi_{011} + \psi_{100}\psi_{010}\psi_{001}\psi_{111}
\ \ .
\label{eq:residualtangle} 
\end{eqnarray}
For reasons explained in Sec.~\ref{sec:hominv} 
(see also Refs.~\cite{Gour2010,Viehmann2012}), we reserve
the name ``three-tangle''~\footnote{%
    The term '3-tangle' was first used,
    to our knowledge, by D\"ur {\em et al.}~\cite{DVC2000}. It adapts to
    the nomenclature by Coffman {\em et al.}~\cite{Coffman2000} to call
    a quantity 'tangle' if it is of degree 4 in the state coefficients.
    Therefore the degree-2 quantity in (\ref{eq:threetangle}) should
    not be called 'three-tangle', in principle.
    However, since the name has become so popular during the past decade
    we continue using it.}  
for the square root of $\tau_{\mathrm{res}}$
\begin{equation}
    \tau_3\ =\ \sqrt{\tau_{\mathrm{res}}} 
\ \ .
\label{eq:threetangle} 
\end{equation}
D{\"u}r {\em et al.} proved that both $\tau_3$ and $\tau_{\mathrm{res}}$
are entanglement monotones.
Intriguingly, it turned out only 
afterwards~\cite{Brylinski2002a,Brylinski2002b}
that these quantities---just as Wootters' concurrence 
(\ref{eq:concurrence})---are invariant under 
SL$(2,\mathbb{C})$ transformations on each qubit. Further it was 
noticed~\cite{Brylinski2002a,Brylinski2002b,Miyake2003} 
that both the concurrence
and the three-tangle are related to Cayley's hyperdeterminant~\cite{Cayley1845}.

It is easily checked that
\begin{equation}
       \tau_3(\mathrm{GHZ_3})=1\ \ , \ \ \ \tau_3(W_3)=0
\end{equation}
and one can conclude that it is impossible to convert 
a single copy of the
$W$ state with nonvanishing probability into a GHZ state
by means of invertible local operations.
For three qubits the three-tangle
(just as the concurrence for two qubits) is the only independent
LSL-invariant polynomial. 
Correspondingly, all pure three qubit states $\psi$
with $\tau_3(\psi)\neq 0$ are locally equivalent to the GHZ state.
We have
\begin{equation}
       0\ \leqq\ \tau_3(\mathrm{\psi})\ \leqq \ 1
\end{equation}
which suggests calling the GHZ state maximally entangled.

We may use the three-tangle to quickly illustrate 
the peculiarities of multipartite entanglement. In section 
\ref{sec:twoqubit} it was mentioned that for two qubits
the concurrence for the {\em superposition} of a 
Bell state and an orthogonal product state
equals the weight of the Bell state~\cite{Abouraddy2001}. 
If we consider an analogous superposition  for three qubits
\begin{equation}
    \ket{\phi(p,\varphi)}=\sqrt{p}\ket{\mathrm{GHZ_3}}-e^{\rmi \varphi}\sqrt{1-p}
                                          \ket{W_3}
\end{equation}
we find~\cite{Lohmayer2006}
\begin{equation}
\tau_{\mathrm{res}}(p,\varphi)\ =\ \left|p^2-\case{8\sqrt{6}}{9}\sqrt{p(1-p)^3}
                                  e^{3\rmi \varphi} \right|
\end{equation}
which vanishes, for example, for $\varphi=0$ and
\begin{equation}
        p_0\ =\ \frac{4\sqrt[3]{2}}{3+4\sqrt[3]{2}}\ \ .
\end{equation}
That is, while for two qubits the effect of the orthogonal unentangled
states is merely to proportionally reduce the weight of the maximally entangled state, 
for three qubits the action of superposing a $W$ state is more complex
-- it may be more harmful with respect to GHZ-type entanglement
(e.g., for $\varphi=0$, $p\geqq p_0$) 
as well as less harmful (for $\varphi=\pi$ and all $p$).

\subsubsection{Four-qubit invariants}
\label{sec:4bitinvs}

While for two and three qubits there is only a single independent
invariant polynomial, for four qubits there are infinitely many.
The invariant polynomials form an algebra (actually a ring),
so there is a set of generating polynomials. In algebraic
geometry, the Hilbert series is a standard tool
to find degrees and dimensions for polynomial spaces.
Its application becomes increasingly difficult for larger systems,
however, in the four-qubit case a complete set of generating polynomials
is known and was first described by Luque and Thibon~\cite{Luque2003}.
It consists of one degree-2 polynomial, three degree-4 polynomials
(among which only two are algebraically independent), and one
degree-6 polynomial. Note that, in order to obtain an entanglement
monotone with nice properties also for the convex-roof extension
(cf.~Section~\ref{sec:hominv}), 
the appropriate power of the polynomials' modulus has 
to be taken, so that the result is of homogeneous degree 2 in
the state coefficients.

We consider four-qubit states $\psi$ in a four-qubit 
Hilbert space $\psi\in \mathcal{H}_{ABCD}=
\left(\mathbb{C}^2\right)^{\otimes 4}$.
The degree-2 polynomial is the  straightforward
generalization of Wootters' concurrence~(\ref{eq:concurrencesigma22})
\begin{equation}
  H(\psi) = \left|\bra{\psi^*}\sigma_y\otimes\sigma_y\otimes
                  \sigma_y\otimes\sigma_y
            \ket{\psi}\right|
\ \ .
\label{eq:H4bit}
\end{equation}
With this definition $H$ has a (physically irrelevant) prefactor of 2
compared to the definitions in \cite{Luque2003} and \cite{Dokovic2009}.
We note that this quantity, in a strict sense, cannot be a measure
of genuine entanglement as it yields 1 on a tensor product of 
two two-qubit Bell states~\cite{Wong2001}. Therefore this kind of measure
needs to be complemented with a measure of genuine multipartite
entanglement. On the other hand, the convex roof of $H$ can
be evaluated exactly by using the procedure due to 
Wootters~\cite{Wootters1998}
and Uhlmann~\cite{Uhlmann2000}.

The degree-4 invariants are denoted $L$, $M$, and $N$ and 
can be written in terms of the pure-state coefficients
$\ket{\psi}=\sum\psi_{i_A i_B i_C i_D }\ket{i_Ai_Bi_Ci_D}$
\begin{equation}
          L(\psi)\ =\ \left|
                \begin{array}{cccc}
                \psi_{0000} & \psi_{0100} & \psi_{1000} & \psi_{1100}\\
                \psi_{0001} & \psi_{0101} & \psi_{1001} & \psi_{1101}\\
                \psi_{0010} & \psi_{0110} & \psi_{1010} & \psi_{1110}\\
                \psi_{0011} & \psi_{0111} & \psi_{1011} & \psi_{1111}
                \end{array}
                \right|
\label{eq:LTL}
\end{equation}
and $M$, $N$ analogous with the second and the third, 
or the second and the fourth qubit exchanged, respectively.
These invariants are not independent, since
\begin{equation}
       L+M+N\ = \ 0
\ \ .
\label{eq:LMNdependence}
\end{equation}
In \cite{Eltschka2012PRA} it was shown that the squared moduli of $L$, $M$, $N$
are nothing but the determinants of the two-qubit reduced density
matrices of the four-qubit state $\psi$:
\numparts
\begin{eqnarray}
\label{eq:LMNdetsa}
|L|^2\ & = & \ \det\left(\tr_{CD}\pi_{\psi}\right)\\
\label{eq:LMNdetsb}
|M|^2\ & = & \ \det\left(\tr_{BD}\pi_{\psi}\right)\\
\label{eq:LMNdets}
|N|^2\ & = & \ \det\left(\tr_{BC}\pi_{\psi}\right)
 \ .
\end{eqnarray}
\endnumparts
One sees that also $L$, $M$, $N$ may be nonzero on 
separable states (for example, on a tensor product
of Bell states). Note also that, according to (\ref{eq:LMNdetsa})--(\ref{eq:LMNdets})
the quartic invariants may be regarded as (powers of) $G$-concurrences on
$4\times 4$ bipartite systems, that is, they are SL$(4,\mathbb{C})$
invariants (cf.\ also \cite{Emary2004}).

To get a complete set of generators, one independent degree-6 invariant
is required. A possible choice is the degree-6 filter $\mathcal{F}^{(4)}_1$
\cite{Osterloh2005} that is discussed in section \ref{sec:invhigher}.
The sextic polynomial $\mathcal{F}^{(4)}_1$ is symmetric under qubit
permutations and defines another permutation-symmetric polynomial $W$ via
\begin{equation}
         \mathcal{F}^{(4)}_1 \ =\ 32W - H^3
\ \ .
\label{eq:F41vsW}
\end{equation}
In \cite{Luque2003} instead, the degree-6 polynomial $D_{xt}$
was used which belongs to a family of three invariants that obey
\begin{equation}
         W\ =\ D_{xy}+D_{xz}+D_{xt}
\label{eq:4bitW}
\end{equation}
and
\numparts
\begin{eqnarray}
         \case12 H(N-M)\ & = &\ 3 D_{xy}- W\ \ ,
\\
         \case12 H(L-N)\ & = &\ 3 D_{xz}- W\ \ ,
\\
         \case12 H(M-L)\ & = &\ 3 D_{xt}- W\ \ .
\label{eq:4bitWGln}
\end{eqnarray}
\endnumparts

There are a few more four-qubit polynomials that deserve particular
interest. There are the degree-8 filter $\mathcal{F}^{(4)}_2$,
and the degree-12 filter $\mathcal{F}^{(4)}_3$
\cite{Osterloh2005,Osterloh2006,Dokovic2009} whose precise
definitions are also given in section \ref{sec:invhigher}.
The peculiarity of the filters $\mathcal{F}^{(4)}_1$,
$\mathcal{F}^{(4)}_2$,$\mathcal{F}^{(4)}_3$ is that they
do vanish on any biseparable state of the Hilbert space $\mathcal{H}_{ABCD}$
and generate an ideal of polynomial invariants~\cite{Dokovic2009}.
Finally, we mention also the degree-24 four-qubit 
hyperdeterminant Det whose relation to four-qubit entanglement
was discussed in \cite{Miyake2003} (we quote the result
from \cite{Dokovic2009}):
\begin{equation}
2^{12} 3^3\ \mathrm{Det}  = - 2\mathcal{F}^{(4)}_1 A+
                   \left( 128\Sigma- H^4)B-(256\Pi+\case18 H^6\right)^2
\label{eq:hyperdet41}
\end{equation}
where
\numparts
\begin{eqnarray}
     \Sigma & = & L^2+M^2+N^2\\
     \Pi & =& (L-M)(M-N)(N-L)\\
     A & = & \case{5}{512} H^9+\case{5}{16}W H^6 - \case{9}{2}\Sigma H^5
             +2(5W^2-24\Pi)H^3-\nonumber\\ 
        &&   -240 W\Sigma H^2 +  768\Sigma^2 H+192W(3W^2+8\Pi) \\
     B & = & \case{1}{256} H^8 -\case{17}{2} \Sigma H^4 -96\Pi H^2 
              +256 \Sigma^2
\ \ .
\label{eq:hyperdet42}
\end{eqnarray}
\endnumparts
It is worthwhile noting that the hyperdeterminant is maximized
by the four-qubit state $\ket{X_4}$ in~\eref{eq:irrbal3}, cf.~\cite{Djokovic2013}.

\subsubsection{Invariants of degree 2 and 4}
\label{sec:inv2and4}

It is evident from the discussion of the four-qubit polynomials
that even for small systems with $N\lesssim 10$ qubits 
it becomes exceedingly complicated to work with invariants of
a higher degree. Therefore, one would hope to extract much
of the relevant physical information from the invariants
of a lower degree, that is, degrees 2 and 4 (there are no
nontrivial invariants of odd-integer degree~\cite{Brylinski2002b}).
Therefore it is useful to present a set of standard rules on how 
to construct such polynomials~\cite{Osterloh2005,Osterloh2006,Eltschka2012PRA}.
In the following we assume that the pure state $\psi$ be always
element of the correct 
Hilbert space corresponding to the quantity under consideration.

First we recall that the two-qubit concurrence (\ref{eq:concurrencesigma22})
can be written as 
\begin{equation}
       C(\psi)=\left| \bra{\psi}\sigma_y\otimes\sigma_y\ket{\psi^*}\right|
              =\left| \bra{\psi^*}\sigma_y\otimes\sigma_y\ket{\psi}\right|
              \equiv |H^{(2)}(\psi)|
\end{equation}
where we introduce a shorthand notation for the expectation value
\begin{equation}
       H^{(2)}(\psi) = \bra{\psi^*}\sigma_y\otimes\sigma_y\ket{\psi}
                 \equiv \langle \sigma_2\sigma_2 \rangle \ \ .
\label{eq:shorthand1}
\end{equation}
Here $\psi^*$ is the state $\psi$ with complex conjugate entries.
Moreover, we enumerate the Pauli operators as 
$\sigma_1\equiv \sigma_x$, $\sigma_2\equiv \sigma_y$, 
$\sigma_3\equiv \sigma_z$ and $\sigma_0\equiv \id_2$.
Straightforward algebra shows that the invariant for the residual tangle
(\ref{eq:residualtangle})
can be expressed as
(where we use the definition $\tau_{\mathrm{res}}\equiv \left| B^{(3)}_1 \right|$)
\begin{eqnarray}
  B^{(3)}_1(\psi) &=& -\bra{\psi}\sigma_0\otimes\sigma_2\otimes\sigma_2\ket{\psi^*}
                \bra{\psi}\sigma_0\otimes\sigma_2\otimes\sigma_2\ket{\psi^*} +
\nonumber\\
   && +\bra{\psi}\sigma_1\otimes\sigma_2\otimes\sigma_2\ket{\psi^*}
                \bra{\psi}\sigma_1\otimes\sigma_2\otimes\sigma_2\ket{\psi^*} +
\nonumber\\
   && +\bra{\psi}\sigma_3\otimes\sigma_2\otimes\sigma_2\ket{\psi^*}
                \bra{\psi}\sigma_3\otimes\sigma_2\otimes\sigma_2\ket{\psi^*} 
\label{eq:umschreib}
\end{eqnarray}
and, if we use the shorthand notation of 
(\ref{eq:shorthand1})
\begin{eqnarray}
       B^{(3)}_1(\psi) \ &=& \ \sum_{\mu=0,1,3} (-1)^{\mu+1}
                 \langle \sigma_{\mu}\sigma_2\sigma_2 \rangle 
                 \langle \sigma_{\mu}\sigma_2\sigma_2 \rangle 
\nonumber\\
                    &\equiv &\
                 \langle \sigma_{\mu}\sigma_2\sigma_2 \rangle 
                 \langle \sigma^{\mu}\sigma_2\sigma_2 \rangle 
\label{eq:shorthand2}
\end{eqnarray}
where, in the last line, we use the summation convention 
with a ``metric'' $\{-1,1,0,1\}$. Why does this work?
This is because, for $S\in \mathrm{SL}(2,\mathbb{C})$
\begin{equation}
      S \sigma_2 S^T \ =\  \sigma_2\ \ , \ \ \
      (S\otimes S) \sigma_{\mu}\otimes\sigma^{\mu}(S\otimes S)^T 
       =  \sigma_{\mu}\otimes\sigma^{\mu}
\end{equation}
that is, the expression in (\ref{eq:umschreib}), (\ref{eq:shorthand2}) is 
local
SL invariant on each qubit by construction. Note that one cannot simply
use $\langle \sigma_2\sigma_2\sigma_2\rangle$ because
combining a real symmetric operator $A$ with an odd number of $\sigma_2$
operators, the antilinear expectation value always vanishes
\begin{eqnarray}
       \bra{\psi^*}\sigma_2^{\otimes(2n+1)}\otimes A \ket{\psi}^*\
      & = &\ - \bra{\psi}\sigma_2^{\otimes(2n+1)}\otimes A \ket{\psi^*}
\nonumber\\
      & = &\ + \bra{\psi}\sigma_2^{\otimes(2n+1)}\otimes A \ket{\psi^*}\  =\  0
\ \ .
\end{eqnarray}
In particular we have for pure single-qubit states $\phi$
\begin{equation}
      \bra{\phi}\sigma_2 \ket{\phi^*}  =  0\ \ , \ \ \
      \left(\bra{\phi}\otimes\bra{\phi}\right)(\sigma_{\mu}\otimes\sigma^{\mu})
      \left(\ket{\phi^*}\otimes\ket{\phi^*}\right) = 0
\ \ .
\label{eq:combs}
\end{equation}
Hence, if a single qubit is separable in the three-qubit state $\psi$ in
(\ref{eq:umschreib}) the expression must vansih -- just as it is the 
case for the three-tangle. The two operators in (\ref{eq:combs})
were called {\em combs} and, accordingly, the technique to systematically
build polynomial invariants for qubit states the {\em invariant-comb
method}~\cite{Osterloh2005,Osterloh2006,Dokovic2009}.
In \cite{Dokovic2009} it was also shown that {\em any} invariant
that can be constructed with a so-called $\Omega$-process
(a standard method in classical invariant theory owed to 
Cayley~\cite{Cayley1846} to systematically obtain invariants)
can be generated also by means of the comb approach.

Now we can write down four-qubit invariants: 
For the degree-2 polynomial $H\equiv H^{(4)}$ we find
\begin{equation}
       H^{(4)}(\psi) 
                 = \langle \sigma_2\sigma_2 \sigma_2\sigma_2 
                 \rangle \ \ .
\end{equation}
As degree-4 invariants we obtain
\numparts
\begin{eqnarray}
  B^{(4)}_{1,2}& = & \langle \sigma_{\mu}\sigma_{\nu} \sigma_2\sigma_2 \rangle
                     \langle \sigma^{\mu}\sigma^{\nu} \sigma_2\sigma_2 \rangle
\\
  B^{(4)}_{1,3}& = & \langle \sigma_{\mu}\sigma_2\sigma_{\nu} \sigma_2 \rangle
                     \langle \sigma^{\mu}\sigma_2\sigma^{\nu} \sigma_2 \rangle
\\
  B^{(4)}_{1,4}& = & \langle \sigma_{\mu}\sigma_2\sigma_2\sigma_{\nu}  \rangle
                     \langle \sigma^{\mu}\sigma_2\sigma_2\sigma^{\nu} \rangle
\ \ .
\end{eqnarray}
\endnumparts
Note that these polynomials are not permutation invariant.
It turns out that~\cite{Dokovic2009}
\numparts
\begin{eqnarray}
       L\ &=&\ \case{1}{48}\left( B^{(4)}_{1,3}-B^{(4)}_{1,4}\right)
\\
       M\ &=&\ \case{1}{48}\left( B^{(4)}_{1,4}-B^{(4)}_{1,2}\right)
\\
       N\ &=&\ \case{1}{48}\left( B^{(4)}_{1,2}-B^{(4)}_{1,3}\right)
\ \ .
\end{eqnarray}
\endnumparts
We mention also the useful identity~\cite{Dokovic2009} 
\begin{equation}
          \left(H^{(4)}\right)^2\ =\ \case13\left(
                     B^{(4)}_{1,2}+B^{(4)}_{1,3}+B^{(4)}_{1,4}
                                            \right)\ \ .
\end{equation}

Obviously this scheme of generating degree-4 invariants
can be extended to any qubit number.
\begin{itemize}
\item
Even qubit number $n=2k$: there are always one degree-2 invariant
\begin{equation}
       H^{(2k)}(\psi) 
                 = \langle \left(\sigma_2\right)^{\otimes 2k}
                 \rangle 
\end{equation}
and $\case12 n(n-1)$ degree-4 polynomials of the type
\begin{equation}
       B^{(2k)}_{a,b}(\psi) 
             = \langle \sigma_2 \ldots \sigma_{\mu}\ldots\sigma_{\nu}
               \ldots \sigma_2 \rangle 
              \langle \sigma_2 \ldots \sigma^{\mu}\ldots\sigma^{\nu}
              \ldots\sigma_2 \rangle 
\end{equation}
where the contractions are located at positions $a$ and $b$
in the brackets, respectively. 
\item For odd qubit number, $n=2k+1$, there is no degree-2 invariant
      and $n$ degree-4 invariants of the type
\begin{equation}
       B^{(2k+1)}_{a}(\psi) 
             = \langle \sigma_2 \ldots \sigma_{\mu} \ldots \sigma_2 \rangle 
              \langle \sigma_2 \ldots \sigma^{\mu} \ldots\sigma_2 \rangle 
\label{eq:Bodd}
\end{equation}
with the contraction at position $a$ in the bracket.
We note the invariants $B^{(2k+1)}_a(\psi)$ are evidently not
invariant under qubit permutations. Therefore one can also
introduce, in addition to those, an explicitly permutation-invariant
polynomial
\begin{equation}
       B^{(2k+1)}_{\mathrm{sym}}(\psi) 
             = \sum_{a=1}^{2k+1}
       B^{(2k+1)}_a(\psi) 
\ \ .
\label{eq:Boddsymm}
\end{equation}
\end{itemize}
In \cite{Eltschka2012PRA} it was shown that 
$B^{(2k+1)}_{2k+1}$ (for odd $n$) 
and the square of $H^{(2k)}$ (for even $n$) 
coincide with the
degree-4 invariants of Wong and Christensen~\cite{Wong2001}.
Clearly, for both even and odd $n$, there are also degree-4 polynomials
with more than two contractions.

\subsubsection{Invariants of higher degree}
\label{sec:invhigher}

With the shorthand notation from the previous section
we can write the precise definitions of the invariants
$\mathcal{F}^{(4)}_j$, $j=1,2,3$, from section \ref{sec:4bitinvs}.
We start with $\mathcal{F}^{(4)}_1$ and recall that a shortcoming
of the invariants $B^{(4)}_{1j}$ ($j=2,3,4$) was that they give
nonzero values for biseparable states. Consider therefore the
definition
\begin{equation}
  \mathcal{F}^{(4)}_1 =  \langle \sigma_{\mu}\sigma_{\nu} \sigma_2\sigma_2 \rangle
                     \langle \sigma^{\mu}\sigma_2\sigma_{\lambda} \sigma_2 \rangle
                     \langle \sigma_2\sigma^{\nu}\sigma^{\lambda} \sigma_2 \rangle
\ \ .
\label{eq:F1}
\end{equation}
First, one notes that separability of any single qubit is excluded
for nonvanishing $\mathcal{F}^{(4)}_1$. This is a very general fact,
non only for qubit invariants~\cite{Verstraete2003}. As to the two-qubit
bipartitions, the expression (\ref{eq:F1}) is constructed in such a way
that for each choice of a two-qubit partition, there is at least
one expectation value, for which a single $\sigma_2$ is paired with a real
symmetric operator, so that it vanishes if this two-qubit bipartition is separable.

This is quite remarkable: We see that the invariant polynomials can be
built in such a way that separability on any bipartition  is excluded.
The price to pay for this is increasing complexity and higher degree
of the polynomial.

In this spirit, also the other filter invariants
$\mathcal{F}^{(4)}_2$ (degree 8) and $\mathcal{F}^{(4)}_3$ (degree 12)
can be defined:
\numparts
\begin{eqnarray}
  \mathcal{F}^{(4)}_2 & = &  
        \langle \sigma_{\mu}\sigma_{\nu} \sigma_2\sigma_2 \rangle
          \langle \sigma^{\mu}\sigma_2\sigma_{\lambda} \sigma_2 \rangle
           \langle \sigma_2\sigma^{\nu}\sigma_2\sigma_{\tau} \rangle
            \langle \sigma_2\sigma_2\sigma^{\lambda} \sigma^{\tau} \rangle
\\
  \mathcal{F}^{(4)}_3 & = &  \case12
        \langle \sigma_{\mu}\sigma_{\nu} \sigma_2\sigma_2 \rangle
          \langle \sigma^{\mu}\sigma^{\nu}\sigma_2 \sigma_2 \rangle
           \langle \sigma_{\rho}\sigma_2\sigma_{\tau}\sigma_2 \rangle
           \langle \sigma^{\rho}\sigma_2\sigma^{\tau}\sigma_2 \rangle\ \times
\nonumber\\
      && \ \ \ \ \ \ \times   
         \langle \sigma_{\kappa}\sigma_2\sigma_2 \sigma_{\lambda} \rangle
         \langle \sigma^{\kappa}\sigma_2\sigma_2 \sigma^{\lambda} \rangle
\ \ .
\label{eq:F2}
\label{eq:F3}
\end{eqnarray}
\endnumparts
In \cite{Dokovic2009} many other examples of higher-degree invariants
also for five qubits can be found.

We mention that it is possible to rewrite the antilinear expectation
values of the comb approach in terms of directly measurable 
quantities which, in general, leads to SL invariants  of higher
degree~\cite{Osterloh2012}. 
However, the approach becomes really
cumbersome even for degree-4 invariants such as the three-tangle
(cf.\ also \cite{Yu2007}).
Therefore, we quote only the simplest possibility, 
the concurrence for pure two-qubit states $\psi$
\begin{equation}
       |C(\psi)|^2\ =\
              \case14 M_{\mu\nu}M_{\kappa\lambda}
\bra{\psi} \sigma_{\mu} \otimes\sigma_{\kappa} \ket{\psi}
              \bra{\psi} \sigma_{\nu} \otimes\sigma_{\lambda} \ket{\psi}
\ \ .
\label{eq:concmodulussquared}
\end{equation}
Curiously, $M_{\mu\nu}$ is given by the full Minkowski metric
$\mathrm{diag}\{1,-1,-1,-1\}$. Since this quantity 
consists of correlation functions of local observables only,
it is---in principle---directly accessible experimentally.
The problem here is that an experiment always deals with mixed
states $\rho$. The correlation functions in (\ref{eq:concmodulussquared})
are then given by $\tr\left(\rho\sigma_{\mu}\otimes\sigma_{\kappa}\right)$,
however, it is not clear whether these correlation functions are
related to the convex roof of the concurrence in a more direct way than
via the fidelity estimate~\ref{sec:concrel}.
Note that the issue of directly measuring entanglement measures
is subtle and requires careful consideration, see, e.g., \cite{vanEnk2007}.

Summarizing this section on polynomial invariants, we have discussed
in some detail one method for systematic generation of SL-invariant
polynomials for multi-qubit states, known as the invariant comb 
approach~\cite{Osterloh2005,Osterloh2006,Dokovic2009}. 
Various other authors proposed generation and investigated properties
of such invariants with respect to entanglement, 
based on mathematical or physical motivation (e.g., 
\cite{Luque2003-170,Luque2006,Levay2006,Jarvis2007,Heydari2008,Sharma2012a,Sharma2012,Sharma2013}.
Moreover, we have not touched upon 
higher-dimensional local systems at all, although
there exist results in the literature also in this direction
(cf., for example, \cite{BriandVerstraete2004,Osterloh2012,Osterloh2013,Jarvis2014}).
In particular, Gour and Wallach~\cite{Gour2013} have developed a method to 
generate invariants systematically also for higher-dimensional systems.

One can conclude that, from the physics point of view,
the relevant polynomials required for the complete characterization
of multi-party entanglement in finite-dimensional systems are probably known.
As it stands at the moment, there is no agreement on which 
of them play a primary role in entanglement quantification, 
and what their precise meaning could be.

\subsubsection{Other local SL invariants}
\label{sec:invother}

In section \ref{sec:qubitmeasures} we have already described
the Lorentz singular values of a two-qubit density matrix,
which are LSL invariant, but cannot obviously expressed in terms
of polynomials. Here we briefly discuss yet another type of 
LSL invariant that can be defined for any $N$-qubit density matrix.

The density matrix $\rho$ of a $d$-dimensional system can be written
in terms of the $d^2-1$ generators $\mathcal{T}_j$ of the SU($d$) algebra
\begin{equation}
       \rho\ =\ \case{1}{d}\id_d+
                 \case{1}{2}\sum_{j=1}^{d^2-1} x_j \mathcal{T}_j
\end{equation}
which we refer to as {\em Bloch representation} of the state\footnote{%
        For this quantity, 
        other names like {\em coherence vector}~\cite{Hioe1981}
        or {\em correlation tensor}~\cite{Brukner2002} are also
        used.}~\cite{Hioe1981,Schlienz1995,Schlienz1996}.
Here, the real entries $x_j$ are defined as 
\begin{equation}
       x_j\ =\ \tr\left(\rho \mathcal{T}_j \right)\ \ \ , \ \
       \tr\mathcal{T}_j\ =\ 0\ \ \ , \ \
          \tr\mathcal{T}_j\mathcal{T}_k\ =\ 2\delta_{jk}\ \ .
\end{equation}
Hence for a single-qubit density matrix
$\rho$ we have (with the Pauli matrices
$\{\mathcal{T}_j\}=\{\sigma_1, \sigma_2, \sigma_3\}$
and $\sigma_0\equiv \id_2$) 
\begin{equation}
       \rho\ =\ \case{1}{2}\sum_{j=0}^{3} x_{j}\ \sigma_j
\ \ .
\label{eq:1bitBloch}
\end{equation}
Now, $x=(x_0,x_1,x_2,x_3)$ 
may be regarded as a four-vector in $\mathbb{R}^4_{1,3}$.
For its Minkowskian length one finds
\begin{equation}
       \|x\|^2 =\ x_0^2-x_1^2-x_2^2-x_3^2\ =\ \det\rho
\ \ .
\label{eq:1bitBlochlength}
\end{equation}
However, we know that $\det\rho$ is an invariant under
SL$(2,\mathbb{C}$) operations.
Hence, application of SLOCC to $\rho$  can be identified
with Lorentz transformations on the four-vector $x$. 
Local unitary operations on $\rho$ correspond to
rotations of $x$ in $\mathbb{R}^3$. The four-vectors $x$
for physical states $\rho$
are located within the ``forward light cone'', with 
the pure states on the surface. 

This idea can be generalized to $N$ qubits~\cite{Teodorescu2003}. Here we
illustrate only the two-qubit case for which, in Bloch representation,
\begin{equation}
       \rho\ =\ \case{1}{4}\sum_{j,k=0}^{3} x_{jk}\ \sigma_j\otimes\sigma_k
\ \ .
\label{eq:2bitBloch}
\end{equation}
The Minkowskian length for the tensor $x$ in (\ref{eq:2bitBloch}) is
\begin{equation}
      \|x\|^2\ =\ (x_{00})^2-  \sum_{j=1}^{3}\left[ (x_{j0})^2+(x_{0j})^2
                                             \right]
                +\sum_{j,k=1}^3 (x_{jk})^2
\label{eq:length2bitBloch}
\end{equation}
which is invariant under Lorentz transformations on each index -- or, in
other words, it is invariant under local SL transformations to the two-qubit
state $\rho$
(cf.\ also~(\ref{eq:concmodulussquared})).
Note that the analog Euclidean length of $x$ is
\begin{equation}
      \tr\rho^2\ =\ \case14  \sum_{j,k=0}^{3} x_{jk}^2\ \ ,
\label{eq:purity2bitBloch}
\end{equation}
that is, the purity of $\rho$.

The Minkowskian length discussed in this
section is an LSL invariant defined directly on the density 
matrix (as opposed to the convex-roof construction, e.g., for the
concurrence). Although this length is homogeneous of degree 1 in the
density matrix, it is not automatically an entanglement monotone.
For that purpose, one needs additional conditions. For example,
it would be sufficient if one could make sure that this
quantity is also a convex function on the state space.
%
%

%
%

\section{Three qubits}
\label{sec:threequbits}

The simplest multipartite quantum system consists of three qubits.
This system is already sufficiently complex that no complete
analytical solution is known, but on the other hand simple enough to
avoid most of the difficulties connected with multipartite
entanglement. Moreover, for the important entanglement measures,
methods to calculate good bounds are known.

\subsection{Pure three-qubit states}
\label{sec:threepure}

One peculiar property of multipartite entanglement can already be seen
for the three-qubit case: Unlike in the two-qubit case, not all pure
states can be generated from a single maximally entangled state. There
are two inequivalent SLOCC classes for genuinely multipartite
entangled classes~\cite{DVC2000}, the GHZ class with the representative
state
\begin{equation}
  \label{eq:ghzstate}
  \ket{\mathrm{GHZ}} = \frac{1}{\sqrt2}(\ket{000} + \ket{111})
\end{equation}
and the $W$ class with the representative state
\begin{equation}
  \label{eq:wstate}
  \ket{W} = \frac{1}{\sqrt3}(\ket{001} + \ket{010} + \ket{100}).
\end{equation}
While conversion between those two classes is not possible in either
direction, as we will see there are still good reasons to consider the
GHZ state the unique (up to local unitary operations) maximally
entangled state.

Any pure three-qubit state is unitarily equivalent to a state of the
form~\cite{Acin2000}
\begin{equation}
  \label{eq:acinform}
  \ket{\psi} = \lambda_0\ket{000} + \lambda_1\rme^{\rmi\varphi}\ket{100} + \lambda_2\ket{101} +
  \lambda_3\ket{110} + \lambda_4\ket{111}
\end{equation}
where $\sum_{k=0}^4 \lambda_k^{2}=1$, all $\lambda_k \geq 0$ and $0 \leq \varphi \leq \pi$. In this form,
several entanglement measures take on a particularly simple form:
\numparts
\begin{itemize}
\item The three-tangle \eref{eq:threetangle} is
  \begin{equation}
    \tau_3 = 2\lambda_0\lambda_4.\label{eq:acinthreetangle}
  \end{equation}
\item The two-qubit concurrences \eref{eq:mixedstateconcurrence} after
  tracing out one qubit are
  \begin{eqnarray}
    C_{AB} = 2\lambda_0\lambda_3, \label{eq:acincab}\\
    C_{AC} = 2\lambda_0\lambda_2, \label{eq:acincac}\\
    C_{BC} = 2\left|\lambda_1\lambda_4\rme^{\rmi \varphi}-\lambda_2\lambda_3\right|. \label{eq:acincbc}
\end{eqnarray}
\item The $I$-concurrence for splitting the first qubit from the rest
  is 
  \begin{equation}
    C_{A|BC} = 2 \lambda_0 \sqrt{\lambda_2^2+\lambda_3^2+\lambda_4^2}.\label{eq:aciniconc}
  \end{equation}
\end{itemize}
\endnumparts
Note that the quantities $J_1$ to $J_4$ of~\cite{Acin2000} are $J_1 =
\case14 C_{BC}^2$, $J_2 = \case14 C_{AC}^2$, $J_3 = \case14 C_{AB}^2$,
$J_4 = \case14 \tau_3^2$.

Using these quantities, the SLOCC classification of pure three-qubit
states is straightforward (following~\cite{Acin2000} with the
criteria, but ignoring their subclasses which are not SLOCC classes;
see~\cite{DVC2000} for a detailed description of the SLOCC classes):
\begin{itemize}
\item The GHZ class consists of all states with $\tau_3 > 0$.
\item The $W$ class consists of states with $\tau_3 = 0$, $C_{AB} > 0$,
  $C_{AC} > 0$ and $C_{BC} > 0$.
\item The three biseparable classes $A-BC$, $B-AC$ and $C-AB$ all have
  the corresponding concurrence $C_{BC}$, $C_{AC}$ or $C_{AB}$
  nonzero, and the others zero.
\item For the completely separable states, all measures vanish.
\end{itemize}

From equations \eref{eq:acinthreetangle} to \eref{eq:aciniconc}, one
can also immediately obtain the Coffman-Kundu-Wootters monogamy
relation~\cite{Coffman2000}
\begin{equation}
  \label{eq:ckwmonogamy}
  C_{A|BC}^2 = C_{AB}^2 + C_{AC}^2 + \tau_3^2.
\end{equation}
from which they originally derived the residual tangle $\tau_{\mathrm{res}}=\tau_3^2$.

There are other quantities which are invariant under local unitary
transformations, but not under local SL transformations on any qubit,
like the Kempe invariant~\cite{Kempe1999,Sudbery2001}. However, they
do not describe SLOCC properties of the state~\cite{Osterloh2010}.

Note that the two-qubit concurrences $C_{AB}$, $C_{AC}$ and $C_{BC}$
are \emph{not} three-qubit monotones, as can be easily seen by the
fact that the GHZ state (all three two-qubit concurrences equal to
$0$) can be converted via SLOCC to the tensor product of a Bell state
(concurrence 1) with a pure one-qubit state, thus increasing the
concurrence on the corresponding two qubits.

\subsection{Hierarchy of mixed states}
\label{sec:threemixed}

\begin{figure}
  \centering
  \includegraphics[width=0.5\linewidth]{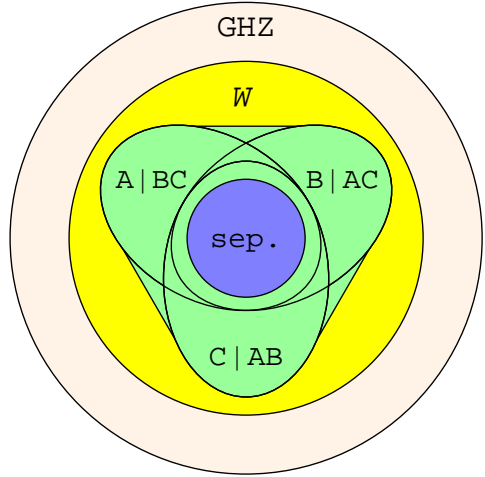}
  \caption{Schematic diagram of the classification of three-qubit
    mixed states. The inner blue circle depicts the convex set of
    separable states. The three enclosing ellipses are the borders of
    the sets which are biseparable on the bipartition written in them.
    While, when including the separable states, each of those sets is
    convex, their union is not; the complete set of biseparable states
    is therefore the convex hull of that union (green
    rounded.triangular shape). Outside that set the $W$ class states
    (yellow circle) and finally the GHZ class states (light brown
    circle) are found.}
  \label{fig:mixedhierarchy}
\end{figure}

The entanglement classification of pure states can be extended to
mixed states by considering the classes of the pure states in the
decomposition. An obvious definition seems: A state is of type X if it
can be decomposed into states of type X, that is, the set of states of
type X is the convex hull of the set of pure states of type X.
However the sets defined in this way have some striking properties which
suggest a different definition: It turns out that with the above
definition, all but one set of zero measure of $W$-type states are also
GHZ-type states, and similarly for biseparable and completely
separable states (those sets of zero measure of course contain all the
pure states of that class). That is, there exists a \emph{hierarchy}
of entanglement types, which leads to a slightly different definition
of mixed state classes, as was defined by Ac\'\i n \etal~\cite{Acin2001}:
\begin{itemize}
\item The set of separable states is, of course, the convex hull of
  the pure separable (i.e. product) states.
\item The set of biseparable states is the convex hull of the set of
  the pure biseparable \emph{and} separable states.
\item The set of $W$-type states is the convex hull of the pure
  $W$-type, biseparable \emph{and} separable states.
\item The set of GHZ-type states is the convex hull of the pure
  GHZ-type, $W$-type, biseparable \emph{and} separable states, that
  is, all states.
\end{itemize}
It is, however, often more useful to define the classes in an
\emph{exclusive} manner, that is, to call a state X-type entangled
only if it is \emph{not} in one of the lower classes (e.g. a state is
$W$-type entangled if it can be decomposed into states of maximally
$W$ type, but \emph{not} into states of maximally biseparable type).
This exclusive classification is what we use below.

Note that the biseparable states have a substructure
\cite{Seevinck2008,Szalay2012}: There are the three classes of states
which are biseparable on a specific decomposition, that is, they can
be decomposed into pure states wich all are biseparable on that
decomposition. The total set of biseparable tates is the convex hull
of the union of those three sets.

Since pure states which are biseparable on each bipartition are fully
separable, one would guess that the same is also true for mixed state;
however this is not the case. There exist mixed states which are
bisearable on each bipartition, yet are not
separable~\cite{Bennett1999}. Those states are PPT entangled.

This hierarchy of entanglement is schematically depictured in figure
\ref{fig:mixedhierarchy}.

The mixed state entanglement classes can be distinguished using
entanglement measures: A state is GHZ-type entangled iff the
three-tangle \eref{eq:threetangle} does not vanish. A state is
$W$-type entangled iff the GME concurrence \eref{eq:GMEconc} does not
vanish, but the three-tangle does. For biseparability, an appropriate
measure is the convex roof of the square root of the global
entanglement \eref{eq:MeyerWallach} (the square root is to get
homogeneous degree $2$ in the state vector). The biseparable states
are exactly those for which this measure does not vanish, but the GME
concurrence does.

Separability on a specific bipartition can be checked using the
concurrence \eref{eq:1partyconc} for that bipartition. Note that
vanishing of all three concurrences is not sufficient for
separability.

\subsection{Exact treatment of GHZ-symmetric states}
\label{sec:ghzsymm}

\begin{figure}
  \centering
  \includegraphics[width=0.7\linewidth]{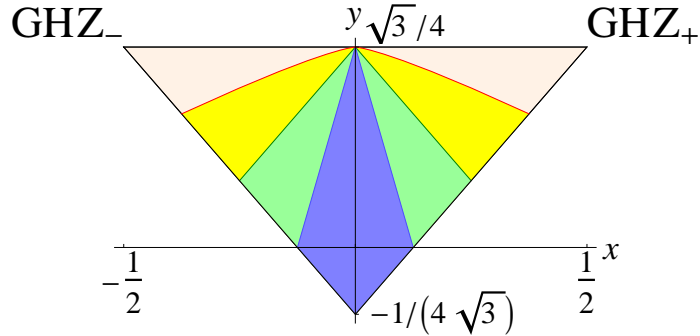}
  \caption{Entanglement classes of the GHZ-symmetric states. On the
    top two corners there are the pure states $\ket{\mathrm{GHZ}_+}$
    (upper-right corner) and $\ket{\mathrm{GHZ}_-}$ (upper-left
    corner). The colours have the same meaning as in figure
    \ref{fig:mixedhierarchy}: The blue kite in the middle are the
    separable states, the two green triangles surrounding it are the
    biseparable states. The roughly triangular yellow areas are the
    $W$-type entangled states, and are bounded by the red, curved
    GHZ-W line. Above the GHZ-W line there lie the GHZ-type entangled
    states.}
  \label{fig:triangle}
\end{figure}
As mentioned in section \ref{sec:axisym}, mixed states of low rank
and/or high symmetry may help considerably to elucidate the structure
and properties of the state space with respect to entanglement. Also
for three-qubit states, important progress could be achieved by
finding the exact solutions for such specific problems. The first
example of this kind was solved by Lohmayer {\em et
  al.}~\cite{Lohmayer2006} who considered rank-2 mixtures
\begin{equation}
     \rho(p) = p \ket{\mathrm{GHZ}}\!\bra{\mathrm{GHZ}}
                     + (1-p) \ket{W}\!\bra{W}\ \ .
\label{eq:GHZW}
\end{equation}
They found the exact residual tangle $\tau_{\mathrm{res}}$ as well as the
concurrence and the 1-tangle for this family. The spirit of the method
behind this calculation (termed the convex characteristic curve) was
applied before (e.g., in \cite{Bennett1996} and
\cite{VollbrechtTerhal2000}) and is outlined in Osterloh {\em et
  al.}~\cite{Osterloh2008}.

Subsequently, Eltschka {\em et al.} extended this result to rank-2
mixtures of generalized GHZ and $W$ states, that is,
$\ket{\mathrm{gGHZ}}=a\ket{000}+b\ket{111}$,
$\ket{\mathrm{g}W}=c\ket{001}+d\ket{010}+e\ket{100}$~\cite{Eltschka2008}.
Jung {\em et al.} provided exact solutions of $\tau_{\mathrm{res}}$ for
the rank-3 problem~\cite{Jung2009}
\begin{equation}
  \rho(p,q) = p \ket{\mathrm{GHZ}}\!\bra{\mathrm{GHZ}}
  + q \ket{W}\!\bra{W}
  + (1-p-q)\ \ket{\overline{W}}\!\bra{\overline{W}}
  \label{eq:GHZWantiW}
\end{equation}
where
\begin{equation}
  \label{eq:Wbar}
  \ket{\overline{W}} = \frac{1}{\sqrt{3}}(\ket{011}+\ket{101}+\ket{110})
\end{equation}
is the bit-flipped $W$ state.
as well as for a family of rank-4 states~\cite{JungParkSon2009}.
Further, He {\em et al.} considered $\tau_{\mathrm{res}}$ for certain
states with up to rank 8~\cite{He2011}. Finally, in
\cite{Viehmann2011}, the three-tangle (i.e.,
$\tau_3=\sqrt{\tau_{\mathrm{res}}}$) for the states \eref{eq:GHZW} and
\eref{eq:GHZWantiW} was found.

In this section we take a closer look at another set of three-qubit
states for which the relevant entanglement properties have been
derived exactly. To our knowledge it is the only completely,
qualitively and quantitatively solved set which covers all
main entanglement classes except PPT entanglement.

Invariance under symmetries has proved to be a valuable tool in
studying entanglement since the seminal work on bipartite states by
Werner~\cite{Werner1989}, and on tripartite states by Eggeling
and Werner~\cite{Eggeling2001}. Since the GHZ state is the maximally
entangled three-qubit state, it is desirable for it to be contained in
the invariant set. To achieve this, it is necessary to use all or a
subset of the symmetries of that state, which are~\cite{Eltschka2012PRL}:\\
(i) qubit permutations, \\
(ii) simultaneous three-qubit flips
(i.e., application of  $\sigma_x\otimes\sigma_x\otimes \sigma_x$), \\
(iii) qubit rotations about the $z$ axis of the form
\begin{equation}
  \label{eq:zrot}
  U(\phi_1,\phi_2) = \rme^{\rmi \phi_1 \sigma_z}\otimes\rme^{\rmi \phi_2 \sigma_z}\otimes\rme^{-\rmi
    (\phi_1+\phi_2) \sigma_z}\ \ .
\end{equation}
Here, $\sigma_x$ and $\sigma_z$ are Pauli operators. Invariance under
the operations (i)-(iii) is called \textit{GHZ symmetry} and states
that are invariant under that symmetry are called
\textit{GHZ-symmetric states}. Except the qubit permutations all those
operations are local and therefore do not change the entanglement type.
Also, the only effect of the qubit permutations on the entanglement is
that they permute the different subclasses of bipartite entanglement,
which implies that the biseparability properties of the GHZ-symmetric
states are the same on all bipartitions.

There are two GHZ-symmetric pure states, the standard GHZ-state
$\ket{\mathrm{GHZ}}\equiv\ket{\mathrm{GHZ}_+}$ \eref{eq:ghzstate} and the
sign-flipped GHZ state $\ket{\mathrm{GHZ}_-} =
(\ket{000}-\ket{111})/\sqrt2$. The complete set of GHZ-symmetric
states consists of mixtures of those two states and the mixed state
$\rho_r=\sum_{klm=001}^{110}\ket{klm}\bra{klm}$.

A GHZ-symmetric state $\rho^{\mathrm{S}}$ is fully specified by two
independent real parameters. A possible choice is
\numparts
\begin{eqnarray}
  \label{eq:rhoparams}
 x(\rho^{\mathrm{S}}) &=& \frac{1}{2}\!
  \left[\bra{\mathrm{GHZ}_+}\rho^{\mathrm{S}}\ket{\mathrm{GHZ}_+} -
        \bra{\mathrm{GHZ}_-}\rho^{\mathrm{S}}\ket{\mathrm{GHZ}_-}\right]\\
 y(\rho^{\mathrm{S}}) &=& \frac{1}{\sqrt{3}}
  \left[\bra{\mathrm{GHZ}_+}\rho^{\mathrm{S}}\ket{\mathrm{GHZ}_+} +
        \bra{\mathrm{GHZ}_-}\rho^{\mathrm{S}}\ket{\mathrm{GHZ}_-} - \frac{1}{4}\right]
\end{eqnarray}
\endnumparts
such that the Euclidean metric in the $(x,y)$ plane coincides with the
Hilbert-Schmidt metric on the density matrices. The completely mixed
state is located at the origin. The three corners of the triangle are
at the points (given in the form $P=(x,y)$)
$P_{\mathrm{GHZ_+}}=(\case12,\sqrt{3}/4)$,
$P_{\mathrm{GHZ_-}}=(-\case12,\sqrt{3}/4)$, $P_r=(0,-1/(4\sqrt{3}))$.

For the GHZ-symmetric states, the entanglement classes have been
determined exactly \cite{Eltschka2012PRL} (see figure
\ref{fig:triangle}). It turns out that all of the permutation
symmetric SLOCC classes except PPT entangled states are present. The
separable states live in the kite with corners given by the four
points $P_r$, $(\case18,0)$, $P_m=(0,\sqrt{3}/4)$ and $(-\case18,0)$.
The biseparable states are those states which are not separable in the
kite with korners $P_r$, $(\case14,1/(4\sqrt{3}))$, $P_m$ and
$(-\case14,1/(4\sqrt{3}))$. The $W$-type states are the remaining
states in the shape between $P_r$ and the ``GHZ-$W$ line'' given in
parametrized form by
\begin{equation}
  \label{eq:ghzwline}
  x^W(v)=\frac{v^5+8v^3}{8(4-v^2)}\,,\qquad
  y^W(v)=\frac{\sqrt{3}}{4}\frac{4-v^2-v^4}{4-v^2}\,.
\end{equation}
All remaining states are of GHZ-type.

Not only is the classification of GHZ-symmetric states known, but also
the most important entanglement measures.

\begin{figure}
  \centering
  \includegraphics[width=0.7\linewidth]{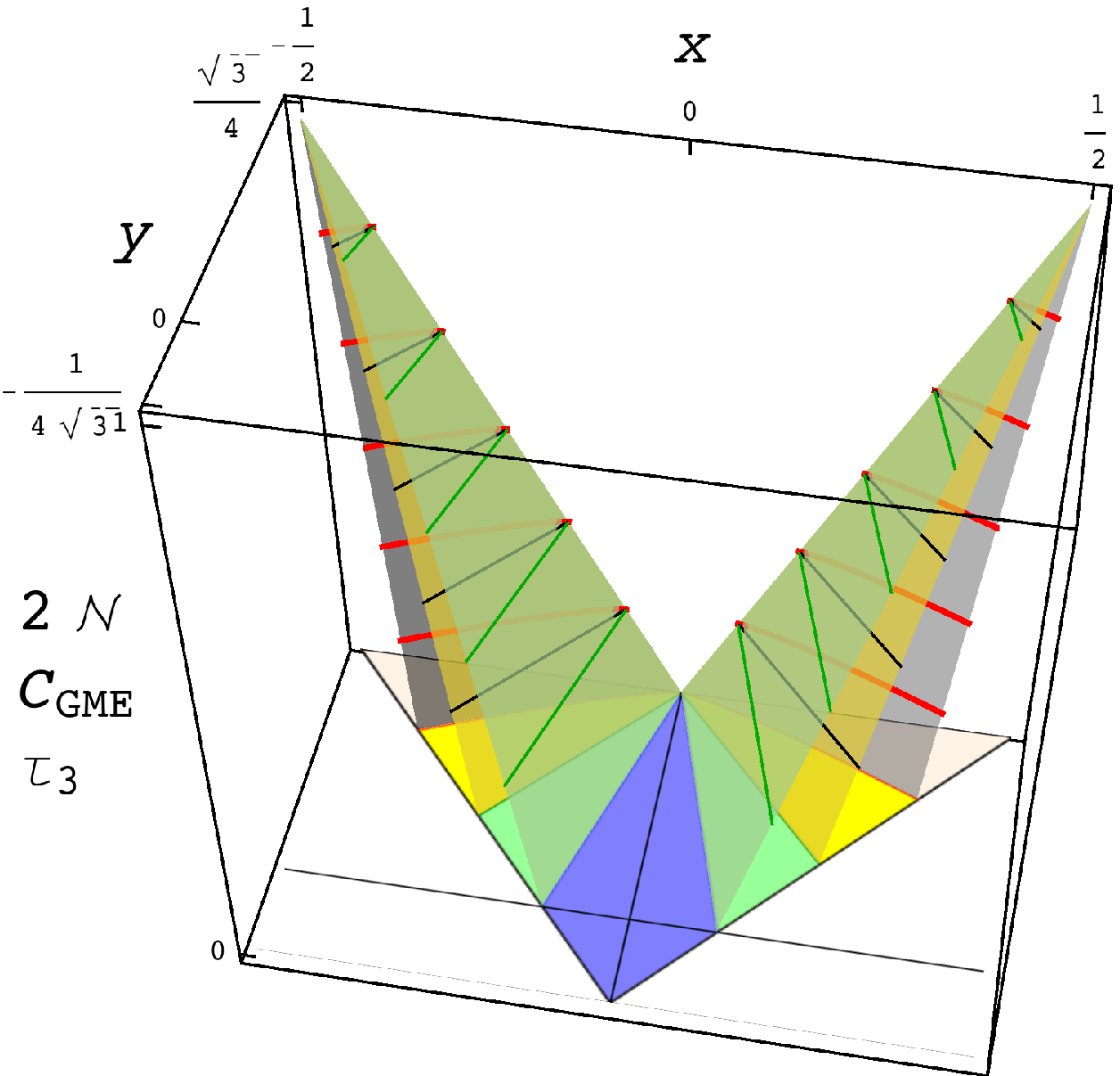}
  \caption{Different measures on the GHZ-symmetric states. The
    graphics show the three-tangle \eref{eq:ghzsymmthreetangle}
    (grey), GME concurrence \eref{eq:ghzsymmgmeconc} (yellow) and two
    times the negativity \eref{eq:ghzsymmnegativity} (green) for
    GHZ-symmetric states. On the base, figure \ref{fig:triangle} is
    reproduced, in order to see the classes the states belong to. It
    can be seen that the entanglement measures are nonzero exactly for
    the states of at least the corresponding class. Note that the GME
    concurrence and the negativity have planar graphs, while the graph
    of the three-tangle is curved.}
  \label{fig:measures}
\end{figure}

As could already be guessed from the GHZ-$W$ line, the most
complicated is the three-tangle. It is calculated as
follows~\cite{Siewert2012}:

Given a GHZ-symmetric three-qubit state $\rho^{\mathrm{S}}$ with
coordinates $(x,y)$, one first determines the straight line which
connects the GHZ$_+$ state at $(1/2,\sqrt{3}/4)$ with the point
$(x,y)$. This line intersects the GHZ-$W$-line at the point
$(x^W,y^W)$. Then the three-tangle
$\tau_3(\rho^{\mathrm{S}}(x,y))$ is given by (cf. figure 4)

\begin{equation}
  \label{eq:ghzsymmthreetangle}
  \tau_3(x,y) =
  \cases{0 & for $x<x^W$, $y<y^W$\\
    \frac{x-x^W}{\case12-x^W}
    & otherwise.
  }
\end{equation}

Since the GHZ-symmetric states are a subset of the GHZ-diagonal
states, the GME concurrence calculated in~\cite{HuberEberly2012} can
be applied. Rewritten in $(x,y)$ coordinates, it is given by
\begin{equation}
  \label{eq:ghzsymmgmeconc}
  C_{\mathrm{GME}}(\rho^S(x,y)) = 2\max \{0,x + \case12\sqrt{3}y - \case38\}
\ \ .
\end{equation}

The permutation symmetry implies that the negativies are equal on all
three bipartitions. The negativity for any bipartition
is~\cite{Eltschka2013}
\begin{equation}
  \label{eq:ghzsymmnegativity}
  \mathcal{N}(\rho^S(x,y)) = \max\{0,
  \frac18-\frac{1}{2\sqrt{3}}y-\left|x\right|\}\ \ ,
\end{equation}
see figure 4.

From this result, the concurrences for the three bipartite splits
(which are, again, equal) can be easily determined from
\eref{eq:concestimate_mixed}, which in this case due to the smaller
dimension $d=2$ simply is $C_{A|BC}(\rho) \geq 2\mathcal{N}_{A|BC}(\rho)$. On
the other hand, for the GHZ states clearly $C_{A|BC}(\rho) =
2\mathcal{N}_{A|BC}(\rho)$, and certainly for the separable states the
concurrence vanishes. Now by direct decomposition, it is clear that
the concurrence cannot be larger than the linear interpolation between
values of the separable states and the GHZ state. But that linear
interpolation agrees with the lower bound given by
$2\mathcal{N}_{A|BC}$, which therefore equals the concurrence.

\subsection{Arbitrary three-qubit mixed states}
\label{sec:threearbitrary}

There have been various attempts to obtain numerical estimates for the
three-tangle (or the residual tangle) in arbitrary mixed states, e.g.,
\cite{Loss2008,Cao2010,Rodriques2013}, but also analytical approaches,
e.g., \cite{Hyllus2010,Gour2010,Osterloh2012}.

The GHZ-symmetric states are useful in deriving a lower bound (in
principle, analytical) of the three-tangle for mixed
states~\cite{Eltschka2012SciRep,Eltschka2014}. The procedure is as
follows:
\begin{enumerate}
\item Calculate the normal form with the algorithm
  in~\cite{Verstraete2003}, remember the trace of the normal form and
  renormalize. If the normal form vanishes, the three-tangle does,
  too, and the procedure finishes.
\item Optimize the normal form obtained in step 1 over local unitary
  transformations in order to minimize the entanglement loss in the
  next step.
\item Project onto the set of GHZ-symmetric states by the twirling
  operation
  \begin{equation}
    \label{eq:twirling}
    P_{\mathrm{GHZ}}(\rho) = \int_{\{U\colon\ U\ket{GHZ}=\ket{GHZ}\}}
    U\rho U^\dagger\,\rmd U\ ,
  \end{equation}
  calculate the three-tangle of the projected state, and multiply with
  the trace from step 1 to obtain the lower bound.
\end{enumerate}
The twirling operation can be described by the two simple
equations
\numparts
\begin{eqnarray}
  \label{eq:twirlexplicit}
  x(\rho) = \case12(\rho_{000,111} + \rho_{111,000})\\
  y(\rho) = \frac{1}{\sqrt3}(\rho_{000,000}+\rho_{111,111}-\case14)
\end{eqnarray}
\endnumparts

Note that the same procedure can be used for any measure which is
known on the GHZ-symmetric states, with the exception that the first
step has to be omitted because it only works for measures that are
invariant under local SL transformation, which for three qubits holds
only for the three-tangle (note that for mixed states, even the
residual tangle is \emph{not} invariant under 
local SL operations).~\footnote{%
    We mention that optimization over arbitrary local operations
    was used also to improve the detection of entangled states
    by witnesses, cf.~\cite{Haeffner2005,Kiesel2007}.
                               }

Rodriques \etal~\cite{Rodriques2013} have developed an algorithm to
obtain an upper bound based on what they call the ``best zero-E
decomposition'', a generalisation of the best separable
approximation~\cite{Lewenstein1998} and best $W$
approximation~\cite{Acin2001}. It is based on decomposing a mixed
state into a pure state with nonzero measure and a mixed state with
zero measure, and optimizing this decomposition.

For the GME-concurrence, a method to calculate the lower bound is
described in section \ref{sec:GMEconc}.

\subsection{Optimal witnesses for three qubits}
\label{sec:threewit}

The knowledge of the exact properties of GHZ-symmetric states also
allows to explicitly derive optimal witnesses for different types of
entanglement~\cite{Eltschka2013}. Due to the convex shape of the
border between GHZ-type and $W$-type states, there exists a complete
continuous family of optimal witnesses for GHZ-type entanglement,
corresponding to the different points of the border. In particular,
the well-known projection witness~\cite{Acin2001}
\begin{equation}
  \label{eq:ghzprojwitness}
  \mathcal{W}_{\mathrm{proj}} = \case34\id - \ket{\mathrm{GHZ}}\bra{\mathrm{GHZ}}
\end{equation}
turns out not to be an optimal witness, but can be impoved
to~\cite{Eltschka2013}
\begin{equation}
  \label{eq:ghzimproved}
  \mathcal{W}_{\mathrm{opt}} = \case34\id - \ket{\mathrm{GHZ}}\bra{\mathrm{GHZ}}
  - \case37\ket{\mathrm{GHZ}_-}\bra{\mathrm{GHZ}_-}
\end{equation}
Those witnesses can also be used as estimators for the actual
three-tangle~\cite{Eltschka2012SciRep,Eltschka2014}. For the witness
\eref{eq:ghzimproved} one obtains the analytical lower
bound~\cite{Eltschka2014}
\begin{eqnarray}
  \label{eq:threetanglebound}
  \tau_3^{\mathrm{approx}}(\rho) &=&
 \max\left(
      0\ , \left[\pm \case{8}{7}(\rho_{000,111}+\rho_{111,000})+
     \right.\right.
   \nonumber \\ && \left.\left.
       +  \case{20}{7}(\rho_{000,000}+\rho_{111,111})
       -3\right] \right)\ \ .
\end{eqnarray}



\section{Towards the general case}
\label{sec:gen}
\subsection{Maximal entanglement in multipartite systems}
\label{sec:gen-max}

Among the many distinctive features of multipartite entanglement
it is worth highlighting the ambiguity of the concept of maximal
entanglement. It was discussed, e.g., in 
\cite{Gisin1998,DVC2000,Braunstein2005,Love2007,Osterloh2010NJP,Gour2010b,Helwig2012}. 
Recall that for bipartite systems the unique (up to local unitaries)
 maximally entangled
state is $\Psi_d$, cf.~(\ref{eq:Bell-d}). This is because
any pure $d\times d$ bipartite state can be produced from $\Psi_d$
by means of SLOCC (including non-invertible operations). The situation
is different for multipartite systems (with more than three parties): 
There are, e.g., many inequivalent types of genuine entanglement that
cannot be converted into one another 
(see section~\ref{sec:gen-class}).~\footnote{%
   If one defines 'maximal entanglement' in the sense that all
   other states can be generated from them by means of LOCC (rather than SLOCC),
   then already for three qubits there are infinitely many maximally entangled 
   states~\cite{Kraus2012,deVicente2013}.
                                            } 

The obvious requirment for maximal entanglement, 
in particular if one advocates the point of view of
characterizing multipartite entanglement by local
SL-invariant polynomials,
is to require maximal mixedness of each local party
(i.e., the state needs to be {\em stochastic}). This is 
exactly the condition to select the state in each orbit that maximizes
all non-vanishing entanglement measures based on local SL-invariant 
polynomials~\cite{Verstraete2003}. This condition can, but need not,
be complemented by additional requirements. Let us first check
what we obtain for four qubits. There are three states that obviously
fulfill that condition 
\numparts
\begin{eqnarray}
         \ket{\mathrm{GHZ}_4} & =&  \case{1}{\sqrt{2}}\left(
                                 \ket{0000}+\ket{1111}\right)
\label{eq:irrbal1} 
\\
         \ket{\mathrm{Cl}_4} & = &  \case12 \left(
                    \ket{0000}+\ket{0111}+\ket{1011}+\ket{1100}\right)
\label{eq:irrbal2} 
\\
         \ket{X_4} & = &  \case{1}{\sqrt{6}} \left(
            \sqrt{2}\ket{1111}+\ket{0001}+\ket{0010}+\ket{0100}+\ket{1000}
                                               \right)
\ \ ,                                     
\label{eq:irrbal3} 
\end{eqnarray}
\endnumparts
the GHZ state, the cluster state, and
 yet another state $X_4$~\cite{Osterloh2005,Love2007,Gour2010,Djokovic2013}.
These are the three ``irreducibly balanced states'' for 
four qubits~\cite{Osterloh2010NJP}.
Clearly, they are not the only stochastic four-qubit states, however,
we can immediately see several interesting facts. These states
have the following Schmidt ranks across their two-qubit bipartitions:
2 (GHZ), 2 or 4 (cluster), 3 ($X_4$). This means they cannot be
locally equivalent. Further, we see that {\em all} the reduced
states are maximally mixed on their span. Hence, requiring this as
an additional property~\cite{Love2007,Osterloh2010NJP,Gour2010}
would yield precisely the states (\ref{eq:irrbal1})--(\ref{eq:irrbal3})
(up to local unitaries and qubit permutations).

One might think that, as an additional condition, one could require
that all $k$-qubit reduced states after tracing out more than half of the
parties (i.e., $N/2\leqq N-k\leqq N-1$) be maximally mixed. However,
this is in general not possible for $N$-qubit 
states~\cite{Gisin1998,Gour2010b}.

\subsection{Classifications of four-qubit states}
\label{sec:gen-class}

D{\"u}r \etal~\cite{DVC2000}  pointed out
 that for multipartite systems
with four or more qubits there are infinitely many SLOCC classes. That is,
there are one or more continuous labels to specify one equivalence class.
On the other hand, it is not generally true that a protocol can be run
only with the states from exactly one SLOCC class. To illustrate this,
assume a protocol that works with the four-qubit $W$ state
\begin{equation}
     \ket{\phi_0}\ =\ \ket{W}\ =\ 
      \case12 (\ket{0001}+\ket{0010}+\ket{0100}+\ket{1000})
\ \ .
\end{equation}
It is conceivable that the protocol still works well if a small component
$\ket{1111}$ is added to the state
\begin{equation}
     \ket{\phi_{\varepsilon}}\ =\ \varepsilon \ket{1111} +
        \case12  \sqrt{1-\varepsilon} (\ket{0001}+\ket{0010}+\ket{0100}+\ket{1000})
\ .
\end{equation}
However, the states are SLOCC inequivalent. Despite this, a quantifier of the
resource for that protocol should measure both $\phi_0$ and $\phi_{\epsilon}$
with a nonzero value. 

This gives rise to the idea that some coarse graining needs to be
applied to the set of SLOCC classes, that is, SLOCC-inequivalent states
get bunched together in families according to some (SLOCC-invariant)
criterion (cf.\ also section \ref{sec:defent}). Usually, this coarse-grained
set of families is referred to as {\em SLOCC classification}.
It also appears evident that a reasonable criterion
for bunching classes will depend on the resource under 
consideration. Consequently,  there is no absolutely
preferable SLOCC classification. Rather, for a given resource
one classification may be more adequate than another.

The first classification for four qubits was presented by 
Verstraete \etal~\cite{Verstraete20024x9} and  
later elaborated on
by Chterental and Djokovic~\cite{Chterental2007}. This approach
essentially classifies the normal forms of four-qubit states.
Further classifications (both for four qubits and also larger
systems) were worked out by 
Miyake and Verstraete~\cite{Miyake2004}, 
Mandilara \etal.~\cite{Mandilara2005},
Lamata \etal~\cite{Lamata2006,Lamata2007},
Bastin \etal~\cite{Bastin2009,Bastin2010},
Borsten \etal~\cite{Borsten2010},
Li \etal~\cite{Li2012}, 
Viehmann \etal~\cite{Viehmann2011},
Sharma and Sharma~\cite{Sharma2012}, and recently
by Huber and de Vicente~\cite{Huber2013-PRL}, 
Walter {\em et al.}~\cite{Walter2013},
and by Gour and Wallach~\cite{Gour2013}. 

Here we want to highlight and compare only two of the above results,
the strikingly elegant
classfication for four-qubit symmetric states~\cite{Bastin2009} 
and a corresponding
polynomial classification~\cite{Viehmann2011}. There are various points
that motivate this choice. First, the symmetric subspace of four qubits
has far fewer parameters than the complete space, however, it displays
many of the essential features that are added when moving on from the
three to the four-qubit case: there is a single continuous label for most
of the SLOCC classes, it contains  
distinct maximally entangled 
states, and there are two independent polynomial invariants
that do not vanish on the symmetric space. Further it appears feasible 
to build a hierarchy also for the mixed states: one defines the hierarchy
for pure states and then extends it, in the spirit of 
section~\ref{sec:threemixed}
to the mixed states. Conversely, this could also provide a clear recipe how to
determine the family of a given mixed state. Note that it
is an important practical requirement for an SLOCC classification
that it be possible to determine the family of a given mixed state. 
So, altogether, with the symmetric four-qubit case we find
ourselves on solid ground without debatable assumptions.

According to \cite{Bastin2009,Martin2010} every symmetric $N$-qubit state
can be written 
\begin{equation}
      \ket{\psi_S}\ =\ \nu \sum_{1\leqq j_1\neq \ldots\neq j_N\leqq N}
                           \ket{\epsilon_{j_1}\ldots
                            \epsilon_{j_N}}
\label{eq:4bitsym}
\end{equation}
where $\nu$ is a normalization factor and $\epsilon_{j}$ denote
one-qubit directions. 
Symmetric $N$-qubit states can be expanded into $N$-qubit Dicke states with $k$ 
excitations~\cite{Toth2007}
\begin{equation}
         \ket{D_N^{(k)}}\ = \ {N\choose k}^{-\case12}
                            \sum_k \mathcal{P}_k\left( \ket{1_11_2\ldots 1_k
                                                            0_{k+1}\ldots 0_N} 
                                                \right)
\label{eq:dicke}
\end{equation}
where the sum is over the permutations $\mathcal{P}_k$ of the $N$ entries.
It turns out that two symmetric states
$\ket{\psi_S}=\nu \sum \ket{\epsilon_{j_1}\ldots\epsilon_{j_N}}$ and
$\ket{\psi_S'}=\nu' \sum \ket{\epsilon_{j_1}'\ldots\epsilon_{j_N}'}$ 
belong to the same SLOCC class if and only if there exists a {\em single}
invertible local operation that converts {\em each} state $\epsilon_j$
into $\epsilon_j'$~\cite{Bastin2009,Bastin2010}. 
Thus, SLOCC equivalence of two states $\psi_S$,
$\psi_S'$ can be excluded already
if they have different degeneracy
configuration $\mathscr{D}_{\{n_j\}}$,
that is, if the multiplicities $n_j$
of the distinct directions $\epsilon_j$ in $\psi_S$ and $\psi_S'$ do not coincide. 

For four qubits there are only five possible degeneracy configurations
and, hence, five SLOCC families which consist of four single orbits
and one continuous family, cf.~table~\ref{tab:table1}. Note that here
the arrangement of all SLOCC classes into five families occurs naturally
because the degeneracy configuration is invariant under SLOCC.

On the other hand, we may check what the invariant polynomials of
section \ref{sec:4bitinvs} yield for the symmetric 
four-qubit states~\cite{Viehmann2011}.
That is, since in table~\ref{tab:table1} there are all SLOCC classes
listed, we just add the values of the polynomials for each
representative to the table. For symmetric four-qubit states,
the polynomials of degree 4 are not independent $\mathcal{B}^{(4)}_{1,j}=H^2$,
therefore we need to consider only $H$ (degree 2) and the sextic
invariant $\mathcal{F}_1^{(4)}$. As expected, all polynomials
vanish on the separable states and the $W$ states. For the orbits
$\mathscr{D}_{2,2}$ (stable) and $\mathscr{D}_{2,1,1}$ (semistable)
only $H\neq 0$
while for the family  $\mathscr{D}_{2,1,1}$ both $H$ and $\mathcal{F}_1^{(4)}$
do not vanish. We see that the resources of the GHZ state and
the Dicke state $D_4^{(4)}$---according to the polynomial measures
%
%
\begin{table}[b]
\begin{tabular}{ccccc}
\hline \hline
 ${\mathscr D}_{\{n_i\}}$  & representative & $H$  & $\mathcal{F}^{(4)}_1$ 
                                                   & type \\[1mm]
\hline
 ${\mathscr D}_{4}$  &    $  D_4^{(0)}$ &   0   &  0           & separable  \\
 ${\mathscr D}_{3,1}$  &    $  D_4^{(1)}$ &   0   &  0         & $W$  \\
 ${\mathscr D}_{2,2}$  &    $   D_4^{(2)}$ &   1   &  -5/9     & 
                                           $D_4^{(2)}$ \\
 ${\mathscr D}_{2,1,1}$  &  $  D_4^{(0)}+D_4^{(2)}$ &  1  & -5/9 & 
                                           $D_4^{(2)} $\\
 ${\mathscr D}_{1,1,1,1}$  &  $  \ket{0000}+\ket{1111}+\lambda D_4^{(2)}$ &   $h(\lambda)$   
                 &  $ f(\lambda)$
                 & $X_4$  \\[1mm]
\hline\hline
\end{tabular}
\caption{Comparison of the polynomial characterization and the SLOCC 
         classification of symmetric four-qubit states~\cite{Bastin2009}.
         The representatives are given in the basis of the symmetric
         four-qubit Dicke states $D_4^{(k)}$.
         For the continuous parameter in the $X$ family we have $\lambda^2 \neq 2/3$ and
 $h(\lambda)=2+\lambda^2$, $f(\lambda)=-8+4\lambda^2-(102\lambda^4+5\lambda^6)/9 $.}
\label{tab:table1}
\end{table}
%
$H$ and $\mathcal{F}_1^{(4)}$---are not strictly distinct. However,
we may equally well find polynomials that measure only the GHZ state
\begin{equation}
         \mu_{\mathrm{GHZ}}\ =\ \mathcal{F}_1^{(4)}+\case{5}{9} H^3
\end{equation}
or only the Dicke state
\begin{equation}
         \mu_{\mathrm{Dicke}}\ =\ \mathcal{F}_1^{(4)}+ H^3\ \equiv\ 32 W
\end{equation}
(for the last identity, cf.~(\ref{eq:F41vsW})).

Hence, the polynomial classification admits a certain freedom
for precise characterization of the resource. Note, however,
that the local SL-invariant polynomials alone are not sufficient for
a complete classification, in particular if one wants
to include the mixed states.  
While for distinguishing the separable
states $\mathscr{D}_4$ from the genuinely entangled
families, the GME concurrence
is appropriate, there is considerable fine structure in the
state space
(even $\mathscr{D}_{2,1,1}$ from $\mathscr{D}_{2,2}$ are not
distinguished by the invariant polynomials).  For this purpose,
the covariants~\cite{Luque2003-170,Holweck2012,Szalay2012,Borsten2013,Holweck2014} 
possibly provide a solution. But this remains to be worked out
in the future.

\subsection{Monogamy}
\label{sec:gen-mon}

An interesting fundamental question is whether there are
strict monogamy relations similar to (\ref{eq:ckwmonogamy})
for systems of four and more parties. The Osborne-Verstraete
monogamy inequality (\ref{eq:OsborneVerstraete}) suggests that 
there are generalizations in terms of quantities derived from
local SL-invariant quantities for the subsystems. 
For example, for pure four-qubit states $\psi$, and denoting
the parties $A$, $B$, $C$, $D$, one might guess for the
difference between 1-tangle and squared concurrences for party $A$
\begin{eqnarray}
\mathcal{R}_{A} & = &\tau_{A}- C^2_{AB}-C^2_{AC}-C^2_{AD}
\nonumber\\
                  & \stackrel{?}{=} & \tau_{3,ABC}^2+\tau_{3,ABD}^2+\tau_{3,ACD}^2+
                            \tau_4^2
\ \ .
\label{eq:ourmonog}
\end{eqnarray}
Indeed one finds states where a monogamy equality in this spirit
holds~\cite{Eltschka2009}. 
For example, for a three-qubit state
$\ket{\phi^{(3)}}$, we have the 
monogamy (\ref{eq:ckwmonogamy}). If we now generate a 
four-qubit state $\phi^{(4)}$ from $\phi^{(3)}$ according to the
rule (`telescoping')
\begin{equation}
\ket{\phi^{(3)}}=\sum_{jkl} \phi^{(3)}_{jkl}\ket{jkl}\ \
\longrightarrow\ \ 
\ket{\phi^{(4)}}=\sum_{jkl} \phi^{(3)}_{jkl}\ket{jkll} \ \ ,
\end{equation}
one finds that the three-tangle of $\phi^{(3)}$ translates 
into a four-tangle that, not so surprisingly, equals one of
the degree-4 invariants $B^{(4)}_{1,j}$ from section \ref{sec:4bitinvs},
cf.~\cite{Eltschka2009}.
However, this relation is not valid for generic
four-qubit states, and counterexamples
to the assumption (\ref{eq:ourmonog}) can be constructed.

On the other hand, there is indeed a relation that holds for
{\em all} four-qubit states, which is {\em not} of the form
(\ref{eq:ourmonog}). To this end, define the global entanglement
\begin{equation}
\tau_1(\psi)=
   \case14\left(\tau_A(\psi)+\tau_B(\psi)+\tau_C(\psi)+\tau_D(\psi)\right) 
\end{equation}
and the average pure-state linear entropy (in analogy with (\ref{eq:1partyconc}))
\begin{equation}
\tau_2(\psi)=\case13\left(C^2_{AB|CD}(\psi)+C^2_{AC|BD}(\psi)+C^2_{AD|BC}(\psi)\right)
\ \ .
\end{equation}
Then one has~\cite{Gour2010b} 
\begin{equation}
    \tau_2(\psi)\ =\ \case13\left(4\tau_1(\psi) - |H(\psi)|^2
                            \right)
\end{equation}
with the degree-2 local SL invariant $H$ from (\ref{eq:H4bit}).

Hence, there are strong indications that general monogamy relations
do exist, however, the precise rules and conditions remain an
open question at this point.

\section{Conclusion}

We have reviewed the topic of quantifying single-copy entanglement
resources of a few finite-dimensional parties. We could witness 
enormous progress in this field in recent years. While a decade
ago, essentially the case of two qubits could be considered solved,
to date the three-qubit problem appears tractable to a large extent.
Also many aspects of $d\times d$ bipartite systems have been understood
at a quantitative level.
In addition, considerable insight into more complex systems has been
gained, in particular, regarding the case of four qubits. 
The general mathematical framework with tools from convex optimization
and algebraic geometry has been identified and applied successfully.
The application of the SLOCC paradigm and its mathematical
model, the representation of SLOCC by local invertible operations,
i.e., the group SL($d,\mathbb{C}$) were instrumental in these developments.

Let us conclude by specifying some of the open challenges for the near
future.  At the moment, there is still insufficient understanding
of the relation between the entanglement measures (the resource
characterization) and the corresponding entanglement families.
Comparing to the three-qubit case, the big step forward was to realize
that GHZ entanglement is a resource that is not contained in all genuinely
entangled three-qubit states, and that it is measured by the three-tangle. 
For four qubits we know---in principle---all possible measures and at 
least some of the interesting families, but we do not know how to 
precisely characterize the relation between them.

We believe that this problem can be solved only by gaining more 
insight into the structure of the space of mixed states, because
then one possibly can decide which type of characterization is relevant,
and which is not.  This means, in turn, that better tools for 
the evaluation of convex-roof extended entanglement measures are
required, possibly also more exact solutions for relevant examples
of mixed states that contain distinct types of multipartite
entanglement resources. Once a more thorough understanding
of the entangled states is obtained, this needs  to be
complemented by an analogous characterization/classification of the states
in the nullcone of the local SL-invariant polynomials.
Finally, we consider it essential to achieve a much better understanding
of Verstraete's normal form and its relation to any kind of entanglement.


\ack

It is a pleasure to thank 
T.\ Bastin, D.\  Braun, R.\ Lohmayer, A.\ Osterloh, and
O.\ Viehmann for our fruitful collaboration and countless
stimulating discussions. In particular, we would like
to thank A.\ Uhlmann for sharing his intuition and deep
physical insight with us.

We gratefully acknowledge illuminating conversations
with D.\ Bru{\ss}, J.-L.\ Chen, Z.-H.\ Chen, J.\ Eisert, O.\ G{\"u}hne, 
      P.\ Horodecki,
         M.\ Huber, P.\ Hyllus, M. Johansson, 
         M.\ Kleinmann, B.\ Kraus, L.\ Lamata, P.J.\ Love, Z.-H.\ Ma, 
         T.\ Moroder,
         D.-K.\ Park, M.B.\ Plenio, N.\ Schuch, C.\ Schwemmer, E.\ Solano, 
         S.\ Szalay, G.\ T{\'o}th, J.\ de Vicente, 
         M.\ Wallquist, M.M.\ Wolf, R.\ Zeier, and Z.\ Zimboras.

This work was funded by the German Research Foundation within 
SPP 1386 (C.E.), by Basque Government grant IT-472-10 
and MINECO grant FIS2012-36673-C03-01 (J.S.). Moreover, the authors thank
J.\ Fabian and K.\ Richter for their invaluable support.

\section*{References}

\bibliographystyle{iopart-num}
\bibliography{biblio_all.bib}

\providecommand{\newblock}{}
\begin{thebibliography}{100}
\expandafter\ifx\csname url\endcsname\relax
  \def\url#1{{\tt #1}}\fi
\expandafter\ifx\csname urlprefix\endcsname\relax\def\urlprefix{URL }\fi
\providecommand{\eprint}[2][]{\url{#2}}

\bibitem{Bell1964}
Bell J~S 1964 {\em Physics\/} {\bf 1} 195--200

\bibitem{Werner1989}
Werner R~F 1989 {\em Phys. Rev. A\/} {\bf 40}(8) 4277--4281

\bibitem{Svetlichny1987}
Svetlichny G 1987 {\em Phys. Rev. D\/} {\bf 35}(10) 3066--3069

\bibitem{GHZ1989}
Greenberger D~M, Horne M~A~H and Zeilinger A 1989 {\em Bell's Theorem, Quantum
  Theory and Conceptions of the Universe\/} (Kluwer Academic, Dordrecht) p~69

\bibitem{Plenio2007}
Plenio M and Virmani S 2007 {\em Quant. Inf. Comp.\/} {\bf 7} 1

\bibitem{Horodecki2009}
Horodecki R, Horodecki P, Horodecki M and Horodecki K 2009 {\em Rev. Mod.
  Phys.\/} {\bf 81}(2) 865--942

\bibitem{Brunner2013}
Brunner N, Cavalcanti D, Pironi S, Scarani V,  and Wehner S 2014 {\em Rev. Mod.
  Phys.\/} {\bf 86} 419

\bibitem{Werner2001}
Werner R~F and Wolf M~M 2001 {\em Quantum Info. Comput.\/} {\bf 1} 1--25 ISSN
  1533-7146

\bibitem{Adesso2007}
Adesso G and Illuminati F 2007 {\em Journal of Physics A: Mathematical and
  Theoretical\/} {\bf 40} 7821

\bibitem{Braunstein2005RMP}
Braunstein S~L and van Loock P 2005 {\em Rev. Mod. Phys.\/} {\bf 77}(2)
  513--577

\bibitem{Tichy2011}
Tichy M~C, Mintert F and Buchleitner A 2011 {\em Journal of Physics B: Atomic,
  Molecular and Optical Physics\/} {\bf 44} 192001

\bibitem{Amico2008}
Amico L, Fazio R, Osterloh A and Vedral V 2008 {\em Rev. Mod. Phys.\/} {\bf
  80}(2) 517--576

\bibitem{Fuentes2012}
Alsing P~M and Fuentes I 2012 {\em Classical and Quantum Gravity\/} {\bf 29}
  224001

\bibitem{Peres2004}
Peres A and Terno D~R 2004 {\em Rev. Mod. Phys.\/} {\bf 76}(1) 93--123

\bibitem{Peres1993}
Peres A 1993 {\em Quantum Theory: Concepts and Methods\/} (Kluwer Academic,
  Dordrecht)

\bibitem{Preskill1998}
Preskill J 1998 Lecture notes on quantum computation
  http://theory.caltech.edu/~preskill/ph229/
  \urlprefix\url{http://theory.caltech.edu/~preskill/ph229/}

\bibitem{Nielsen2000}
Nielsen M~A and Chuang I~L 2000 {\em Quantum Computation and Quanum
  Information\/} (Cambridge University Press) ISBN 0-521-63503-9

\bibitem{MMW}
Wolf M~M 2012 Quantum channels \& operations, a guided tour
  \urlprefix\url{http://www-m5.ma.tum.de/foswiki/pub/M5/Allgemeines/MichaelWol%
f/QChannelLecture.pdf}

\bibitem{Bengtsson2006}
Bengtsson I and {\.Z}yczkowski K 2006 {\em Geometry of Quantum States: An
  Introduction to Quantum Entanglement\/} (Cambridge University Press, New
  York) ISBN 9781139453462

\bibitem{Bennett1996}
Bennett C~H, DiVincenzo D~P, Smolin J~A and Wootters W~K 1996 {\em Phys. Rev.
  A\/} {\bf 54}(5) 3824--3851

\bibitem{Plenio1998}
Plenio M~B and Vedral V 1998 {\em Contemporary Physics\/} {\bf 39} 431--446

\bibitem{Bruss2002}
Bru{\ss} D 2002 {\em Journal of Mathematical Physics\/} {\bf 43} 4237--4251

\bibitem{Lo1999}
Lo H~K and Popescu S 1999 {\em Phys. Rev. Lett.\/} {\bf 83}(7) 1459--1462

\bibitem{DiVincenzo1995}
DiVincenzo D~P 1995 {\em Science\/} {\bf 270} 255--261

\bibitem{Gisin2007}
Gisin~N T~R 2007 {\em Nature Photonics\/} {\bf 1} 165--171

\bibitem{Gisin2002}
Gisin N, Ribordy G, Tittel W and Zbinden H 2002 {\em Rev. Mod. Phys.\/} {\bf
  74}(1) 145--195

\bibitem{Raussendorf2003}
Raussendorf R, Browne D~E and Briegel H~J 2003 {\em Phys. Rev. A\/} {\bf 68}(2)
  022312

\bibitem{Vidal2000}
Vidal G {2000} {\em {Journal of Modern Optics}\/} {\bf {47}} {355--376}

\bibitem{Schlienz1995}
Schlienz J and Mahler G 1995 {\em Phys. Rev. A\/} {\bf 52}(6) 4396--4404

\bibitem{Vedral1997}
Vedral V, Plenio M, Rippin M and Knight P {1997} {\em {Phys. Rev. Lett.}\/}
  {\bf {78}} {2275--2279}

\bibitem{VidalTarrach1999}
Vidal G and Tarrach R 1999 {\em Phys. Rev. A\/} {\bf 59}(1) 141--155

\bibitem{Bennett2000}
Bennett C~H, Popescu S, Rohrlich D, Smolin J~A and Thapliyal A~V 2000 {\em
  Phys. Rev. A\/} {\bf 63}(1) 012307

\bibitem{DVC2000}
D\"ur W, Vidal G and Cirac J~I 2000 {\em Phys. Rev. A\/} {\bf 62}(6) 062314

\bibitem{Verstraete2001}
Verstraete F, Dehaene J and DeMoor B 2001 {\em Phys. Rev. A\/} {\bf 64}(1)
  010101

\bibitem{Bennett1992}
Bennett C~H and Wiesner S~J 1992 {\em Phys. Rev. Lett.\/} {\bf 69}(20)
  2881--2884

\bibitem{Bennett1993}
Bennett C~H, Brassard G, Cr\'epeau C, Jozsa R, Peres A and Wootters W~K 1993
  {\em Phys. Rev. Lett.\/} {\bf 70}(13) 1895--1899

\bibitem{Jozsa1994}
R J and Schumacher B 1994 {\em J. Modern Opt.\/} {\bf 41} 2343

\bibitem{Barnum1996}
Barnum H, Fuchs C~A, Jozsa R and Schumacher B 1996 {\em Phys. Rev. A\/} {\bf
  54}(6) 4707--4711

\bibitem{Bennett1996c}
Bennett C~H, Brassard G, Popescu S, Schumacher B, Smolin J~A and Wootters W~K
  1996 {\em Phys. Rev. Lett.\/} {\bf 76}(5) 722--725

\bibitem{Ji2004}
Ji Z, Duan R and Ying M 2004 {\em Physics Letters A\/} {\bf 330} 418 -- 423
  ISSN 0375-9601

\bibitem{Chitambar2008}
Chitambar E, Duan R and Shi Y 2008 {\em Phys. Rev. Lett.\/} {\bf 101}(14)
  140502

\bibitem{Yu2014}
Yu N, Guo C and Duan R 2014 {\em Phys. Rev. Lett.\/} {\bf 112}(16) 160401

\bibitem{Davies1970}
Davies E and Lewis J 1970 {\em Communications in Mathematical Physics\/} {\bf
  17} 239--260 ISSN 0010-3616

\bibitem{Ziman2008}
Heinosaari T and Ziman M 2008 {\em Acta Physica Slovaca\/} {\bf 58} 487

\bibitem{Hellwig1970}
Hellwig K~E and Kraus K 1970 {\em Communications in Mathematical Physics\/}
  {\bf 16} 142--147 ISSN 0010-3616

\bibitem{Kraus1983}
Kraus K 1983 {\em {States, Effects, and Operations}\/} (Springer-Verlag,
  Berlin)

\bibitem{Gisin1996}
Gisin N 1996 {\em Physics Letters A\/} {\bf 210} 151 -- 156 ISSN 0375-9601

\bibitem{VerstraeteWolf2002}
Verstraete F and Wolf M~M 2002 {\em Phys. Rev. Lett.\/} {\bf 89}(17) 170401

\bibitem{Plenio2005}
Plenio M~B 2005 {\em Phys. Rev. Lett.\/} {\bf 95} 090503

\bibitem{Nielsen1999}
Nielsen M~A 1999 {\em Phys. Rev. Lett.\/} {\bf 83}(2) 436--439

\bibitem{Vidal1999}
Vidal G 1999 {\em Phys. Rev. Lett.\/} {\bf 83}(5) 1046--1049

\bibitem{Vedral1998}
Vedral V and Plenio M~B 1998 {\em Phys. Rev. A\/} {\bf 57}(3) 1619--1633

\bibitem{Uhlmann1998}
Uhlmann A 1998 {\em Open Syst. Inf. Dyn\/} {\bf 5} 209

\bibitem{Uhlmann2010}
Uhlmann A 2010 {\em Entropy\/} {\bf 12} 1799--1832 ISSN 1099-4300

\bibitem{Viehmann2012}
Viehmann O, Eltschka C and Siewert J 2012 {\em Appl. Phys. B\/} {\bf 106} 533

\bibitem{Verstraete2003}
Verstraete F, Dehaene J and De~Moor B {2003} {\em {Phys. Rev. A}\/} {\bf {68}}
  {012103}

\bibitem{Eltschka2012PRA}
Eltschka C, Bastin T, Osterloh A and Siewert J 2012 {\em Phys. Rev. A\/} {\bf
  85} 022301

\bibitem{Osterloh2005}
Osterloh A and Siewert J 2005 {\em Phys. Rev. A\/} {\bf 72} 012337

\bibitem{Gour2013}
Gour G and Wallach N~R 2013 {\em Phys. Rev. Lett.\/} {\bf 111}(6) 060502

\bibitem{Araki1970}
Araki H and Lieb E 1970 {\em Communications in Mathematical Physics\/} {\bf 18}
  160--170 ISSN 0010-3616

\bibitem{Horodecki1996}
Horodecki M, Horodecki P and Horodecki R 1996 {\em Physics Letters A\/} {\bf
  223} 1 -- 8 ISSN 0375-9601

\bibitem{Terhal2000PLA}
Terhal B~M 2000 {\em Physics Letters A\/} {\bf 271} 319 -- 326 ISSN 0375-9601

\bibitem{Lewenstein2000}
Lewenstein M, Kraus B, Cirac J~I and Horodecki P 2000 {\em Phys. Rev. A\/} {\bf
  62}(5) 052310

\bibitem{GuhneTothReview2009}
G{\"u}hne O and T{\'o}th G 2009 {\em Physics Reports\/} {\bf 474} 1 -- 75 ISSN
  0370-1573

\bibitem{Sanpera2001}
Sanpera A, Bru\ss{} D and Lewenstein M 2001 {\em Phys. Rev. A\/} {\bf 63}(5)
  050301

\bibitem{Acin2001}
Ac\'\i{}n A, Bru\ss{} D, Lewenstein M and Sanpera A 2001 {\em Phys. Rev.
  Lett.\/} {\bf 87}(4) 040401

\bibitem{Hulpke2004}
Hulpke F, Bruss D, Lewenstein M and Sanpera A 2004 {\em Quantum Info.
  Comput.\/} {\bf 4} 207--221 ISSN 1533-7146

\bibitem{Eltschka2013}
Eltschka C and Siewert J {2013} {\em {QUANTUM INFORMATION \& COMPUTATION}\/}
  {\bf {13}} {210--220} ISSN {1533-7146}

\bibitem{JianWeiPan8bits}
Yao X~C, Wang T~X, Xu P, Lu H, Pan G~S, Bao X~H, Peng C~Z, Lu C~Y, Chen Y~A and
  Pan J~W 2012 {\em Nat Photon\/} {\bf 6} 225--228 ISSN 1749-4885

\bibitem{Blatt2011}
Monz T, Schindler P, Barreiro J~T, Chwalla M, Nigg D, Coish W~A, Harlander M,
  H\"ansel W, Hennrich M and Blatt R 2011 {\em Phys. Rev. Lett.\/} {\bf
  106}(13) 130506

\bibitem{Blatt2013}
Lanyon B~P, Jurcevic P, Zwerger M, Hempel C, Martinez E~A, D\"ur W, Briegel
  H~J, Blatt R and Roos C~F 2013 {\em Phys. Rev. Lett.\/} {\bf 111}(21) 210501

\bibitem{Horodecki-Jaynes1999}
Horodecki R, Horodecki M and Horodecki P 1999 {\em Phys. Rev. A\/} {\bf 59}(3)
  1799--1803

\bibitem{Audenaert2006}
Audenaert K~M~R and Plenio M~B 2006 {\em New Journal of Physics\/} {\bf 8} 266

\bibitem{Brandao2005}
Brand\~ao F~G~S~L 2005 {\em Phys. Rev. A\/} {\bf 72}(2) 022310

\bibitem{Eisert2007}
Eisert J, ao F~G~S~L~B and Audenaert K~M~R 2007 {\em New Journal of Physics\/}
  {\bf 9} 46

\bibitem{GuehneReimpellWerner2007}
G\"uhne O, Reimpell M and Werner R~F 2007 {\em Phys. Rev. Lett.\/} {\bf 98}(11)
  110502

\bibitem{GuehneReimpellWerner2008}
G\"uhne O, Reimpell M and Werner R~F 2008 {\em Phys. Rev. A\/} {\bf 77}(5)
  052317

\bibitem{Eltschka2012SciRep}
Eltschka C and Siewert J {2012} {\em {Scientific Reports}\/} {\bf {2}} {942}

\bibitem{Schroedinger1936}
Schr{\"o}dinger E 1936 {\em Proc. Camb. Phil. Soc.\/} {\bf 32} 446--452

\bibitem{HJW1993}
Hughston L~P, Jozsa R and Wootters W~K 1993 {\em Physics Letters A\/} {\bf 183}
  14 -- 18 ISSN 0375-9601

\bibitem{Terhal2000PRA}
Terhal B~M and Horodecki P 2000 {\em Phys. Rev. A\/} {\bf 61}(4) 040301

\bibitem{Sperling2011}
Sperling J and Vogel W 2011 {\em Phys. Scr.\/} {\bf 83} 045002

\bibitem{Leinaas2006}
Leinaas J~M, Myrheim J and Ovrum E 2006 {\em Phys. Rev. A\/} {\bf 74}(1) 012313

\bibitem{Eisert2001}
Eisert J and Briegel H~J 2001 {\em Phys. Rev. A\/} {\bf 64}(2) 022306

\bibitem{Gour2005}
Gour G 2005 {\em Phys. Rev. A\/} {\bf 71}(1) 012318

\bibitem{Rungta2001}
Rungta P, Bu\ifmmode~\check{z}\else \v{z}\fi{}ek V, Caves C~M, Hillery M and
  Milburn G~J 2001 {\em Phys. Rev. A\/} {\bf 64}(4) 042315

\bibitem{Rungta2003}
Rungta P and Caves C~M 2003 {\em Phys. Rev. A\/} {\bf 67}(1) 012307

\bibitem{Zyczkowski1998}
\ifmmode~\dot{Z}\else \.{Z}\fi{}yczkowski K, Horodecki P, Sanpera A and
  Lewenstein M 1998 {\em Phys. Rev. A\/} {\bf 58}(2) 883--892

\bibitem{Lee2000}
Lee J, Kim M~S, Park Y~J and Lee S 2000 {\em J. Mod. Opt.\/} {\bf 47} 2151

\bibitem{Eisert2001PhD}
Eisert J 2001 {\em {Entanglement in quantum information theory}\/} Ph.D. thesis
  University of Potsdam

\bibitem{Vidal2002}
Vidal G and Werner R~F 2002 {\em Phys. Rev. A\/} {\bf 65}(3) 032314

\bibitem{Peres1996}
Peres A 1996 {\em Phys. Rev. Lett.\/} {\bf 77}(8) 1413--1415

\bibitem{Horodecki1997}
Horodecki P 1997 {\em Physics Letters A\/} {\bf 232} 333 -- 339 ISSN 0375-9601

\bibitem{Horodecki1998b}
Horodecki M, Horodecki P and Horodecki R 1998 {\em Phys. Rev. Lett.\/} {\bf
  80}(24) 5239--5242

\bibitem{Lee2003}
Lee S, Chi D~P, Oh S~D and Kim J 2003 {\em Phys. Rev. A\/} {\bf 68}(6) 062304

\bibitem{Audenaert2003}
Audenaert K, Plenio M~B and Eisert J 2003 {\em Phys. Rev. Lett.\/} {\bf 90}(2)
  027901

\bibitem{Shimony1995}
Shimony A {1995} {\em {Fundamental problems in quantum theory: A conference
  held in honor of professor John A. Wheeler}\/} ({\em {Annals of the New York
  Academy of Sciences}\/} vol {755}) ed {Greenberger, DM and Zeilinger, A} {New
  York Acad Sci} ({2 EAST 63RD ST, NEW YORK, NY 10021}: {NEW YORK ACAD
  SCIENCES}) pp {675--679} ISBN {0-89766-921-5} ISSN {0077-8923} {Conference on
  Fundamental Problems in Quantum Theory, A Conference held in Honor of
  Professor John A. Wheeler, BALTIMORE, MD, JUN 18-22, 1994}

\bibitem{Barnum2001}
Barnum H and Linden N 2001 {\em Journal of Physics A: Mathematical and
  General\/} {\bf 34} 6787

\bibitem{Audenaert2004}
Audenaert K~M and Braunstein S~L 2004 {\em Communications in Mathematical
  Physics\/} {\bf 246} 443--452 ISSN 0010-3616

\bibitem{Hastings2009}
Hastings M~B 2009 {\em Nat Phys\/} {\bf 5} 255--257 ISSN 1745-2473

\bibitem{Horodecki1998}
Horodecki P, Horodecki R and Horodecki M 1998 {\em Acta Phys. Slov.\/} {\bf 48}
  141--156

\bibitem{Hayden2001}
Hayden P~M, Horodecki M and Terhal B~M 2001 {\em Journal of Physics A:
  Mathematical and General\/} {\bf 34} 6891

\bibitem{Bennett1996b}
Bennett C~H, Bernstein H~J, Popescu S and Schumacher B 1996 {\em Phys. Rev.
  A\/} {\bf 53}(4) 2046--2052

\bibitem{Christandl2004}
Christandl M and Winter A 2004 {\em J. Math. Phys.\/} {\bf 45} 829--840

\bibitem{Vollbrecht2000}
Vollbrecht K~G~H and Werner R~F 2000 {\em Journal of Mathematical Physics\/}
  {\bf 41} 6772--6782

\bibitem{Hill1997}
Hill S and Wootters W~K 1997 {\em Phys. Rev. Lett.\/} {\bf 78}(26) 5022--5025

\bibitem{Wootters1998}
Wootters W~K 1998 {\em Phys. Rev. Lett.\/} {\bf 80}(10) 2245--2248

\bibitem{Albeverio2001}
Albeverio S and Fei S~M 2001 {\em Journal of Optics B: Quantum and
  Semiclassical Optics\/} {\bf 3} 223

\bibitem{Abouraddy2001}
Abouraddy A~F, Saleh B~E~A, Sergienko A~V and Teich M~C 2001 {\em Phys. Rev.
  A\/} {\bf 64}(5) 050101

\bibitem{Verstraete2001JPA}
Verstraete F, Audenaert K, Dehaene J and Moor B~D 2001 {\em Journal of Physics
  A: Mathematical and General\/} {\bf 34} 10327

\bibitem{Wei2003}
Wei T~C and Goldbart P~M 2003 {\em Phys. Rev. A\/} {\bf 68}(4) 042307

\bibitem{Verstraete2002}
Verstraete F, Dehaene J and De~Moor B 2002 {\em Phys. Rev. A\/} {\bf 65}(3)
  032308

\bibitem{Liang2008}
Liang Y~C, Masanes L and Doherty A~C 2008 {\em Phys. Rev. A\/} {\bf 77}(1)
  012332

\bibitem{Audenaert2001}
Audenaert K, Verstraete F and De~Moor B 2001 {\em Phys. Rev. A\/} {\bf 64}(5)
  052304

\bibitem{Badziag2002}
Badziag P, Deuar P, Horodecki M, Horodecki P and Horodecki R 2002 {\em Journal
  of Modern Optics\/} {\bf 49} 1289--1297

\bibitem{Akhtarshenas2005}
Akhtarshenas S~J 2005 {\em Journal of Physics A: Mathematical and General\/}
  {\bf 38} 6777

\bibitem{Li2008}
Li Y~Q and Zhu G~Q 2008 {\em Frontiers of Physics in China\/} {\bf 3} 250--257
  ISSN 1673-3487

\bibitem{Ma2012-QIC}
Ma Z~H, Chen Z~H, Han S, Fei S~M and Severini S 2012 {\em Quantum Info.
  Comput.\/} {\bf 11} 983--988

\bibitem{Eltschka2013-PRL}
Eltschka C and Siewert J 2013 {\em Phys. Rev. Lett.\/} {\bf 111}(10) 100503

\bibitem{Vogel2011}
Sperling J and Vogel W 2011 {\em Phys. Rev. A\/} {\bf 83} 042315

\bibitem{Audenaert2001-JMA}
Verstraete F, Audenaert K, Dehaene J and De~Moor B 2001 {\em Journal of Physics
  A-Mathematical and General\/} {\bf 34} 10327--10332 ISSN 0305-4470

\bibitem{Grudka2004}
Miranowicz A and Grudka A 2004 {\em Journal of Optics B: Quantum and
  Semiclassical Optics\/} {\bf 6} 542

\bibitem{Albeverio2005}
Chen K, Albeverio S and Fei S~M 2005 {\em Phys. Rev. Lett.\/} {\bf 95}(21)
  210501

\bibitem{Breuer2006}
Breuer H~P 2006 {\em Journal of Physics A: Mathematical and General\/} {\bf 39}
  11847

\bibitem{BuchMintert2007}
Mintert F and Buchleitner A 2007 {\em Phys. Rev. Lett.\/} {\bf 98}(14) 140505

\bibitem{Zhang2007}
Zhang C~J, Zhang Y~S, Zhang S and Guo G~C 2007 {\em Phys. Rev. A\/} {\bf 76}(1)
  012334

\bibitem{Liu2009}
Li-Guo L, Cheng-Lin T, Ping-Xing C and Nai-Chang Y 2009 {\em Chinese Physics
  Letters\/} {\bf 26} 060306

\bibitem{Augusiak2009}
Augusiak R and Lewenstein M 2009 {\em Quantum Information Processing\/} {\bf 8}
  493--521 ISSN 1570-0755

\bibitem{Gittsovich2010}
Gittsovich O and G\"uhne O 2010 {\em Phys. Rev. A\/} {\bf 81}(3) 032333

\bibitem{ZhaoFei2011}
Zhao M~J, Zhu X~N, Fei S~M and Li-Jost X 2011 {\em Phys. Rev. A\/} {\bf 84}(6)
  062322

\bibitem{MaGuhne2012}
Chen Z~H, Ma Z~H, G\"uhne O and Severini S 2012 {\em Phys. Rev. Lett.\/} {\bf
  109}(20) 200503

\bibitem{GuhneSeevinck2010}
G{\"u}hne O and Seevinck M 2010 {\em New Journal of Physics\/} {\bf 12} 053002

\bibitem{Huber2010}
Huber M, Mintert F, Gabriel A and Hiesmayr B~C 2010 {\em Phys. Rev. Lett.\/}
  {\bf 104}(21) 210501

\bibitem{MaHuber2011}
Ma Z~H, Chen Z~H, Chen J~L, Spengler C, Gabriel A and Huber M 2011 {\em Phys.
  Rev. A\/} {\bf 83}(6) 062325

\bibitem{WuHuber2012}
Wu J~Y, Kampermann H, Bru\ss{} D, Kl\"ockl C and Huber M 2012 {\em Phys. Rev.
  A\/} {\bf 86}(2) 022319

\bibitem{Huber2013-PRA}
Huber M, Perarnau-Llobet M and de~Vicente J~I 2013 {\em Phys. Rev. A\/} {\bf
  88}(4) 042328

\bibitem{Horodecki1999}
Horodecki M and Horodecki P 1999 {\em Phys. Rev. A\/} {\bf 59}(6) 4206--4216

\bibitem{Vollbrecht2001}
Vollbrecht K~G~H and Werner R~F 2001 {\em Phys. Rev. A\/} {\bf 64}(6) 062307

\bibitem{Osborne2005}
Osborne T~J 2005 {\em Phys. Rev. A\/} {\bf 72}(2) 022309

\bibitem{HarrowNielsen2003}
Harrow A~W and Nielsen M~A 2003 {\em Phys. Rev. A\/} {\bf 68}(1) 012308

\bibitem{Rungta2001Book}
Rungta P, Munro W, Nemoto K, Deuar P, Milburn G and Caves C 2001 {\em
  Directions in Quantum Optics\/} ({\em Lecture Notes in Physics\/} vol 561) ed
  Carmichael H~J, Glauber R~J and Scully M~O (Springer Berlin Heidelberg) pp
  149--164

\bibitem{VollbrechtTerhal2000}
Terhal B~M and Vollbrecht K~G~H 2000 {\em Phys. Rev. Lett.\/} {\bf 85}(12)
  2625--2628

\bibitem{Lang1987}
Lang S 1987 {\em Linear Algebra\/} Undergraduate texts in mathematics (New
  York: Springer)

\bibitem{ManneCaves2008}
Manne K~K and Caves C~M 2008 {\em Quantum Info. Comput.\/} {\bf 8} 295--310
  ISSN 1533-7146

\bibitem{ES-unpublished}
Eltschka C and Siewert J Unpublished

\bibitem{Coffman2000}
Coffman V, Kundu J and Wootters W~K 2000 {\em Phys. Rev. A\/} {\bf 61}(5)
  052306

\bibitem{MeyerWallach2002}
Meyer D~A and Wallach N~R 2002 {\em Journal of Mathematical Physics\/} {\bf 43}
  4273--4278

\bibitem{Brennen2003}
Brennen G~K 2003 {\em Quantum Info. Comput.\/} {\bf 3} 619--626 ISSN 1533-7146

\bibitem{Emary2004}
Emary C 2004 {\em Journal of Physics A: Mathematical and General\/} {\bf 37}
  8293

\bibitem{Osterloh2002}
Osterloh A, Amico L, Falci G and Fazio R 2002 {\em Nature\/} {\bf 416} 608--610

\bibitem{Osborne2002}
Osborne T~J and Nielsen M~A 2002 {\em Phys. Rev. A\/} {\bf 66}(3) 032110

\bibitem{Osborne2006}
Osborne T~J and Verstraete F 2006 {\em Phys. Rev. Lett.\/} {\bf 96}(22) 220503

\bibitem{Ou2007}
Ou Y~C and Fan H 2007 {\em Phys. Rev. A\/} {\bf 75}(6) 062308

\bibitem{Fei-Monog2008}
Ou Y~C, Fan H and Fei S~M 2008 {\em Phys. Rev. A\/} {\bf 78} 012311

\bibitem{Oliveira2014}
de~Oliveira T~R, Cornelio M~F and Fanchini F~F 2014 {\em Phys. Rev. A\/} {\bf
  89} 034303

\bibitem{Bai2014}
Bai Y~K, Xu Y~F and Wang Z~D 2014 {\em Phys. Rev. Lett.\/} {\bf 113} 100503

\bibitem{Laustsen2003}
Laustsen T, Verstraete F and van Enk S~J 2003 {\em Quantum Information and
  Computation\/} {\bf 3} 64

\bibitem{VerstraetePopp2004}
Verstraete F, Popp M and Cirac J~I 2004 {\em Phys. Rev. Lett.\/} {\bf 92}(2)
  027901

\bibitem{Popp2005}
Popp M, Verstraete F, Mart{\'\i}n-Delgado M~A and Cirac J~I 2005 {\em Phys.
  Rev. A\/} {\bf 71}(4) 042306

\bibitem{Gour2006}
Gour G 2006 {\em Phys. Rev. A\/} {\bf 74}(5) 052307

\bibitem{PopeMilburn2003}
Pope D~T and Milburn G~J 2003 {\em Phys. Rev. A\/} {\bf 67}(5) 052107

\bibitem{Scott2004}
Scott A~J 2004 {\em Phys. Rev. A\/} {\bf 69} 052330

\bibitem{Love2007}
Love P, Brink A, Smirnov A, Amin M, Grajcar M, Il’ichev E, Izmalkov A and
  Zagoskin A 2007 {\em Quantum Information Processing\/} {\bf 6} 187--195 ISSN
  1570-0755

\bibitem{Facchi2006}
Facchi P, Florio G and Pascazio S 2006 {\em Phys. Rev. A\/} {\bf 74} 042331

\bibitem{Parisi2010}
Facchi P, Florio G, Marzolino U, Parisi G and Pascazio S 2010 {\em Journal of
  Physics A: Mathematical and Theoretical\/} {\bf 43} 225303

\bibitem{Gao2012}
Hong Y, Gao T and Yan F 2012 {\em Phys. Rev. A\/} {\bf 86} 062323

\bibitem{Huber2012-PRA}
Huber M, Erker P, Schimpf H, Gabriel A and Hiesmayr B 2011 {\em Phys. Rev. A\/}
  {\bf 83}(4) 040301

\bibitem{HuberEberly2012}
Hashemi~Rafsanjani S~M, Huber M, Broadbent C~J and Eberly J~H 2012 {\em Phys.
  Rev. A\/} {\bf 86}(6) 062303

\bibitem{vEnk2014}
Gao T, Yan F and van Enk S~J 2014 {\em Phys. Rev. Lett.\/} {\bf 112} 180501

\bibitem{Zyczkowski2011}
Rudnicki L, Horodecki P and \ifmmode~\dot{Z}\else \.{Z}\fi{}yczkowski K 2011
  {\em Phys. Rev. Lett.\/} {\bf 107} 150502

\bibitem{Jungnitsch2011}
Jungnitsch B, Moroder T and G\"uhne O 2011 {\em Phys. Rev. Lett.\/} {\bf
  106}(19) 190502

\bibitem{Hofmann2014}
Hofmann M, Moroder T and G{\"u}hne O 2014 {\em Journal of Physics A:
  Mathematical and Theoretical\/} {\bf 47} 155301

\bibitem{Boyd1996}
Vandenberghe L and Boyd S 1996 {\em SIAM Review\/} {\bf 38} 49--95

\bibitem{Briegel2001}
Briegel H~J and Raussendorf R 2001 {\em Phys. Rev. Lett.\/} {\bf 86}(5)
  910--913

\bibitem{Schlienz1996}
Schlienz J and Mahler G 1996 {\em Physics Letters A\/} {\bf 224} 39 -- 44 ISSN
  0375-9601

\bibitem{Grassl1998}
Grassl M, R\"otteler M and Beth T 1998 {\em Phys. Rev. A\/} {\bf 58}(3)
  1833--1839

\bibitem{Linden1998}
Linden N and Popescu S 1998 {\em Fortschritte der Physik\/} {\bf 46} 567--578
  ISSN 1521-3978

\bibitem{Linden1999}
Linden N, Popescu S and Sudbery A 1999 {\em Phys. Rev. Lett.\/} {\bf 83}(2)
  243--247

\bibitem{Carteret1999}
Carteret H, Linden N, Popescu S and Sudbery A 1999 {\em Foundations of
  Physics\/} {\bf 29} 527--552 ISSN 0015-9018

\bibitem{Carteret2000}
Carteret H~A and Sudbery A 2000 {\em Journal of Physics A: Mathematical and
  General\/} {\bf 33} 4981

\bibitem{Rains2000}
Rains E 2000 {\em IEEE Trans. on Information Theory\/} {\bf 46}(1) 54--59

\bibitem{Sudbery2001}
Sudbery A 2001 {\em Journal of Physics A: Mathematical and General\/} {\bf 34}
  643

\bibitem{Kus2001}
Ku\ifmmode~\acute{s}\else \'{s}\fi{} M and \ifmmode~\dot{Z}\else
  \.{Z}\fi{}yczkowski K 2001 {\em Phys. Rev. A\/} {\bf 63}(3) 032307

\bibitem{Brylinski2002a}
Brylinski J~L 2002 {\em Algebraic measures of entanglement\/} (Chapman \&
  Hall/CRC Press, Boca Raton, FL) chap~1 "Mathematics of Quantum Computation"

\bibitem{Brylinski2002b}
Brylinski J~L and Brylinski R 2002 {\em Invariant polynomial functions on k
  qudits\/} (Chapman \& Hall/CRC Press, Boca Raton, FL) chap~11 "Mathematics of
  Quantum Computation"

\bibitem{Klyachko2002}
Klyachko A 2002 Coherent states, entanglement, and geometric invariant theory
  eprint arXiv:quant-ph/0206012

\bibitem{Kraus2010}
Kraus B 2010 {\em Phys. Rev. Lett.\/} {\bf 104}(2) 020504

\bibitem{deVicente2013}
de~Vicente J~I, Spee C and Kraus B 2013 {\em Phys. Rev. Lett.\/} {\bf 111}(11)
  110502

\bibitem{Johansson2012}
Johansson M, Ericsson M, Singh K, Sj\"oqvist E and Williamson M~S 2012 {\em
  Phys. Rev. A\/} {\bf 85}(3) 032112

\bibitem{Johansson2013}
Johansson M, Khoury A~Z, Singh K and Sj\"oqvist E 2013 {\em Phys. Rev. A\/}
  {\bf 87}(4) 042112

\bibitem{Johansson2014}
Johansson M, Ericsson M, Sj\"oqvist E and Osterloh A 2014 {\em Phys. Rev. A\/}
  {\bf 89}(1) 012320

\bibitem{Wu2001}
Wu S and Zhang Y 2000 {\em Phys. Rev. A\/} {\bf 63}(1) 012308

\bibitem{Hein2006}
Hein M, D{\"u}r W, Eisert J, Raussendorf R, den Nest M~V and Briegel H~J 2006
  {\em Entanglement in graph states and its applications\/} ({\em Proceedings
  of the International School of Physics "Enrico Fermi"\/} vol 162) (IOS Press)
  pp 115--218

\bibitem{Gour2010}
Gour G 2010 {\em Phys. Rev. Lett.\/} {\bf 105}(19) 190504

\bibitem{Miyake2003}
Miyake A 2003 {\em Phys. Rev. A\/} {\bf 67}(1) 012108

\bibitem{Cayley1845}
Cayley A 1845 {\em Cambridge Math.\ J.\/} {\bf 4} 193

\bibitem{Lohmayer2006}
Lohmayer R, Osterloh A, Siewert J and Uhlmann A 2006 {\em Phys. Rev. Lett.\/}
  {\bf 97}(26) 260502

\bibitem{Luque2003}
Luque J~G and Thibon J~Y 2003 {\em Phys. Rev. A\/} {\bf 67}(4) 042303

\bibitem{Dokovic2009}
Djokovi{\'c} D~Z and Osterloh A 2009 {\em Journal of Mathematical Physics\/}
  {\bf 50} 033509

\bibitem{Wong2001}
Wong A and Christensen N 2001 {\em Phys. Rev. A\/} {\bf 63}(4) 044301

\bibitem{Uhlmann2000}
Uhlmann A 2000 {\em Phys. Rev. A\/} {\bf 62}(3) 032307

\bibitem{Osterloh2006}
Osterloh A and Siewert J 2006 {\em International Journal of Quantum
  Information\/} {\bf 04} 531--540

\bibitem{Djokovic2013}
Chen L and Djokovi\'{c} D~i~c~v 2013 {\em Phys. Rev. A\/} {\bf 88}(4) 042307

\bibitem{Cayley1846}
Cayley A 1846 {\em Cambridge and Dublin Math.\ J.\/} {\bf 1} 104--122

\bibitem{Osterloh2012}
Osterloh A and Siewert J 2012 {\em Phys. Rev. A\/} {\bf 86}(4) 042302

\bibitem{Yu2007}
Yu C~s and Song H~s 2007 {\em Phys. Rev. A\/} {\bf 76}(2) 022324

\bibitem{vanEnk2007}
van Enk S~J, L\"utkenhaus N and Kimble H~J 2007 {\em Phys. Rev. A\/} {\bf
  75}(5) 052318

\bibitem{Luque2003-170}
Briand E, Luque J~G and Thibon J~Y 2003 {\em Journal of Physics A: Mathematical
  and General\/} {\bf 36} 9915

\bibitem{Luque2006}
Luque J~G and Thibon J~Y 2006 {\em Journal of Physics A: Mathematical and
  General\/} {\bf 39} 371

\bibitem{Levay2006}
L{\'e}vay P 2006 {\em Journal of Physics A: Mathematical and General\/} {\bf
  39} 9533

\bibitem{Jarvis2007}
King R~C, Welsh T~A and Jarvis P~D 2007 {\em Journal of Physics A: Mathematical
  and Theoretical\/} {\bf 40} 10083

\bibitem{Heydari2008}
Heydari H 2008 {\em Quantum Information Processing\/} {\bf 7} 43--50 ISSN
  1570-0755

\bibitem{Sharma2012a}
Sharma S and Sharma N 2012 {\em Quantum Information Processing\/} {\bf 11}
  1695--1714 ISSN 1570-0755

\bibitem{Sharma2012}
Sharma S~S and Sharma N~K 2012 {\em Phys. Rev. A\/} {\bf 85}(4) 042315

\bibitem{Sharma2013}
Sharma S~S and Sharma N~K 2013 {\em Phys. Rev. A\/} {\bf 87}(2) 022335

\bibitem{BriandVerstraete2004}
Briand E, Luque J~G, Thibon J~Y and Verstraete F 2004 {\em Journal of
  Mathematical Physics\/} {\bf 45} 4855--4867

\bibitem{Osterloh2013}
Osterloh A 2013 Eprint arXiv:1309.6235

\bibitem{Jarvis2014}
Jarvis P~D 2014 {\em Journal of Physics A: Mathematical and Theoretical\/} {\bf
  47} 215302

\bibitem{Hioe1981}
Hioe F~T and Eberly J~H 1981 {\em Phys. Rev. Lett.\/} {\bf 47}(12) 838--841

\bibitem{Brukner2002}
\ifmmode~\dot{Z}\else \.{Z}\fi{}ukowski M and Brukner C 2002 {\em Phys. Rev.
  Lett.\/} {\bf 88}(21) 210401

\bibitem{Teodorescu2003}
Teodorescu-Frumosu M and Jaeger G 2003 {\em Phys. Rev. A\/} {\bf 67}(5) 052305

\bibitem{Acin2000}
Ac\'\i{}n A, Andrianov A, Costa L, Jan\'e E, Latorre J~I and Tarrach R 2000
  {\em Phys. Rev. Lett.\/} {\bf 85}(7) 1560--1563

\bibitem{Kempe1999}
Kempe J 1999 {\em Phys. Rev. A\/} {\bf 60}(2) 910--916

\bibitem{Osterloh2010}
Osterloh A 2010 {\em Applied Physics B\/} {\bf 98} 609--616 ISSN 0946-2171

\bibitem{Seevinck2008}
Seevinck M and Uffink J 2008 {\em Phys. Rev. A\/} {\bf 78}(3) 032101

\bibitem{Szalay2012}
Szalay S and K\"ok\'enyesi Z 2012 {\em Phys. Rev. A\/} {\bf 86}(3) 032341

\bibitem{Bennett1999}
Bennett C~H, DiVincenzo D~P, Mor T, Shor P~W, Smolin J~A and Terhal B~M 1999
  {\em Phys. Rev. Lett.\/} {\bf 82}(26) 5385--5388

\bibitem{Osterloh2008}
Osterloh A, Siewert J and Uhlmann A 2008 {\em Phys. Rev. A\/} {\bf 77}(3)
  032310

\bibitem{Eltschka2008}
Eltschka C, Osterloh A, Siewert J and Uhlmann A 2008 {\em New Journal of
  Physics\/} {\bf 10} 043014

\bibitem{Jung2009}
Jung E, Hwang M~R, Park D and Son J~W 2009 {\em Phys. Rev. A\/} {\bf 79}(2)
  024306

\bibitem{JungParkSon2009}
Jung E, Park D and Son J~W 2009 {\em Phys. Rev. A\/} {\bf 80}(1) 010301

\bibitem{He2011}
Shu-Juan H, Xiao-Hong W, Shao-Ming F, Hong-Xiang S and Qiao-Yan W 2011 {\em
  Communications in Theoretical Physics\/} {\bf 55} 251

\bibitem{Viehmann2011}
Viehmann O, Eltschka C and Siewert J 2011 {\em Phys. Rev. A\/} {\bf 83}(5)
  052330

\bibitem{Eggeling2001}
Eggeling T and Werner R~F 2001 {\em Phys. Rev. A\/} {\bf 63} 042111

\bibitem{Eltschka2012PRL}
Eltschka C and Siewert J 2012 {\em Phys. Rev. Lett.\/} {\bf 108}(2) 020502

\bibitem{Siewert2012}
Siewert J and Eltschka C 2012 {\em Phys. Rev. Lett.\/} {\bf 108}(23) 230502

\bibitem{Loss2008}
R\"othlisberger B, Lehmann J, Saraga D~S, Traber P and Loss D 2008 {\em Phys.
  Rev. Lett.\/} {\bf 100}(10) 100502

\bibitem{Cao2010}
Cao K, Zhou Z~W, Guo G~C and He L 2010 {\em Phys. Rev. A\/} {\bf 81}(3) 034302

\bibitem{Rodriques2013}
Rodriques S, Datta N and Love P 2014 {\em Phys. Rev. A\/} {\bf 90}(1) 012340

\bibitem{Hyllus2010}
Osterloh A and Hyllus P 2010 {\em Phys. Rev. A\/} {\bf 81}(2) 022307

\bibitem{Eltschka2014}
Eltschka C and Siewert J 2014 {\em Phys. Rev. A\/} {\bf 89}(2) 022312

\bibitem{Haeffner2005}
Haffner H, Hansel W, Roos C~F, Benhelm J, Chek-al kar D, Chwalla M, Korber T,
  Rapol U~D, Riebe M, Schmidt P~O, Becher C, Guhne O, Dur W and Blatt R 2005
  {\em Nature\/} {\bf 438} 643

\bibitem{Kiesel2007}
Kiesel N, Schmid C, T\'oth G, Solano E and Weinfurter H 2007 {\em Phys. Rev.
  Lett.\/} {\bf 98}(6) 063604

\bibitem{Lewenstein1998}
Lewenstein M and Sanpera A 1998 {\em Phys. Rev. Lett.\/} {\bf 80}(11)
  2261--2264

\bibitem{Gisin1998}
Gisin N and Bechmann-Pasquinucci H 1998 {\em Physics Letters A\/} {\bf 246}
  1--6

\bibitem{Braunstein2005}
Brown I~D~K, Stepney S, Sudbery A and Braunstein S~L 2005 {\em Journal of
  Physics A: Mathematical and General\/} {\bf 38} 1119

\bibitem{Osterloh2010NJP}
Osterloh A and Siewert J 2010 {\em New Journal of Physics\/} {\bf 12} 075025

\bibitem{Gour2010b}
Gour G and Wallach N~R 2010 {\em Journal of Mathematical Physics\/} {\bf 51}
  112201

\bibitem{Helwig2012}
Helwig W, Cui W, Latorre J~I, Riera A and Lo H~K 2012 {\em Phys. Rev. A\/} {\bf
  86} 052335

\bibitem{Kraus2012}
de~Vicente J~I, Carle T, Streitberger C and Kraus B 2012 {\em Phys. Rev.
  Lett.\/} {\bf 108} 060501

\bibitem{Verstraete20024x9}
Verstraete F, Dehaene J, De~Moor B and Verschelde H 2002 {\em Phys. Rev. A\/}
  {\bf 65}(5) 052112

\bibitem{Chterental2007}
Chterental O and Djokovic D~Z 2007 {\em Normal Forms and Tensor Ranks of Pure
  States of Four Qubits\/} (Nova Science Publishers New York) chap~4 Linear
  Algebra Research Advances

\bibitem{Miyake2004}
Miyake A and Verstraete F 2004 {\em Phys. Rev. A\/} {\bf 69}(1) 012101

\bibitem{Mandilara2005}
Mandilara A, Akulin V~M, Smilga A~V and Viola L 2006 {\em Phys. Rev. A\/} {\bf
  74}(2) 022331

\bibitem{Lamata2006}
Lamata L, Le\'on J, Salgado D and Solano E 2006 {\em Phys. Rev. A\/} {\bf
  74}(5) 052336

\bibitem{Lamata2007}
Lamata L, Le\'on J, Salgado D and Solano E 2007 {\em Phys. Rev. A\/} {\bf
  75}(2) 022318

\bibitem{Bastin2009}
Bastin T, Krins S, Mathonet P, Godefroid M, Lamata L and Solano E 2009 {\em
  Phys. Rev. Lett.\/} {\bf 103}(7) 070503

\bibitem{Bastin2010}
Mathonet P, Krins S, Godefroid M, Lamata L, Solano E and Bastin T 2010 {\em
  Phys. Rev. A\/} {\bf 81}(5) 052315

\bibitem{Borsten2010}
Borsten L, Dahanayake D, Duff M~J, Marrani A and Rubens W 2010 {\em Phys. Rev.
  Lett.\/} {\bf 105}(10) 100507

\bibitem{Li2012}
Li X and Li D 2012 {\em Phys. Rev. Lett.\/} {\bf 108}(18) 180502

\bibitem{Huber2013-PRL}
Huber M and de~Vicente J~I 2013 {\em Phys. Rev. Lett.\/} {\bf 110}(3) 030501

\bibitem{Walter2013}
Walter M, Doran B, Gross D and Christandl M 2013 {\em Science\/} {\bf 340} 1205

\bibitem{Martin2010}
Martin J, Giraud O, Braun P~A, Braun D and Bastin T 2010 {\em Phys. Rev. A\/}
  {\bf 81} 062347

\bibitem{Toth2007}
T\'{o}th G 2007 {\em J. Opt. Soc. Am. B\/} {\bf 24} 275--282

\bibitem{Holweck2012}
Holweck F, Luque J~G and Thibon J~Y 2012 {\em Journal of Mathematical
  Physics\/} {\bf 53} 102203

\bibitem{Borsten2013}
Borsten L 2013 {\em Journal of Physics A: Mathematical and Theoretical\/} {\bf
  46} 455303

\bibitem{Holweck2014}
Holweck F, Luque J~G and Thibon J~Y 2014 {\em Journal of Mathematical
  Physics\/} {\bf 55} 012202

\bibitem{Eltschka2009}
Eltschka C, Osterloh A and Siewert J 2009 {\em Phys. Rev. A\/} {\bf 80}(3)
  032313

\end{thebibliography}


\end{document}